\documentclass[final,11pt]{article}
\pdfoutput=1
\usepackage{MRnotes}

\newcommand{\slt}{\mathfrak{sl}(2)}
\newcommand{\sut}{\mathfrak{su}(2)}

\newcommand{\ts}{\mathfrak{t}}
\newcommand{\tsb}{\overline{\mathfrak{t}}}
\newcommand{\zs}{\mathfrak{z}}
\newcommand{\zsb}{\overline{\mathfrak{z}}}
\newcommand{\us}{\mathfrak{u}}
\newcommand{\usb}{\overline{\mathfrak{u}}}
\newcommand{\vs}{\mathfrak{v}}

\newcommand{\qb}{\overline{q}}

\newcommand{\Kcft}{K_{_\text{CFT}}}
\newcommand{\Tcft}{T_{_\text{CFT}}}
\newcommand{\Ls}{L^{\text{\tiny{CFT}}}}
\newcommand{\Lsb}{\overline{L}^\text{\,\tiny{CFT}}}
\newcommand{\Js}{J^{\text{\tiny{CFT}}}}
\newcommand{\Gs}{G^{\text{\tiny{CFT}}}}

\newcommand{\hx}{h_{\text{int}}}
\newcommand{\hxb}{\bar{h}_{\text{int}}}
\newcommand{\hs}{h_{_\text{CFT}}}
\newcommand{\hsb}{\bar{h}_{_\text{CFT}}}
\newcommand{\js}{j_{_\text{CFT}}}

\newcommand{\ks}{\mathsf{k}}

\newcommand{\asx}[1]{$\mathrm{AdS}_3 \times \mathbf{S}^3 \times #1$}
\newcommand{\schol}{\bm{\chi}_{_\text{g}}}
\newcommand{\scholb}{\overline{\bm{\chi}}_{_\text{g}}}
\newcommand{\etchar}{\bm{\chi}_{_\text{NS}}}

\newcommand{\sccft}{\bm{\xi}}
\newcommand{\Nwx}{\mathsf{N}^{^\text{int}}_{w}}

\renewcommand{\=}{\; = \;}
\renewcommand{\order}[1]{\text{O}(#1)}

\title{Strings in \texorpdfstring{AdS$_3$}{AdS3}: one-loop partition function and near-extremal BTZ thermodynamics}
\author{Christian Ferko${}^a$, Sameer Murthy${}^{b,c}$, Mukund Rangamani$^{a}$}

\affiliation[a]{
	Center for Quantum Mathematics and Physics (QMAP)\\
	Department of Physics \& Astronomy, University of California, Davis, CA 95616 USA}
\affiliation[b]{
	Department of Mathematics, King’s College London, The Strand, London WC2R 2LS, UK}
\affiliation[c]{
	School of Natural Sciences, Institute for Advanced Study, Princeton, NJ, USA}

\emailAdd{caferko@ucdavis.edu}
\emailAdd{sameer.murthy@kcl.ac.uk}
\emailAdd{mukund@physics.ucdavis.edu}

\abstract{
We revisit the computation of the string partition function in \AdS{3} focussing on the appearance of spacetime (super) symmetries.
We show how the asymptotic symmetries of the \AdS{3} spacetime, which generate the boundary (super) Virasoro currents, are captured by the one-loop partition sum.
We use this to argue that the recent understanding of near-extremal black hole thermodynamics based on the gravitational path integral continues to hold for finite string length.
Along the way we clarify some aspects of the AdS$_3$/CFT$_2$ duality and, in particular,
deduce which bulk gauge fields lead to boundary currents.
We also explain how one can interpolate between supersymmetric and thermal (Atick-Witten) fermion boundary conditions in the target space by suitably tuning rotational chemical potentials in the string partition function.
}

\begin{document}
\maketitle


\section{Introduction}\label{sec:intro}

In this paper we study the one-loop partition function, i.e., the torus amplitude, of
(super) strings in~\AdS{3} backgrounds. This is a well-studied subject, and the elements of string perturbation theory in \AdS{3} based on the~$\slt$ WZW theory on the worldsheet~\cite{Maldacena:2000hw,Maldacena:2000kv,Maldacena:2001km} have led to many applications.
Our motivations to revisit this subject are threefold:
\begin{enumerate}[left=0pt]
\item[(i)] Investigations of near-extremal black holes have shown the existence of nearly-gapless (Schwarzian) modes in
the gravitational spectrum near the horizon~\cite{Ghosh:2019rcj,Iliesiu:2020qvm}.
It is natural to ask whether such modes are seen in the corresponding spectrum of string theory.
The string partition function in \AdS{3} allows us to probe these modes in BTZ black holes.
\item[(ii)] Different aspects of perturbative superstring theory and supergravity in \asx{T^4/K3}
have been emphasized in various studies over the years, including the subtleties of the GSO projection~\cite{Giveon:1998ns},
the periodicity conditions of~\cite{Atick:1988si} in the thermal theory,
the spacetime symmetry superalgebra~\cite{Giveon:1998ns, deBoer:1998gyt,Kutasov:1999xu},
and the low-energy effective supergravity~\cite{Deger:1998nm,Larsen:1998xm,Kutasov:1998zh}.
Our goal is to demonstrate how all these aspects relate to the one-loop partition function (cf.~\cite{Israel:2003ry,Raju:2007uj,Ashok:2020dnc,Dabholkar:2023tzd} for earlier work) and recover the BPS spectrum discussed in~\cite{Kutasov:1998zh,Dabholkar:2007ey}.
\item[(iii)] There has been recent progress in understanding subtle features of the spectrum of the CFT$_2$ dual to \asx{T^4}~\cite{Aharony:2024fid}. We show how some of these features can be understood from the string worldsheet perspective.
\end{enumerate}
As we see below, these three motivations are tied together by the
appearance, in the spectrum of worldsheet excitations, of currents corresponding to the spacetime Virasoro symmetry
and its  generalizations. In the rest of the introduction we expand on the above motivations, focussing on the predictions for the \AdS{3} string spectrum from the point of view of gapless modes near extremal black holes, and from the point of view of spacetime symmetries.

\bigskip

\noindent {\bf Strings in near-extremal black hole backgrounds}

\smallskip

The conventional picture of black hole thermodynamics is based on the
semiclassical evaluation of the black hole entropy.
In the extremal limit of a non-supersymmetric black hole, this naively leads to a degeneracy of states that grows exponentially with the mass of the black hole, even at zero temperature.
However, this conclusion is puzzling:
not only is there no reason to expect macroscopic degeneracy in a system
without a symmetry constraint, but  there is also a question as to how black holes
very close to extremality  would Hawking-radiate~\cite{Preskill:1991tb}.
A more recent understanding of this situation paints a more prosaic picture:
the non-trivial degeneracy is illusory, and in fact near-extremal black holes have a vanishing degeneracy in the full quantum theory~\cite{Iliesiu:2020qvm, Iliesiu:2020zld,Heydeman:2020hhw,Iliesiu:2022onk}.
In effect, they  behave like a conventional quantum mechanical system with a few low-lying excitations.

The essential point is that the aforementioned semiclassical analysis receives large quantum corrections
due to nearly-zero modes localized in the near-horizon region of the near-extremal geometry.
These nearly-zero modes are governed by the Schwarzian action that appears in the effective nearly-\AdS{2}
geometry~\cite{Almheiri:2014cka,Sachdev:2015efa,Maldacena:2016upp,Almheiri:2016fws,Nayak:2018qej,Moitra:2018jqs,Sachdev:2019bjn},
and the result of integrating over them is to
dramatically reduce the density of states at very low
temperatures.\footnote{The conclusion
is different for supersymmetric black holes, for which the zero-temperature degeneracy equals the exponentially large microscopic index~\cite{Iliesiu:2022kny}.}
As the temperature~$T \to 0$, the partition function of the black hole scales as~$T^{\frac{3}{2}}\, e^{S_0}$, where $S_0$ is the semiclassical entropy of the extremal black hole.

These calculations of the gravitational functional integral constitute a robust prediction for any sensible theory of quantum gravity.
A natural question is whether these results can be verified explicitly in string theory.
One of the goals of this paper is to show that the one-loop string partition function reproduces the above low-temperature result.
Note that, although the black hole geometry is a saddle point of the gravitational path integral by virtue of the fact that it solves the gravitational equations of motion,
the leading Bekenstein-Hawking entropy obtained as the saddle point value of the gravitational action is yet to be reliably reproduced from a string worldsheet computation.
The reason is that the string partition function at genus-$0$ (sphere) with no insertions needs to
be regulated in non-compact target spacetimes, and the associated boundary terms are not properly understood in string theory.
For the question at hand, however, we need the 1-loop partition function, which arises from a genus-$1$ (torus)
worldsheet.  This part of the result is well-defined and technically within reach, and forms the focus of our analysis.

The BTZ black hole in asymptotically \AdS{3} spacetime~\cite{Banados:1992wn} provides a particularly good laboratory to investigate such questions, for two known reasons:
firstly, string theory on \AdS{3} with pure NS-NS flux is an exactly solvable sigma model for string perturbation theory~\cite{Giveon:1998ns,Maldacena:2000hw} and, secondly, string perturbation theory around the BTZ background exactly maps to string
perturbation theory on thermal \AdS{3} space.
The second statement arises from the fact that the Euclidean BTZ and thermal \AdS{3} both correspond to the same three-dimensional Euclidean geometry (the hyperbolic solid torus), and the two physical theories differ only by a different identification of the time and space circles.

The map relating \AdS{3} and the BTZ black hole is simply the S-modular transformation in the holographic dual CFT$_2$ that exchanges the two cycles of the boundary torus.
By applying this modular transformation to the partition function of the CFT$_2$, the authors of~\cite{Ghosh:2019rcj} showed that any two-dimensional CFT with a twist gap has a universal low-temperature, high spin-partition function that scales as $T^{\frac{3}{2}}\, e^{S_0}$ as consistent with the behavior of the black hole at low temperatures.
The essence of the argument of~\cite{Ghosh:2019rcj} is that the lowest excitation levels of any CFT$_2$ are captured by the vacuum Virasoro character, whose modular transform precisely reproduces the desired low-temperature behavior. The twist gap is then used to parametrically separate the excited states from the vacuum character.\footnote{This statement can be robustly established  using light-cone modular bootstrap techniques~\cite{Pal:2023cgk}.}

Extending the above logic into the bulk theory, we see that
the  effect of Schwarzian modes in the BTZ background can be reproduced by obtaining the vacuum Virasoro character (and a twist gap) in the one-loop determinant of the excitations around thermal \AdS{3}. In semiclassical gravitational calculations, the vacuum Virasoro character is precisely reproduced by the contribution of the boundary gravitons in \AdS{3}~\cite{Maloney:2007ud,Giombi:2008vd}.
In this paper, we isolate these modes in the one-loop string amplitude.

\bigskip

\noindent {\bf Spacetime symmetries of string theory on \AdS{3}}

\smallskip

Independent of their relevance for the low-temperature analysis for BTZ black holes, explaining the origin of the boundary
gravitons---and the closely related asymptotic symmetries---in the \AdS{3} worldsheet functional integral is an interesting problem.\footnote{For an interesting analysis of asymptotic symmetries from the worldsheet, see~\cite{Kraus:2002cb}.} Recall that semiclassical gravity in \AdS{3} has an enhanced asymptotic symmetry group~\cite{Brown:1986nw}:
the $\mathrm{SL}(2,\mathbb{R}) \times\mathrm{SL}(2,\mathbb{R})$ isometry group of \AdS{3} extends to two copies of Virasoro algebra with central charge $c=\frac{3\,\lads}{2\,G_N}$.  One important consequence of the extended Virasoro symmetry in~\AdS{3} is the existence of non-trivial modes---the boundary gravitons---which do not propagate as local excitations in the bulk, but are nevertheless present in the spectrum of excitations.
These modes are clearly visible in the semiclassical fluctuation analysis around the \AdS{3} background~\cite{Maloney:2007ud,Giombi:2008vd}.
In particular, the one-loop gravitational partition function of Einstein-Hilbert gravity without any matter (pure 3d gravity) reproduces the vacuum character of the Virasoro algebra, precisely as expected from the boundary CFT$_2$.

The generators of the asymptotic Virasoro symmetry of \AdS{3} spacetimes have been constructed from the string worldsheet in~\cite{Giveon:1998ns} (see also~\cite{deBoer:1998gyt,Kutasov:1999xu}). Specifically,
the worldsheet theory, which is described by a $\slt_{k}$ WZW model, has a set of vertex operators that correspond to the asymptotic Virasoro generators. Here $k$ is the AdS length scale in string units,
\begin{equation}\label{eq:kdef}
k^2 = \frac{\lads}{\ell_s} \,.
\end{equation}
The existence of these operators relies on the non-compact nature of the target space: operators that are naively BRST exact are, in fact, physical due to the non-normalizability of the gauge parameter, and should be retained in the spectrum.

The above considerations make it clear that the 1-loop torus partition function of \AdS{3} string theory~\cite{Maldacena:2000kv}, ought to unambiguously contain the boundary graviton contribution. Somewhat surprisingly, this has not been addressed in the literature directly so far.
Our analysis in this paper makes this manifest from the worldsheet partition function, thereby showing that the boundary gravitons persist in string theory even away from the limit where the string scale is hierarchically separated from the AdS length scale.
We also see a state corresponding to the spacetime dilaton, which is a twist-4 operator, for any~$k \geq 3$. This state arises from a worldsheet current-current bilinear.

A comment about the Schwarzian result is in order. Since the low-temperature analysis in the dual channel holds for any non-zero twist gap~\cite{Ghosh:2019rcj, Pal:2023cgk}, the relation~\eqref{eq:kdef} shows that it should hold
for finite string length even away from the asymptotic semiclassical regime $k\gg1$. The one place where we could expect to see a breakdown is when the twist gap goes to zero, which happens at the tensionless point of string theory~\cite{Gaberdiel:2017oqg}.
In the following, we first establish that the bosonic string partition function computed in~\cite{Maldacena:2000kv} does, in fact, contain the boundary graviton states, and reproduces the Virasoro vacuum character (up to the zero-point energy). This result holds for all $k>3$, where the bosonic theory has a non-vanishing twist gap. It then immediately follows that one recovers the Schwarzian prediction for the near-extremal BTZ partition function for this regime.

\bigskip

\noindent {\bf Superstrings in \AdS{3} and extended spacetime symmetry algebras}

\smallskip

Revisiting the worldsheet superstring theory on \AdS{3} turns out to even more enriching.  In this paper, we consider Type II theories on \asx{T^4} with pure NS-NS flux.\footnote{There is no distinction between the IIA and IIB theories, since they are T-dual to each other.} The super-isometry algebra $\mathfrak{psu}(1,1|2) \oplus \mathfrak{psu}(1,1|2)$ is enhanced to the small $\mathcal{N} = (4,4)$ superconformal symmetry.
 The characters of interest were derived in~\cite{Eguchi:1987wf} using superconformal representation theory, but hitherto have not been obtained in the literature from a 1-loop calculation in 3d supergravity.
The relevant worldsheet analysis, on the other hand, turns out to be a simple extension of the bosonic theory. We show how the worldsheet torus partition function recovers the vacuum character of this superconformal algebra.\footnote{
    It is reasonable that one does not need such a heavy hammer to crack open the one-loop determinant, and that one should be able to calculate it directly using the appropriate supergravity action~\cite{Turiaci:2024xyz,Murthy:2024xyz}.}

The vacuum character of the~$\mathcal{N}=4$ superconformal algebra is naturally written as an infinite sum, which can be interpreted as gravitational saddles~\cite{Heydeman:2020hhw}. These saddles are labelled by allowed large gauge transformations of the $\mathrm{SU}(2)_R$ background gauge field. This can be equivalently explained by the fact that the symmetry algebra has an affine  $\mathrm{SU}(2)_R$ R-symmetry factor at level $\mathsf{k} =c/6$, whose modules include non-trivial null states whose conformal weights scale as $c$~\cite{Eguchi:1987wf}.\footnote{Another perspective is that charge quantization implies that one should sum over configurations with $\mathrm{SU}(2)$ chemical potential shifted by multiples of $2\pi i$ in units of inverse temperature.}  Around each saddle, we have a one-loop determinant of the small fluctuations. The result around any saddle is given by the action of large gauge (elliptic) transformations (at level determined by~$\mathsf{k}$)~\cite{Dabholkar:2012nd} on the determinant around pure thermal \AdS{3}. The worldsheet analysis agrees precisely with this seed, i.e., the expected fluctuation determinant around thermal \AdS{3}.

\smallskip

In addition to the boundary supergravitons, the worldsheet partition function beautifully captures the full spectrum of the BPS states of the theory that have been computed using Type II supergravity on $\text{AdS}_3 \times \textbf{S}^3$~\cite{Deger:1998nm,Larsen:1998xm}. It is worth noting that all these results follow naturally from the rules of the worldsheet theory \cite{Maldacena:2000hw, Maldacena:2000kv}.
In particular, it recovers the full spacetime CFT spectrum (primaries and descendants) without a priori having to use the spacetime symmetry
algebra.\footnote{In the AdS/CFT context the \emph{spacetime} CFT and its symmetries can be identified with the \emph{boundary} CFT$_2$. We make this identification here and in the rest of the text.}
We also see that the multiplet containing the dilaton has a fixed, order~1 twist gap for any~$k \geq 2$. States in this multiplet arise from worldsheet current-current bilinears.

\medskip

The discussion so far can be understood as verifying and summarizing various features of the \AdS{3} theory that could have been anticipated from the low-energy theory. In fact, the analysis of the full string theory turns out to reveal some further surprises. While some elements of our discussion are present in previous treatments of the superstring~\cite{Kutasov:1998zh,Israel:2003ry,Raju:2007uj,Dabholkar:2007ey, Ashok:2020dnc,Dabholkar:2023tzd}, we encounter certain subtleties that are worth clarifying.
\begin{enumerate}[wide,left=0pt]
    \item One has to choose boundary conditions for the fermions around the Euclidean time circle in  thermal \AdS{3}. The supersymmetry-preserving choice of periodic fermions and the supersymmetry-breaking choice of anti-periodic fermions require a different sum over worldsheet spin structures. The former naturally works with the standard (chiral) GSO projected trace over the worldsheet fermions, while the latter requires the introduction of the phases introduced in~\cite{Atick:1988si}. The two boundary conditions, however,
    can be related by exploiting the spacetime $\mathfrak{su}(2)$ R-symmetry---by a suitable shift of the chemical potential one should be able to interpolate between periodic and antiperiodic boundary conditions for the fermions in target spacetime. We  demonstrate
    that this can indeed be realized using a theta-function identity, providing an alternate perspective on the phases inferred for the  spin structure sum in~\cite{Atick:1988si}.

    \item Our results clarify the nature of spacetime conserved currents in the \asx{T^4} string. In addition to the generators of the $\mathcal{N} =(4,4)$ superconformal algebra, we demonstrate the existence of 4 right-moving and 4 left-moving $\mathrm{U}(1)$ currents, and their fermionic superpartners, corresponding to an overall $T^4$ of the boundary theory.\footnote{We shall also see states corresponding to spacetime current-current deformations and their superpartners, which includes the dilaton, as anticipated in~\cite{Giveon:2017nie}.} The currents arise as the NS-NS sector states associated with the $T^4$ excitations on the worldsheet.
    It has been suggested that the string theory should additionally have 8 more $U(1)$ currents (4 left and 4 right) arising from the
    RR-sector~\cite{Kutasov:1999xu}.
    We do not see a signature of these from the torus partition function.\footnote{Harmonic analysis in supergravity suggests that there are abelian Chern-Simons gauge fields in three dimensions arising from the RR fields.
    However, these do not seem to have associated vertex operators on the worldsheet~\cite{Kutasov:1999xu}. We thank Ofer Aharony for raising this question and for useful discussions about this point. }
    This is consistent with the recent analysis of the boundary theory in~\cite{Aharony:2024fid}, who suggested that the additional currents are a decoupled sector.

    \item There is a tension between the GSO projection used to construct superstrings on \asx{T^4}~\cite{Giveon:1998ns} and the modular invariance of the torus partition function. This tension originates in the fact that the GSO projection does not arise from an $\mathcal{N}=2$ current on the string worldsheet~\cite{Giveon:1998ns}. Nevertheless, assembling
    the worldsheet NS-R fermions into characters as in 10-dimensional flat space leads to a modular invariant torus amplitude (see~\cite{Raju:2007uj,Ashok:2020dnc}
    for a similar analysis of the partition function). We show, further, that the resulting low-lying BRST cohomology agrees with~\cite{Kutasov:1998zh}, \cite{Dabholkar:2007ey} that is obtained using the GSO projection of~\cite{Giveon:1998ns}.

    \item A final and somewhat serious subtlety we encounter, but do not resolve, involves the superstring on \asx{\mathrm{K3}}. One expects to be able to analyze this theory straightforwardly at the
    orbifold point $\mathrm{K3} = T^4/\mathbb{Z}_2$, where the $\mathbb{Z}_2$ action reflects the four free bosons and fermions characterizing the $T^4$ CFT, preserving the $\mathrm{SU}(2)$ holonomy. This indeed works at the level of covariant quantization,
    as one can check by examining the BRST cohomology. One can write down a modular invariant worldsheet partition function following the usual rules of orbifold constructions. Perplexingly, the resulting answer breaks spacetime supersymmetry!\footnote{One expects that the orbifold projects out the $\mathrm{U}(1)$ currents associated with the $T^4$ (the NS-NS sector states) and their fermionic
    partners (which are R-NS states). The former are indeed projected out, but only half of the latter are, along with two of the supercurrents of the $\mathcal{N}=4$ algebra.}
    This is manifested by the fact that the combination of the orbifold and GSO projection breaks the $\mathrm{SU}(2)$ R-symmetry.
    This is, of course, absurd, suggesting that the orbifold action is more involved.\footnote{This puzzle has also been noticed by Lorenz Eberhardt~\cite{Eberhardt:2024xyz}. We thank him for discussions about this issue.} We outline some of our findings, but leave the resolution of this important issue to a future
    analysis. It is worth noting that there is no such issue with backgrounds like \asx{\mathbf{S}^3 \times \mathbf{S}^1}, as we shall explain elsewhere~\cite{Murthy:2024s3s1}.
\end{enumerate}

The outline of the paper is as follows. We begin with a review of some basic aspects of the bosonic string theory on \AdS{3} in~\cref{sec:bosonic}. In~\cref{sec:bosonicZ} we reanalyze the bosonic string partition function, and demonstrate its key features, such as the boundary gravitons and a non-vanishing twist gap for $k>3$. We then review salient aspects of the superstring in \asx{T^4}
in~\cref{sec:superT4}, describing both the worldsheet theory, and the dual spacetime CFT symmetries. In~\cref{sec:superZ} we reanalyze the superstring partition function on \asx{T^4}, and show that it correctly captures the aforesaid symmetry generators, and reproduces the spectrum of chiral primary states which are visible in supergravity.  In~\cref{sec:nhbtz} we describe the implications of our results for the near-extremal BTZ thermodynamics, and conclude with a set of future directions in~\cref{sec:discuss}. The~\cref{sec:appreviews,sec:superpart} contain some technical details for the evaluation of bosonic and superstring partition functions, respectively.

A note on conventions: we work in the regime $k\geq 3$ for the bosonic string and $k\geq 2$ for the superstring. In particular, we will not attempt to access the tensionless limit of the superstring~\cite{Eberhardt:2018ouy} which has seen much progress recently.

\section{Review: Worldsheet theory of strings on \texorpdfstring{\AdS{3}}{AdS3}}
\label{sec:bosonic}

In this section we give a quick overview of the worldsheet sigma model for string propagation on \AdS{3} backgrounds with NS-NS flux.
We begin with the bosonic $\slt_k$ sigma model, highlighting salient features from the analysis of~\cite{Giveon:1998ns,Kutasov:1999xu,Maldacena:2000hw}.

\subsection{The bosonic \texorpdfstring{$\slt_k$}{sl2k} sigma model}\label{sec:bosls2}

Consider Euclidean \AdS{3} ($\mathbb{H}_3^+$) with the line-element given by
\begin{equation} \label{eq:ads3metricWaki}
ds^2 \= \lads^2\, (d\phi^2 + e^{2\phi}\, d\gamma\, d\bar{\gamma}) \,.
\end{equation}
The worldsheet action takes the form
\begin{equation}\label{eq:wslag}
S_\text{ws} \= k\, \int\, d^2\sigma\, \Bigl( \partial \phi \, \bar{\partial}\phi + e^{2\phi}\, \bar{\partial}\gamma\, \partial\bar{\gamma} \Bigr) \,,
\end{equation}
which also takes into account the presence of a Neveu-Schwarz B-field.
This sigma model has an $\slt\times \overline{\slt}$ affine current algebra, with generators $J^a_n$, $\overline J^a_n$,
$a\in \{\pm,3\}$, $n\in \mathbb{Z}$.
The zero modes of this algebra $J^a_0$, $\overline J^a_0$ can be represented in two equivalent ways.
One is to use a representation in terms of the worldsheet fields $\{\phi,\gamma,\bar{\gamma}\}$,
and the other is to employ a representation space
parameterized by $x,\bar{x}$, on which the generators act as differential operators (so that~$x^m$ carries~$J_0^3$ charge~$m$).
The variables $(x,\bar{x})$ can be thought of as spacetime  coordinates of the dual boundary CFT.
It is convenient to assemble the three currents into one object as
\begin{equation}
J(x;z) \; \coloneqq \; - J^-(z) + 2\, x\, J^3(z) -x^2\, J^+(z)
\end{equation}
The worldsheet stress tensor is obtained from the Sugawara construction
\begin{equation}\label{eq:Twssl2}
T_\text{ws}(x;z)
\=
	\frac{1}{2\,(k-2)} \, \Bigl( J(x;z)\, \partial_x^2\, J(x;z) - \frac{1}{2}\, \left(\partial_x J(x;z)\right)^2 \Bigr) \, .
\end{equation}

To characterize the string spectrum, we require knowledge of the representations of $SL(2,\mathbb{R})$.
Of relevance are the discrete series (lowest/highest weight) representations $\mathcal{D}_j^\pm$,
$j \in \mathbb{R}$, and the continuous series $\mathcal{C}_j^\alpha$, $j \in \frac{1}{2} + i\,\mathbb{R}$, $0 < \alpha < 1$.
The quantum number $j$ determines the quadratic Casimir $c_2 = -j(j-1)$. For instance, $\mathcal{D}_j^+$ has states
\begin{equation}
\ket{j,m}\,, \qquad J_0^3 \ket{j,m} \= m\, \ket{j,m}\,, \qquad m \= j + \mathsf{n}\,, \qquad \mathsf{n} \in \mathbb{Z}_{\geq0} \,.
\end{equation}
Unitarity, i.e.,~the no-ghost theorem restricts $\frac{1}{2}< j < \frac{k-1}{2}$ for the discrete series~\cite{Maldacena:2000hw}.\footnote{
	Originally, the upper bound from the no-ghost theorem was deduced to be $j < \frac{k}{2}$ in~\cite{Evans:1998qu} from the sigma model analysis. We are quoting the stronger bound argued for by~\cite{Maldacena:2000hw} for consistency with the $\slt$ spectral flow.}
The state $\ket{j,j}$ is annihilated by $J_n^{3,\pm}$,  $n\geq 1$, and by $J_0^-$. States at a fixed level have their $J_0^3$ eigenvalue raised by $J_0^+$.
Acting by the raising operators $J_n^{3,\pm}$ with $n\leq -1$ on $\ket{j,m}$ we  raise the $L_0$ eigenvalue, but can lower the  $J_0^3$ eigenvalue.
States at affine level $N$ are obtained by the action of raising operators $\prod_i\,( J_{-n_i}^{3,\pm})$ with~$\sum_i n_i = N$.
Starting from $\ket{j,j}$ the lowest $J_0^3$ eigenvalue, one obtains is $j-N$, while the $L_0$ eigenvalue shifted to $-\frac{j(j-1)}{k-2} +N$.

The result of~\cite{Maldacena:2000hw}
is that the full spectrum of string theory consists of the aforementioned ones along with their spectral flowed versions, viz., $\mathcal{D}_{j,w}^{\pm}$ and $\mathcal{C}_{j,w}^\alpha$ with~$w \in \mathbb{Z}$ parameterizing the spectral flow.
The spectral flow operation changes the moding of the current algebra generators, but unlike the case of
compact WZW models (e.g.,~$\sut_k$) generates new representations, owing to the non-compactness of the target space.

\subsection{The bosonic string spectrum in \texorpdfstring{\AdS{3}}{AdS3}}\label{sec:ssspectrum}

We are now in a position to discuss the canonical quantization of bosonic strings propagating on \AdS{3} $\times \, X$. Taking the states described above for the $\slt_k$ WZW model, one imposes the Virasoro constraints on the string worldsheet
\begin{equation}
\left(L_n + L_n^\text{int} - \delta_{n,0} \right) \ket{\Psi} \=0 \,, \qquad n\geq 0\,.
\end{equation}
Focusing on the sector with no spectral flow, take a state $\ket{j,m,N}$ at descent level $N$ in the $\slt_k$ part, tensored with~$\ket{\hx}$ in the internal CFT. The physical state condition leads to
\begin{equation}
(L_0 + L_0^\text{int} -1) \ket{j,m,N,\hx} \= 0  \quad \Longrightarrow \quad
-\frac{j(j-1)}{k-2} + N + \hx -1 \= 0 \,.
\end{equation}
This fixes $j$ in terms of $N$ and $\hx$,
\begin{equation}
j \= \frac{1}{2} + \sqrt{(k-2) (N+\hx-a_k)} \;,
\end{equation}
with $a_k$ being the net zero-point energy
\begin{equation}\label{eq:zeroptB}
a_k \=  1-\frac{1}{4\,(k-2)} \,.
\end{equation}

The spacetime CFT Virasoro generators are the zero modes of the $\slt$ current algebra,
\begin{equation}
\Ls_0 \= J_0^3 \,, \qquad   \Lsb_0 \=  \overline{J}_0^3 \,,
\end{equation}
which shows that the spacetime weights are given by the $J_0^3$ eigenvalue~$m$. Putting together left and right movers, and using $\mathsf{n}, \overline{\mathsf{n}}$ to denote the contribution to the $J_0^3, \overline{J}_0^3$ quantum numbers from the oscillators, we have
\begin{equation}
\begin{split}
\Delta &\=
	\hs + \hsb
	\= m+\overline{m} \= j + \overline{j} + \mathsf{n}+ \overline{\mathsf{n}} \,, \\
\ell &\= \hs - \hsb
\=
	m-\overline{m} \= j-\overline{j} + \mathsf{n} -  \overline{\mathsf{n}} \,.
\end{split}
\end{equation}
Using the solution for $j$, we obtain
\begin{equation}
\Delta \= 1 + \mathsf{n}+ \overline{\mathsf{n}} + \sqrt{(k-2)\,(N+\hx-a_k)} \;,
\qquad \ell \= \mathsf{n} -  \overline{\mathsf{n}}\,,
\end{equation}
where we have used level matching to fix $N+\hx = \overline{N} + \hxb$. Note that $\mathsf{n} \geq -N$, given the raising operation of the affine algebra. The unitarity bound on the representations is
\begin{equation}
\frac{1}{2} \,\leq \, j \,\leq \, \frac{k-1}{2} \quad \Longrightarrow \quad
\frac{4}{k-2} \, \bigl(N+ \hx -a_k \bigr)\,\in \,  [0,1]  \,,
\end{equation}
which guarantees that $\hs \geq j-N = \frac{1}{2} + \sqrt{(k-2)\,(N+ \hx -a_k)} - N  >0$.

In addition to the states above, the proposal of~\cite{Maldacena:2000hw} was to include spectrally flowed representations in the $\slt_k$ WZW model. In particular, once we include spectral flow parameterized by~$w \geq 0$,
the unitarity bound, which seems to impose an artificial cut-off on the oscillator excitations disappears. For the discrete representation, the spectral flow results in:
\begin{equation}
\hs \= \frac{1}{2} + w + \sqrt{\frac{1}{4} + (k-2) \left(N + \hx - w \, \mathsf{n} -\frac{w\,(w+1)}{2} -1 \right)} + \mathsf{n} \,.
\end{equation}
Here $N$ is the oscillator level before spectral flow, and $\mathsf{n}$, as before, is the relative number of $J_\pm$ oscillators that are excited.
Unitarity implies the modified constraint
\begin{equation}
N + \hx -a_k - w\, \mathsf{n} - \frac{w(w+1)}{2} \;\in \; \frac{k-2}{4} \, [w^2,(w+1)^2] \,.
\end{equation}
Finally, we also have the states in the continuum arising from $\mathcal{C}^\alpha_{j,w}$, which correspond to CFT states with conformal dimensions:
\begin{equation}
\hs \= \frac{w\,k}{4} + \frac{1}{w} \left(\frac{s^2 + \frac{1}{4}}{k-2} + N + \hx -1 \right)  , \qquad s \in \mathbb{R}\,.
\end{equation}
\paragraph{Comments on the spectrum:}  Let us fix $k > \frac{9}{4}$, so that $a_k \in (0,1)$ and examine low-lying values of $N+ \hx$:
\begin{itemize}[wide,left=0pt]
\item  $N+\hx =0$: disallowed by unitarity bound.
\item $N+\hx =1$: requires $k \geq 3$. Note that we have two choices $N=1$, $\hx =0$ which gives $\hs =0$, or $N=0$, $\hx =1$ which gives $\hs =1$. The latter statement is the fact that the vacuum of the $\slt$ WZW model tensored with a holomorphic conserved current, stays a conserved current in the spacetime theory.
\item For general $N$ and $\hx$ we need $k \geq 2\, ( N+\hx) + \sqrt{1+ 4\, ( N+\hx-1)^2}$. Once we satisfy this constraint, it is guaranteed that $\hs \geq 0$.
However, for any such choice, one notes that $\dv{\hs}{N} \geq 0$, so the conformal dimensions are monotone increasing with $N$ (for fixed $\hx$). Apart from the
special case, $N=1$, $\hx=0$, it then follows that all states in the discrete representation have $\hs \sim \order{\sqrt{k}}$ in the semiclassical limit $k\gg1$.
\item Without spectral flow, the only state in the continuous series is the tachyon. To see this, note that the Virasoro constraint reads:
\begin{equation}
\frac{s^2 + \frac{1}{4}}{k-2} + N + \hx -1 =0\,.
\end{equation}
Since $s\in \mathbb{R}$, this only has a solution when $N =0$. If we fix $\hx =0$ then we completely fix $s = \pm\sqrt{k- \frac{9}{4}}$. Alternately, we can take $\frac{1}{4} + (k-2) \left(\hx -1\right) \leq 0$ and solve for $s$. The spacetime CFT dimension is however,
\begin{equation}
\hs \= \alpha + n \,, \qquad n \in \mathbb{Z}_{\geq0} \,, \quad \alpha \, \in \, [0,1)\,.
\end{equation}
\item Spectrally flowed continuous series states have $\hx \sim \order{k}$ in the semiclassical limit.
\end{itemize}

\subsection{Spacetime currents}\label{sec:chiralcurrents}

There are natural objects in the representation space, which transform in the spin $(j-1,j-1)$, representation of the global $\slt\times \overline{\slt}$ algebra (NB: restriction $h =\bar{h}$). They are $(j,j)$ tensors $\Phi_j\, dx^j\, d\bar{x}^j $, with wavefunctions
\begin{equation}
\Phi_j (x,\bar{x};z,\bar{z}) \= \frac{1}{\pi} \left(\frac{1}{\abs{\gamma-x}^2\, e^{\phi}+ e^{-\phi}}\right)^{2j}\,.
\end{equation}
One can view these as the bulk-boundary propagators of a massive field $m^2 = \frac{1}{k}\, j(j-1)$ in the E\AdS{3} geometry. In the quantum theory, the OPE of these primaries and the currents reads
\begin{equation}
J(x; z) \, \Phi_j(y,\bar{y}; w,\bar{w}) \; \sim \; \frac{1}{z-w}\, \Bigl((y-x)^2\,\partial_y + 2\, j\, (y-x)\Bigr) \,\Phi_j(y,\bar{y};w,\bar{w}) \,.
\end{equation}
One can similarly deduce that $T_\text{ws}$ defined in~\eqref{eq:Twssl2} transforms as a $(0,0)$ tensor, i.e., it is a spacetime scalar.

There is one special operator of interest with $j=\bar{j}=1$. The operator $\Phi_1$, which is a $(1,1)$ tensor in spacetime,  naively corresponds to a massless particle. It obeys:
\begin{equation}
\begin{split}
\bar{J}(\bar{x};\bar{z}) \, \Phi_1(x,\bar{x};z,\bar{z})
& \= \frac{k}{\pi}\, \partial_{\bar{z}}\, \Lambda(x,\bar{x};z,\bar{z}) \,,\qquad
\Lambda(x,\bar{x};z,\bar{z})
\= -\frac{(\bar{\gamma}-\bar{x})\,e^{2\phi}}{(\gamma-x)(\bar{\gamma}-\bar{x}) e^{2\phi} +1}\,.
\end{split}
\end{equation}
The operator $\Lambda$ is not a well-behaved operator on the boundary, but  plays a role in defining the asymptotic symmetries. For one, it has a well-defined $\overline{\slt}$ charge ($\overline{j}=0$), but does not carry definite $\slt$ charge.

A curious operator in the theory is the spacetime identity operator $I$, defined as
\begin{equation}\label{eq:Idef}
I(x,\bar{x}) \= \frac{1}{k^2}\, \int\, d^2z\, J(x;z)\, \bar{J}(\bar{x};\bar{z}) \,\Phi_1(x,\bar{x}; z, \bar{z}) \,.
\end{equation}
This operator is a spacetime scalar, i.e., a $(0,0)$ tensor. Naively, when one evaluates the
OPEs of the spacetime currents~\eqref{eq:SpK} and~\eqref{eq:SpT}, one does not get a simple number, but rather an integrated object on the worldsheet.
To confirm the intuition that $I$ acts as the spacetime CFT identity, one can at least check that $\partial_x I = \partial_{\bar{x}} I =0$.
That said, the operator is more complicated than indicated; in correlation functions it acts as a constant, with a value that depends on which set of operators are involved~\cite{Giveon:2001up}.

The spacetime currents associated with the large gauge transformations or diffeomorphisms in the asymptotic \AdS{3} geometry are built using $\Phi_1$.
Corresponding to worldsheet current, $k^a(z)$ at level $k_{_G}$ one constructs spacetime current algebra generators $\Kcft^a$ in terms of worldsheet
integrals\footnote{The~$\bar{x}$-dependence of the right-hand side of this equation decouples from all physical correlators, which is indicated by the pure~$x$-dependence of the left-hand side~\cite{Kutasov:1999xu}.}
\begin{equation}\label{eq:SpK}
\Kcft^a(x)
	\= \int d^2z\, k^a(z)\, \partial_{\bar{z}}\, \Lambda \= -\frac{1}{k}\, \int\, d^2z\, k^a(z) \, \bar{J}(\bar{x};\bar{z})\, \Phi_1(x,\bar{x};z,\bar{z}) \,.
\end{equation}
Similarly, the spacetime CFT Virasoro generators $\Tcft$ are given as
\begin{equation}\label{eq:SpT}
\Tcft(x) \= \frac{1}{2k}\int d^2z\, \left(\partial_x J\, \partial_x\Phi_1 + 2\, \partial_x^2\, J\, \Phi_1\right) \bar{J}(\bar{x};\bar{z}) \= \frac{1}{2}\, \oint \frac{dz}{2\pi i} \left(\partial_x \, J\, \partial_x\,\Lambda + 2\, \Lambda\, \partial_x^2 J\right) \,.
\end{equation}
%
%
They obey
\begin{equation}\label{eq:Svirasoroalg}
\begin{split}
\Kcft^a(x)\, \Kcft^b(y)
&\; \sim \;
	\frac{f^{abc}}{x-y}\, \Kcft^c(y) + \frac{1}{2}\, \frac{p(g_s)\, k_{_G}\, \delta^{ab}}{(x-y)^2} \,,\\
\Tcft(x)\, \Tcft(y)
& \; \sim\;
	\frac{1}{2}\, \frac{6\,k\, p(g_s)}{(x-y)^4} + \frac{2\, \Tcft(y)}{(x-y)^2} + \frac{\partial_y \Tcft(y)}{x-y} \,,
\end{split}
\end{equation}
where $p(g_s)$ can be identified as the expectation value of the operator $I$ introduced above.
Note that the spacetime CFT currents are at a different level; the spacetime central charge for instance is $p(g_s)\, k$.  This is needed for consistency with semiclassical intuition. The Brown-Henneaux asymptotic symmetry algebra has central charge given by measuring AdS curvatures relative to the Planck scale,  $c \sim \frac{\lads}{\ell_P}$, whereas  $k$ is a measure of curvature in string units as noted in~\eqref{eq:kdef}.  Recently, using the three-point function of the operator~$I$, the authors of~\cite{Eberhardt:2023lwd}
were able to derive the spacetime CFT central charge in accord with the Brown-Henneaux expectations.

\section{Bosonic partition function on \texorpdfstring{\AdS{3} $\times\; X $}{AdS3 x X}}\label{sec:bosonicZ}

We now turn to the one-loop partition function for strings around the thermal \AdS{3} geometry $H_3^+/\mathbb{Z}$.
It is convenient to parameterize $H_3^+$ by the line element
\begin{equation} \label{eq:EucAdS3}
ds^2 \= k \, \bigl(\cosh^2\rho\, d\tE^2 + d\rho^2 + \sinh^2\rho\, d\varphi^2\bigr) \,,
\end{equation}
with~$\varphi \sim \varphi+2\pi$.
Thermal \AdS{3} is obtained by
implementing the identification
\begin{equation}
\label{eq:thermalid}
(\tE,\varphi)
	\; \sim \; (\tE + \beta, \varphi +i\, \beta\,\mu) \,.
\end{equation}
The asymptotic boundary of this space is~$T^2$  whose modular parameter~$\tau$ and the related~$q$-parameter are given by
\begin{equation}\label{eq:qtau}
\tau \=  \frac{\beta \mu+ i\,\beta}{2\pi} \,, \qquad
 q \= e^{2\pi i\, \tau} \, .
\end{equation}
Since we reserve the standard parameterization of the complex structure for the spacetime torus, we will use $\ts, \zs$ for the corresponding worldsheet torus, where $\ts$ is the worldsheet modular parameter, and
\begin{equation}\label{eq:wsz}
\ts \= \ts_1 + i\, \ts_2 \,,\qquad \zs \= e^{2\pi i \,\ts} \,.
\end{equation}

The one-loop partition function of free bosonic strings propagating in \AdS{3} $\times X$ was computed in~\cite{Maldacena:2000kv},
using the $H_3^+/\mathbb{Z}$ presentation of thermal \AdS{3}.
We give a brief synopsis in~\cref{sec:appreviews} for completeness. The resulting 1-loop partition function of the $\slt_k$ bosonic sigma model is\footnote{We use the conventions in~\citep[Chapter 7]{Polchinski:1998rq}  for the theta functions.}
\begin{equation}\label{eq:Zbosads3}
\begin{split}
\mathcal{Z}_{\slt_k}(\ts,\tsb; \tau, \overline{\tau})
\=
	\frac{\beta\, \sqrt{k-2}}{2\pi\, \sqrt{\ts_2}}\,
	 \sum_{n,m \in \mathbb{Z}}\,  \frac{\exp \left( -\frac{k\, \beta^2}{4\pi\,\ts_2} \, \abs{m- n\,\ts}^2 + \frac{2\pi}{\ts_2} \, (\Im(\us_{n,m}))^2\right)}{\abs{\vartheta_1(\us_{n,m},\ts)}^2}\,,
\end{split}
\end{equation}
where we introduced the worldsheet holonomy
\begin{equation}\label{eq:unmdef}
\us_{n,m} \= \tau\, (n\, \ts - m)\,.
\end{equation}
The factor proportional to $k\,\beta^2 \abs{m-n\,\ts}^2$ is the contribution from the winding of the worldsheet cycles, labeled by integers $m$ and $n$,  around the non-contractible thermal circle in spacetime. The theta function and  $\Im(\us_{n,m})$ are the contribution of the three bosonic degrees of freedom constituting the $\slt_k$ WZW model. The bosons associated with $J^\pm$ carry charge $\pm 1$ and are thus sensitive to the background holonomy~\eqref{eq:unmdef}. The pre-factor is the zero-mode contribution from the non-compact direction.  This partition function is modular invariant on the worldsheet (cf.~\cref{sec:appreviews}).

\subsection{The worldsheet partition function}\label{sec:boswsZ}

Folding in the partition function of the internal space $\mathcal{Z}_X$, and the $bc$ ghost system (which we recall  gives $\abs{\eta(\ts)}^4$) we can write the final result for the 1-loop string partition function as follows,
\begin{equation}\label{eq:Zwsads}
\mathcal{Z}_\text{ws}(\tau,\bar{\tau})
\=
	\int_{\mathcal{F}} \, \frac{d^2\ts}{\ts_2} \, \mathcal{Z}_{\slt_k}(\ts,\tsb ;\tau, \overline{\tau}) \, \mathcal{Z}_X(\ts,\tsb) \, \abs{\eta(\ts)}^4 \,.
\end{equation}
Here $\mathcal{F}$ is the fundamental domain of the worldsheet modular parameter.

We would like to carry out the integral and express the result in terms of the spacetime CFT characters.
To do so, we will assume that the internal CFT $X$ is generic, so that the spacetime theory
has no conserved currents apart from the Virasoro currents.\footnote{
	We make this assumption at this point of the presentation for convenience to illustrate the basic aspects. When we discuss the superstring in the following, the internal CFT will have conserved currents, which we will take into account.}
We denote the contribution of the internal CFT as
\begin{equation}
\mathcal{Z}_X(\ts,\tsb)  \= \sum_{\hx, \, \hxb} \,  D(\hx,\hxb)\,  \zs^{\hx}\, \zsb^{\hxb} \,.
\end{equation}

Following~\cite{Polchinski:1985zf} the modular invariance of the worldsheet integrand can be exploited
to simplify the double sum over the winding labels $(m,n)$ into a single sum over $m \in \mathbb{Z}_{>0}$,
using the Rankin-Selberg unfolding method  (see e.g.,~\cite{Zagier:RanSel}).
In particular, the term with $(m,n) = (0,0)$ drops out. This uses the fact that these labels transform as a doublet under $\mathrm{SL}(2,\mathbb{Z})$ modular
transformations, and is achieved at the expense of unfolding the domain of the integral to the strip in the $\ts$ plane. One can implement this as explained
in~\cite{Maldacena:2000kv} and write the worldsheet answer as
\begin{equation}\label{eq:Zwsfock}
\mathcal{Z}_\text{ws}(\tau,\bar{\tau}) \= -\beta\, \sum_{m=1}^\infty \, f(m\tau, m\overline{\tau}) \,,
\end{equation}
where
\begin{equation}
    f(\tau, \overline{\tau}) \= 	\frac{\beta\, \sqrt{k-2}}{2\pi}
   \int_{-\frac{1}{2}}^{\frac{1}{2}} \, d\ts_1\, \int_0^\infty\,
 	\frac{d\ts_2}{\ts_2^{\frac{3}{2}}}\,
 \frac{e^{-(k-2)\, \frac{\beta^2}{4\pi\, \ts_2}}}{\abs{\vartheta_1(\tau,\ts)}^2}\,
 \mathcal{Z}_X(\ts,\tsb) \, \abs{\eta(\ts)}^4  \,.
\end{equation}
The quantity $f(\tau, \overline{\tau})$ may be interpreted as the free-energy computed in the single-string Hilbert space,
\begin{equation}\label{eq:fwssingle}
f(\tau, \overline{\tau}) \=
\frac{1}{\beta}\,\sum_{\mathcal{H}_\text{single-string}} \, e^{-\beta\, E - i\,\beta\,\mu \, \ell}\,.
\end{equation}

Using the expressions~\eqref{eq:Zbosads3}, \eqref{eq:Zwsads},
and exploiting the following identity obtained by Gaussian integration,
\begin{equation}\label{eq:auxint}
e^{-(k-2)\, \frac{\beta^2}{4\pi\, \ts_2}} \= -\frac{8\pi i}{\beta}\, \left(\frac{\ts_2}{k-2}\right)^\frac{3}{2} \,
\int\limits_{-\infty}^\infty\, d\zeta\, \zeta\, e^{-\frac{4\pi\,\ts_2}{k-2}\, \zeta^2 +2 i\,\beta\,\zeta} \,,
\end{equation}
we obtain the following expression for the single-string free energy,
\begin{equation}\label{eq:fsingleB}
\begin{split}
f(\tau, \overline{\tau})
&\=
	\frac{4}{i\beta\,(k-2) } \,
	\int\limits_{-\infty}^\infty\, d\zeta\,\zeta\,
	\int_{-\frac{1}{2}}^{\frac{1}{2}} \, d\ts_1\, \int_0^\infty\, d\ts_2 \,
	e^{4\pi\, \left(a_k - \frac{\zeta^2}{k-2}\right) \ts_2 + 2i\,\beta\,\zeta} \, \mathcal{P}_\text{bos}(\zs,\zsb;q,\qb) \,, \\
\mathcal{P}_\text{bos}(\zs,\zsb;q,\qb)
&\=
	\frac{Z_X(\zs,\zsb)}{\abs{q^{\frac{1}{2}} - q^{-\frac{1}{2}}}^{2}}
	\;\abs{ \prod_{n=1}^\infty\frac{1-\zs^n}{(1- q\,\zs^n)\,(1- q^{-1} \,\zs^n)}}^2 \,.
\end{split}
\end{equation}
The parameter $a_k$ defined earlier in~\eqref{eq:zeroptB} is the net zero-point energy
of the $\slt_k$ sigma model, the internal CFT, and the ghost sectors.

\smallskip

The contribution of multi-string states to the spacetime partition function is obtained by exponentiating the
worldsheet partition function~\eqref{eq:Zwsfock}~\cite{Maldacena:2000kv}.
This correctly takes into account the statistics of identical particles in the quantum theory.
The result captures the one-loop contribution of string fluctuations around the thermal \AdS{3} saddle
of the full spacetime CFT.
We write this as
\begin{equation}\label{eq:Zwssp}
\mathscr{Z}_{_\text{CFT}}(\tau,\overline{\tau}) \,\Big{|}_\text{thermal \AdS{3}} \; \= \;
\exp \Bigl( S_\text{tree} + \mathcal{Z}_\text{ws}(\tau,\overline{\tau}) + \dots \Bigr) \,.
\end{equation}

The tree-level contribution, denoted $S_\text{tree}$,  should capture the Casimir energy of the spacetime CFT. This should be obtained as a genus-$0$  worldsheet (sphere) with no insertions. There is thus far no direct computation of this in the bosonic string theory.\footnote{
	In the bosonic theory~\cite{Eberhardt:2023lwd} report a result that does not match with the semi-classical expectation. }
Given these complications, we will simply focus on evaluating the worldsheet contribution and re-express it in terms of building blocks for the spacetime CFT characters.

\subsection{Spectral decomposition of worldsheet partition function}\label{sec:spectrum}

In order to evaluate the integral, we expand the infinite product $\mathcal{P}_\text{bos}(\zs,\zsb)$ in a series in~$\zs, \zsb$ and integrate over the worldsheet moduli.
This can be done along the lines described
in~\cite{Maldacena:2000kv}.
We systematize this expansion by exploiting a result due to D.~Zagier which gives a clean series expansion for the~$\slt$ partition sum.
Using this, we evaluate the worldsheet modular integral and organize the expansion in terms of spacetime CFT primaries.
We outline the procedure explicitly in~\cref{sec:bosZeval} for completeness.

The basic observation, as in~\cite{Maldacena:2000kv}, is that while $\abs{q\,\zs} < 1$, and thus $\abs{q\, \zs^n} <1$ for~$n>0$, the quantity~$\abs{q^{-1}\, \zs}$, which appears in the one-loop determinant,
can be less than or greater than~1, depending on the point in moduli space.
The strategy to take this into account~\cite{Maldacena:2000kv} is to break up the domain $\ts_2 \in [0,\infty]$
into a series of subdomains $\frac{\beta}{2\pi} [(w+1)^{-1}, w^{-1}]$ with $w\in \mathbb{Z}_{\geq 0}$.
At the boundary of the subdomains, $q^{-1}\, \zs^n =1$ for some value of~$n$ and there is a consequent pole in integrand of the partition function.
In each subdomain one expands out the infinite product in the smaller of~$q^{-1}\,\zs^n$ and $q\,\zs^{-n}$, collects terms, and carries out the integral.
The precise set of terms of the two kinds are demarcated by the value of~$w$, which plays the role of the spectral flow parameter in this analysis.
The contributions with~$w=0$ come from the excitations of the short strings,
while the contributions with~$w \neq 0$ include spectral-flowed versions of the short strings and long strings that wind $w$ times around the boundary of \AdS{3}.\footnote{
	In this analysis, $w \in \mathbb{Z}_{\geq 0}$, while in the canonical quantization discussed in~\cite{Maldacena:2000hw} $w \in \mathbb{Z}$. One way to understand the difference is to realize that states with $w \geq0$ are state vectors (kets), while those with $w<0$ can be mapped to conjugate state vectors (bras)~\cite{Eberhardt:2019qcl}. This interpretation is consistent with the fact that one can attain similar states by spectrally flowing from highest (lowest) weight representations by positive (negative) amounts.}

Carrying out this exercise, the single string free energy can be shown to be
\begin{equation}\label{eq:bosfss1}
\begin{split}
f(\tau, \overline{\tau})
&\=
     \frac{1}{\beta\, \abs{1-q}^2 }  \,
     \sum_{w=0}^\infty\,
	(q\,\qb)^{\frac{1}{2}+w}\sum\limits_{\substack{N,\overline{N} \in \mathbb{Z}\\ \hx,\hxb}} \,
	\delta_{N+\hx, \overline{N} + \hxb}  \, D(\hx,\hxb)\,
	\mathbf{P}^w_N(q)\, \mathbf{P}^w_{\overline{N}}(\qb) \,\mathfrak{I}_w(q,\qb) \,, \\
\mathfrak{I}_w(q,\qb)
&\= 	 (q\,\qb)^w \,\int\limits_{-\infty}^\infty\,\frac{d\zeta}{i\pi}\, \frac{\zeta\, e^{2i\beta \,\zeta} }{\zeta^2 + (k-2)\, \Nwx } \, \Bigg[ e^{-\frac{2\beta}{1+w} \left(\frac{\zeta^2}{k-2} + \Nwx\right)} -e^{-\frac{2\beta}{w} \,\left(\frac{\zeta^2}{k-2} + \Nwx\right) } \Bigg] \,.
\end{split}
\end{equation}
Here $\Nwx$ is a combination of the $\slt$ oscillator level, the internal CFT weight $\hx$, and the spectral flow parameter $w$, shifted by the vacuum energy, viz.,
\begin{equation}\label{eq:NXdef}
\Nwx =  N + \hx -a_k - \frac{1}{2}\, w(w+1) \,.
\end{equation}
The summand with $w=0$ is the contribution from the sector with no spectral flow,
while \hbox{$w>0$} connotes the spectrally flowed sectors. The delta
function is the level-matching constraint imposed by the~$\ts_1$ integral.
The coefficient $\mathbf{P}^w_N(q)$, which only depends on the
spacetime CFT modular parameter, originates from expanding the infinite product in~\eqref{eq:auxint}.  We discuss these coefficients in~\cref{sec:bosZeval}.
We can view $N$ as the
effective level for the $\slt_k$ and $bc$ ghost oscillators in the $w^\mathrm{th}$ spectrally flowed sector.
For instance, in the $w=0$ sector,  which should pick up the short string representations,  we have
\begin{equation}\label{eq:PNlvals}
\begin{aligned}
\mathbf{P}_0^0(q)
&\=
	1 \,, & \qquad
\mathbf{P}^0_1(q)
&\=
	\frac{1-q+q^2}{q}\,, \\
\mathbf{P}^0_2(q)
&\=
	\frac{1+ q^4}{q^2}  \,, & \qquad
\mathbf{P}^0_3(q)
&\=
	\frac{1+ q^3+ q^6}{q^3}  \,,
\end{aligned}
\end{equation}
and so on.  We have restricted to the vacuum sector of the internal CFT to focus on the universal part of the \AdS{3} spectrum. Depending on the details of the internal CFT we will also have contributions for $\hx, \hxb \neq 0$.

The integral over the auxiliary parameter can be carried out along the lines described in~\cite{Maldacena:2000kv}. Relegating the details to~\cref{sec:bosZeval}, we note that the single string free energy can be expressed as a combination of a discrete and a continuum set of states
\begin{equation}
f(\tau, \overline{\tau})
\=
	 f_\text{tachyon}(\tau, \overline{\tau})+  f_\text{disc}(\tau, \overline{\tau})
	+ f_\text{cont}(\tau, \overline{\tau})\,.
\end{equation}
In writing this, we have isolated the contribution from the bosonic string tachyon. It contributes in the sector with no spectral flow, and has the explicit form
\begin{equation}
f_\text{tachyon}(\tau, \overline{\tau})
\=
	\frac{\sqrt{q \,\qb}}{2\,\abs{1-q}^2}
	\Bigl( e^{- 2\beta\, i\,\sqrt{k- \frac{9}{4}} } + e^{ 2\beta\, i\,\sqrt{k- \frac{9}{4}} } \Bigr) \,.
\end{equation}
We will ignore this contribution henceforth.

The discrete states across all spectral flow sectors lead to
\begin{equation}\label{eq:bosshort}
\begin{split}
f_\text{disc}(\tau, \overline{\tau})
&\=
	\sum\limits_{\substack{N,\overline{N}\in\mathbb{Z}\\ \hx,\hxb}}\,\delta_{N+\hx, \overline{N} + \hxb} \,
	\sum_{w=0}^\infty\,
	 D(\hx,\hxb)\,
	\mathbf{P}^w_N(q) \,\mathbf{P}^w_{\overline{N}}(\qb)
	\times \,
	\frac{(q\,\qb)^{\frac{1}{2} +w+ \sqrt{(k-2) \, \Nwx}}}{\abs{1-q}^2} \,,
\end{split}
\end{equation}
with the condition that
\begin{equation}\label{eq:NXrange}
\Nwx  \;\in\; \frac{k-2}{4} [w^2,(w+1)^2]
\end{equation}
in the $w^{\mathrm{th}}$ spectrally flowed sector.
The continuum states are captured by a density of states $\bm{\varrho}(s;q,\qb)$ and can be expressed as
\begin{equation}\label{eq:fcontbos}
f_\text{cont}(\tau, \overline{\tau})
\=
	\sum_{w=1}^\infty\,
	  \int_{-\infty}^\infty\, \frac{ds}{i\pi}\,\frac{\bm{\varrho}(s;q,\qb)}{\abs{1-q}^2}\, \, (q\,\qb)^{\frac{k}{4}\,w + \frac{1}{w} \left(\frac{s^2+\frac{1}{4}}{k-2} + N + \hx -1 \right)} \,.
\end{equation}
This density of states will not play a role in our analysis, so we refer to~\cite{Maldacena:2000kv} for the details of its derivation (see also~\cref{sec:bosZeval}).

\subsection{The spectrum from the spacetime CFT perspective}\label{sec:Zwsspace}

We now have the basic results, which we can interpret in terms of the spacetime CFT.
For the moment, we ignore the contribution from the closed string tachyon $f_\text{tachyon}$, as it is an artifact of the bosonic string theory and is  absent in the superstring (for a suitable choice of GSO projection).

We therefore focus on the discrete states, the contribution captured in~\eqref{eq:bosshort}, and we let $k >3$ be arbitrary.  Notice that each term in the summand is of the form
\begin{equation}
\frac{q^{\hs}\; \qb^{\hsb}}{\abs{1-q}^2}\,.
\end{equation}
The numerator can clearly be interpreted as corresponding to primaries of the spacetime CFT with scaling dimensions $\hs$ and $\hsb$.
The denominator also has a simple interpretation: it corresponds to the action of derivatives $\partial_x$ and $\overline{\partial}_{\overline{x}}$
in the spacetime CFT. Thus, one can interpret the
typical contribution from the worldsheet in terms of spacetime primaries.\footnote{We thank Shiraz Minwalla for a helpful discussion on this point.}

However, there are some subtleties at low-levels. To see this, let us look at the sector with no spectral flow $w=0$, and turn off the excitations in the internal CFT,  i.e., focus on the identity operator $\hx = \hxb =0$. This sector contributes\footnote{ The $N=0$ term is absent here, and in fact is captured by the tachyon contribution described above. }
\begin{equation}\label{eq:boslow}
\begin{split}
f_\text{disc}(\tau, \overline{\tau})
& \;\supset \;
	\sum_{N=1}^{\frac{k-2}{4}}
	\mathbf{P}^0_N(q) \, \mathbf{P}^0_{\overline{N}}(\qb)
	\times \,
	\frac{(q\,\qb)^{\frac{1}{2} + \sqrt{(k-2) \,(N-a_k)}}}{\abs{1-q}^2}\,, \\
&\=
	\underbrace{1 + \frac{q^2}{1-q} + \frac{\qb^2}{1-\qb} + \frac{(q\,\qb)^2}{\abs{1-q}^2}}_{N=1} \; + \; \underbrace{\frac{1+ q^4}{1-q}\, \frac{1+ \qb^4}{1-\qb} \, (q \qb)^{-\frac{3}{2}+ \sqrt{(k-2)+\frac{1}{4}}}}_{N=2} \; + \; \cdots \,.
\end{split}
\end{equation}
In the second line, we explicitly employed~\eqref{eq:PNlvals} to write out the first few terms.

Of interest to us  is the contribution from the first oscillator level $N=1$. This has states that are chiral in the spacetime CFT.
In fact, these are the boundary graviton excitations expected from the semiclassical Brown-Henneaux analysis.
The first three terms in the $N=1$ piece are to be interpreted as the identity, the action of a single instance $\Ls_{-n}$ and $\Lsb_{-n}$ ($n\geq 2$), respectively.
In other words, these three terms constitute the vacuum Virasoro module. The fourth term corresponds to a primary with $\hs = \hsb = 2$.

For $N\geq 2$ we simply get a set of primaries with non-trivial spacetime weights, and can interpret the factor of $\abs{1-q}^2$ as the action of boundary derivatives, as before.
In the semiclassical limit, $k \gg 1$, these states have weights of $\order{\sqrt{k}}$ and are heavy. This is to be expected; these states are obtained by the action of string oscillators,
so should have spacetime energy proportional to $\ell_s^{-1} \sim \sqrt{k}$.

Note, however, that there \emph{is} one operator that is light, which arises at level $N=1$, with spacetime conformal weights $\hs = \hsb = 2$.
This operator can be identified with the spacetime dilaton, as noted originally in~\cite{Kutasov:1999xu}.
Analogously to~\eqref{eq:SpT}, we can write it in terms of the worldsheet fields as
\begin{equation}\label{eq:TTbarop}
D(x,\overline{x}) \= \int\, d^2z \, (\partial_x J + 2\,\partial_x^2 J)(\partial_{\overline{x}} \bar{J} + 2\,\partial_{\overline{x}}^2 \bar{J}) \Phi_1 \,.
\end{equation}
This operator originates from the $\slt$ WZW model, and exists for any choice of internal worldsheet CFT $X$.
It arises as a product of a left-moving and right-moving current, and has been interpreted as the `single-trace' $T\overline{T}$ operator in the spacetime CFT~\cite{Giveon:2017nie}.

This interpretation is borne out from our spectrum: the single string Hilbert space contains exactly one such operator, independent of the details of $\mathcal{Z}_X$.
Moreover, the $N=1$ level states in the discrete part of the string free energy nicely factorize,
\begin{equation}\label{eq:boscurrents}
f_\text{disc}(\tau, \overline{\tau}) \Big|_{N=1, \hx = \hxb = 0}  \= \biggl( 1+ \frac{\qb^2}{1-\qb} \biggr) \biggl( 1+ \frac{q^2}{1-q} \biggr)\,,
\end{equation}
showing the presence of the currents and the spacetime dilaton operator.

Consider the contribution to the discrete spectrum from level $N=1$ of $\slt$ WZW model, and the
identity sector of the $X$ CFT. The leading multi-string  partition function\footnote{The sum over the free string Fock space needs to be regulated, since the identity
operator can contribute an arbitrary number of times. We implicitly assume that this has been done with a suitable regulating
scheme.} in~\eqref{eq:Zwssp}
is simply the product of Virasoro vacuum characters, and the character for a weight $\hs  = \hsb =2$ state, i.e.,
\begin{equation}\label{eq:bosexpZws}
\exp \bigl(\mathcal{Z}_\text{ws}(\tau,\overline{\tau}) \bigr) \=
\abs{\prod\limits_{n=2}^\infty\, \frac{1}{(1-q^n)}}^2 \times
 \prod_{n_1,n_2=0}^\infty \frac{1}{1-q^{2+n_1}\, \qb^{2+n_2}} \times \dots \,.
\end{equation}
As expected, we reproduce the contribution from the boundary multi-graviton states as we would anticipate from semiclassical gravity calculations~\cite{Maloney:2007ud,Giombi:2008vd}.
The spacetime dilaton contribution shows up as a separate factor.

In the semiclassical gravity limit, $k \gg 1$, in addition to the spacetime currents,
we only have the one state with
conformal dimensions of $\order{1}$, which corresponds  to the spacetime operator~\eqref{eq:TTbarop}. All other states are heavy, with dimensions of $\order{\sqrt{k}}$.
Retreating from the semiclassical limit, for $\lads \sim \ell_s$, we note that the spectrum retains the
twist gap of~$\order{1}$ due to this state corresponding to the dilaton, as long as $k >3$.
As $k \to 3$, the allowed states are constrained to obey $ N + \hx \in \frac{1}{4} [3+2\,w + 3\,w^2,4 + 4\, w + 3\, w^2]$, which constrains the oscillator modes depending on the details of the internal CFT.
For $k=3$, it was demonstrated in~\cite{Gaberdiel:2017oqg} there are (chiral) higher spin currents which sit at the bottom of the continuum representations.

To summarize, the bosonic string partition function on \AdS{3} $\, \times X$ has states naturally organized in terms of the boundary Virasoro characters.
In particular, the final result contains the Virasoro vacuum character, which was one of the main aspects we wanted to check.
Furthermore, for any $k>3$ we see that the spectrum has a twist gap -- for generic $X$, there are no other chiral currents in the theory.

\section{Superstring theory on \texorpdfstring{\asx{T^4}}{AdS3 x S3 x T4}}\label{sec:superT4}

We now turn to the superstring, and focus on the Type II string on \asx{T^4}. The specific choice of chiral GSO projection (IIA versus IIB) is immaterial here, since the two theories are related by T-duality on the torus. We phrase our analysis in a language adapted to the Type IIB presentation, and work at a generic point on the $T^4$ moduli space. We realize the geometry in terms of the near-horizon limit of a bound state of $N_1$ fundamental strings and $N_5 \equiv k$ NS5-branes in the near-horizon decoupling limit. The radius of \AdS{3} and $\mathbf{S}^3$, as well as the value of the string coupling, are frozen in terms of these parameters as\footnote{Here we quote the three-dimensional string coupling to facilitate comparison with the Brown-Henneaux formula~\eqref{eq:cN4}. The six-dimensional string coupling, which is more relevant since the \AdS{3}  and $\mathbf{S}^3$ length scales are commensurate, is instead $g_{s,6d}^2 = \frac{k}{N_1}$. }
\begin{equation}\label{eq:kn1fix}
\frac{\lads}{\ell_s} \= \sqrt{k} \,, \qquad \qquad
g_s^2  \= \frac{1}{N_1\,\sqrt{k} }\,.
\end{equation}

The 32 supercharges of IIA/B are unbroken upon
compactification on~$T^4$.
However, the  \AdS{3} $\times \,\mathbf{S}^3$ background
only realizes two copies
of  $\mathrm{PSU}(1,1|2)  \supset \mathrm{SL}(2,\mathbb{R}) \times \mathrm{SU}(2)$ as the super-isometry group.
This super-isometry group is
enhanced by the \AdS{3} asymptotics to the small  $\mathcal{N} = (4,4)$ superconformal algebra, with central charge
\begin{equation}\label{eq:cN4}
c \= \frac{3\,\lads}{2\,\ell_P} \= 6\, N_1\, k \,.
\end{equation}
The dual holographic spacetime 2d CFT should realize this symmetry.

On general grounds, we expect that the computation of the gravitational path integral on the above
background should organize itself into characters of the $\mathcal{N} = (4,4)$ superconformal algebra. In addition, owing to the presence of the $T^4$ factor, we should also see the character of a free supermultiplet of $\mathrm{U}(1)^4$ currents and their superpartners.
The task ahead of us is to realize these expectations  from the worldsheet partition function.

\subsection{Canonical quantization of the superstring}\label{sec:cqss}

We briefly summarize the analysis of the superstring in~\cite{Giveon:1998ns},
and comment on the chiral primary spectrum obtained from supergravity. The latter was matched with the worldsheet computation in~\cite{Kutasov:1998zh} (for later discussions, see also~\cite{Dabholkar:2007ey} and~\cite{Ferreira:2017pgt}).

The worldsheet theory on \asx{T^4} is described by an $\mathcal{N}=(1,1)$ supersymmetric sigma model coupled to worldsheet supergravity. The sigma model comprises:
\begin{itemize}[wide,left=0pt]
\item  An $\slt_k$ super-WZW model
with~$c=\frac{9}{2} + \frac{6}{k}$
corresponding to the \AdS{3} part, with currents $J^A$. By a field
redefinition~\cite{DiVecchia:1984nyg}, we can obtain three decoupled fermions $\psi^A$, $A \in \{\pm, 3\}$.
The algebra then decomposes into a bosonic
subalgebra, with currents $j^A$ at level $k+2$ and a fermionic piece, with currents $\widehat{j}^A$ at level $-2$.
\item An $\sut_k$ super-WZW model
with~$c=\frac{9}{2} - \frac{6}{k}$ corresponding to the $\mathbf{S}^3$, with currents $K^a$. Again, by field redefinition, we obtain three decoupled fermions $\chi^a$, $a \in \{\pm, 3\}$.
The algebra decomposes into a bosonic
subalgebra, with currents $k^a$ at level $k-2$ and a fermionic piece, with currents $\widehat{k}^z$ at level $2$.
\item Additionally, we have four free bosons and fermions $Y^i$ and $\lambda^i$, $i\in \{1,2,3,4\}$,
with~$c=6$ corresponding to the $T^4$ directions.
\end{itemize}

The vertex operators on the worldsheet are constructed from the primaries of these WZW models. The operators of interest are as follows:
\begin{itemize}[wide,left=0pt]
\item  For the decoupled bosonic $\slt_{k+2}$ model, we have the primaries  $\Phi_j(z,x,\overline{z},\overline{x})$, which have worldsheet conformal dimension $\Delta = \overline{\Delta} = -\frac{j(j-1)}{k}$. The unitarity bound demands~\cite{Maldacena:2000hw}
\begin{equation}
\frac{1}{2} \leq j \leq \frac{k+1}{2} \,.
\end{equation}
Here $z,\overline{z}$ are the worldsheet coordinates, while $x,\overline{x}$ are auxiliary variables which can be viewed as parameterizing the spacetime CFT Riemann sphere, as in the bosonic theory.
\item Similarly, for the  $\sut_{k-2}$ model, the primaries of interest are $V_l(z,y,\overline{z},\overline{y})$. Their worldsheet conformal dimension is $\Delta = \overline{\Delta}= \frac{l\,(l+1)}{k}$ and $y,\overline{y}$ are coordinates conjugate to the discrete azimuthal quantum number.
\item In the NS sector, we have the fermionic operators $\psi$ and $\chi$ defined by
\begin{equation}
\begin{split}
\psi(x;z) &\=
	-\psi^+(z) + 2\, x\, \psi^3(z) - x^2\, \psi^-(z) \,, \\
\chi(y;z)
&\=
	-\chi^+(z) + 2\, y\, \chi^3(z) + y^2\, \chi^-(z) \,.
\end{split}
\end{equation}
The former is a  primary of weight $\Delta=-1$ and $\slt_k$ spin~$1$, while the latter has weight $\Delta=+1$ and $\sut_k$ spin~$1$.

\item For the R-sector vertex operators, we introduce the bosonized representation:
\begin{equation}
\begin{aligned}
&\partial H_1  = \frac{2}{k}\, \psi^2\, \psi^1  \,, \qquad
& \partial H_2  = \frac{2}{k}\, \chi^2\, \chi^1  \,, &\quad  &
\partial H_3 =  \frac{2i}{k}\, \chi^3\, \psi^3  \,, \\
&\partial H_4  = \lambda^2\, \lambda^1  \,, \qquad
&\partial H_5  =  \lambda^4\, \lambda^3  \,. &\quad &
\end{aligned}
\end{equation}
In writing these we are eliding over the cocycle factors; for explicit formulae, see~\cite{Dabholkar:2007ey}.
\end{itemize}

To define the superstring we need to impose the GSO projection. As discussed in~\cite{Giveon:1998ns}, the natural projection is different from the usual one employed in flat spacetime string constructions.
Although there is an accidental $\mathcal{N}=2$ superconformal symmetry in the worldsheet theory,
its $\mathrm{U}(1)$ R-symmetry is not used to define a $\mathbb{Z}_2$ involution for the worldsheet fermion
number operator, as that leads to a spacetime superalgebra  different from the boundary theory~\cite{Giveon:1998ns}.

Naively, the spacetime supersymmetry generators are built out of the spin-fields, and one expects them to be given by contour integrals, viz.,
\begin{equation}\label{eq:Qdefs}
Q \= \oint \,dz \,  e^{-\frac{\varphi}{2}}\, \exp\biggl(\frac{i}{2}\,\sum_{I=1}^5 \epsilon_I\,H_I \biggr) \,,
\end{equation}
with $\varphi$  being the superconformal ghost contribution.
The $\slt$ and $\sut$ super-WZW models have cubic terms involving the fermions in the worldsheet superconformal
generators $G_{\text{ws}} \supset \psi^+ \psi^- \psi^3 + \chi^+ \chi^- \chi^3$.  Locality with these fields
imposes constraints on the spacetime supercharges; only the $Q$s satisfying the constraints
\begin{equation}\label{eq:gksgso}
\prod_{i=1}^5\, \epsilon_I \= 1 \,, \qquad
\epsilon_1\, \epsilon_2 \, \epsilon_3 \= 1\,,
\end{equation}
generate admissible spacetime symmetries. Note that the cubic terms in the worldsheet current are a feature of
the curved target spacetime (and lack of enhancement of worldsheet supersymmetry).
Nevertheless, the GSO projection constrains the spin-fields associated with the $T^4$ part,
for~\eqref{eq:gksgso} implies $\epsilon_4 \,  \epsilon_5 = 1$.
The net result is that only a quarter of the supersymmetry generators survive and realize the spacetime $\mathcal{N}=4$ superalgebra.

To give the explicit expressions for the spacetime supercharges and the physical vertex operators, it is useful to introduce spin fields for the $\slt \oplus \sut$ part as
\begin{equation}
\begin{split}
S^+_\alpha(x,y)
&\=
    -i\, x\,y\, e^{\frac{i}{2} (-H_1 - H_2 + H_3)} - x\,  e^{\frac{i}{2} (-H_1 + H_2 - H_3)}
    + i\,y\, e^{\frac{i}{2} (H_1 - H_2 - H_3)} + e^{\frac{i}{2} (H_1 + H_2 + H_3)} \,, \\
S^-_{\alpha}(x,y)
&\=
    i\, x\,y\, e^{\frac{i}{2} (-H_1 - H_2 - H_3)} + x\,  e^{\frac{i}{2} (-H_1 + H_2 + H_3)}
    + i\,y\, e^{\frac{i}{2} (H_1 - H_2 + H_3)} + e^{\frac{i}{2} (H_1 + H_2 - H_3)} \,. \\
\end{split}
\end{equation}
The GSO projection~\eqref{eq:gksgso} picks out $S^+_\alpha(x,y)$ for the supercharges, which when combined with the torus fermions leads to the following expressions
\begin{equation}
Q\indices{_\alpha^a} \= \oint \, dz\, e^{-\frac{\varphi}{2}} \, S_\alpha^+(x,y)\, e^{\pm \frac{i}{2} (H_4 + H_5)} \,.
\end{equation}
The supercharges transform under $(\mathbf{2},\mathbf{2},\mathbf{2},\mathbf{1})$ of the
global $\mathrm{SL}(2) \times \mathrm{SU}(2)_R \times \mathrm{SU}(2)_1 \times \mathrm{SU(2)}_2$ where the last
two groups refer to the local rotation group along the $T^4$ directions. Specifically, the spin fields $e^{\pm \frac{i}{2} (H_4 + H_5)}$ make up a chiral spinor of $\mathrm{SO}(4)$ which transforms in the $(\mathbf{2},\mathbf{1})$ representation for the
decomposition into $\mathrm{SU}(2)_1 \times \mathrm{SU(2)}_2$.

The physical vertex operators are constructed in terms of the primary operators given above. For later reference, it will be helpful to collate the information about the chiral primary operators of the CFT~\cite{Kutasov:1998zh}. Following~\cite{Dabholkar:2007ey}, we define a combination of the $\slt_{k+2}$ and $\sut_{k-2}$ bosonic primaries
\begin{equation}
\mathcal{O}_j(x,\overline{x},y,\overline{y}) \= \Phi_{j+1}(x,\overline{x})\, V_j(y,\overline{y}) \,, \qquad
j \=  0, \frac{1}{2},1,\dots  \,.
\end{equation}
Here we have suppress the dependence on the worldsheet coordinates for convenience.

Focusing for the moment on the left-moving part,\footnote{ Strictly speaking the $\slt$ vertex operators do not
admit left-right factorization, but it is useful to pretend that they do to see the basic structure.} we build chiral states in terms of vertex operators
\begin{equation}\label{eq:wsvops}
\begin{aligned}
\mathcal{A}_{j}
&\=
    e^{-\varphi}\, \mathcal{O}_j(x,y)\, \psi(x) \,,&
 \hs
&\=
    j\,, \\
\mathcal{B}_{j+1}
&\=
    e^{-\varphi}\, \mathcal{O}_j(x,y)\, \chi(y) \,, &
\hs
&\=
    j+1\,, \\
\left( \mathcal{R}\indices{_\alpha^{\dot{a}}} \right)_{j+\frac{1}{2}}
&\=
    e^{-\frac{\varphi}{2}}\, \mathcal{O}_j(x,y)\, S^-_{{\alpha}} (x,y)\, e^{\pm\frac{i}{2}\, (H_4- H_5)} \,, &
\quad \hs
&\=
    j+ \frac{1}{2}\,.
\end{aligned}
\end{equation}
These represent spacetime states with spacetime conformal dimension equal to the R-charge $\hs = \js $, with the conformal dimension as specified above,  and form the holomorphic half of a chiral primary operator. We will shortly put together the left and right movers and write down the full representation content.

Note that the R-sector vertex operators that are in the BRST cohomology are projected into $\epsilon_1 \,\epsilon_2 \, \epsilon_3 = \epsilon_4 \, \epsilon_5 =-1$. This is reflected in the spin-field contribution to $\mathcal{R}\indices{_{\dot{\alpha}}^{\dot{a}}}$, which transforms as
$(\mathbf{1},\mathbf{2})$ of $\mathrm{SU}(2)_1 \times \mathrm{SU(2)}_2$.
In writing the above, we are partially eliding over the Clebsch-Gordon decomposition under the $\slt$ and $\sut$ algebras. The operators need to be projected into the correct representations, which the interested reader can find described in~\cite{Kutasov:1998zh,Dabholkar:2007ey}.

\subsection{The spacetime \texorpdfstring{$\mathcal{N}=4$}{N=4} supercharacters} \label{sec:n4super}

In order to discuss the chiral primary spectrum of the spacetime CFT,
it is useful to first take a short detour and introduce the characters of the $\mathcal{N} =4$ superconformal algebra.
These characters were computed in~\cite{Eguchi:1987wf} directly by examining the structure of the modules.
The algebra is characterized by a level $\ks$ which determines the spacetime central charge $c=6\, \ks$.

We work in the NS sector, where the left-moving states are labeled as $\ket{h,\ell}$, with $h$
being the conformal weight and $\ell$ the spin under $\mathrm{SU(2)}_R$. Let $q = e^{2\pi i\,\tau}$
and $r = e^{2\pi i\,\rho}$.
The characters for both long and short representations can be uniformly written as follows,\footnote{ In writing these characters we will not include the vacuum energy contribution $q^{-c/24}$ for later convenience.}
\begin{equation} \label{eq:genericchars}
\begin{split}
\Tr_{_\text{NS}} \Bigl( y^F\,q^{\Ls}\, r^{\Js} \Bigr)
&\=
	q^{h+\frac{1}{4}}\, F_{_\text{NS}}(r,q,y)\, \Bigl(\mu_h(r,q,y) - \mu_h(r^{-1},q,y)\Bigr) \,, \\
F_{_\text{NS}}(r,q,y)
&\=
	-i \,\frac{1}{\eta(\tau)^3\, \vartheta_1(\rho, \tau)} \, \Theta^{(y)}\left(\tfrac{\rho}{2},\tau\right)\, \Theta^{(y^{-1})}\left(\tfrac{\rho}{2},\tau\right) \,, \\
\Theta^{(y)}\left(\tfrac{\rho}{2},\tau\right)
&\; \equiv \;
	 \prod_{n=1}^{\infty}
		(1-q^n)\, (1+y\,r^{\frac{1}{2}}\,q^{n-\frac{1}{2}})\,  (1+y\,r^{-\frac{1}{2}}\,q^{n-\frac{1}{2}}) \,.
\end{split}
\end{equation}
The non-trivial information about the module is in the function $\mu_h$. It   depends on whether the representation is long or short, and is given by
\begin{equation}
\mu_h(r,q,y) =
\begin{dcases}
&
	\sum_{m\in \mathbb{Z} } \,
		r^{ (\ks+1)\, m + \ell + \frac{1}{2} }\, q^{(\ks+1)\, m^2 + (2\ell+1)\,m} \,, \qquad\qquad  h > \ell \,,\\
&
	\sum_{m\in \mathbb{Z} } \,
	\frac{r^{(\ks+1)\, m +  \ell + \frac{1}{2}}\, q^{(\ks+1)\, m^2 + (2\ell+1)\,m}}{(1+r^{\frac{1}{2}}\,y\, q^{m+\frac{1}{2}})\, (1+r^{\frac{1}{2}}\,y^{-1}\, q^{m+\frac{1}{2}}) }\,, \qquad h = \ell \,.
\end{dcases}
\end{equation}
It is useful to realize that $\Theta^{(1)}(\frac{\rho}{2},\tau) = \vartheta_3(\frac{\rho}{2},\tau)$ and $\Theta^{(-1)}(\frac{\rho}{2},\tau) = \vartheta_4(\frac{\rho}{2},\tau)$ arise from the fermionic generators of the superalgebra, while the denominator factor is the bosonic contribution.

The character of a long multiplet can be expressed in terms of characters of the bosonic $\sut_{\ks-1}$.\footnote{This is an observation about the $\mathcal{N} =4$ superconformal algebra characters, cf.~\cite{Eguchi:1987wf}, which is unrelated to the shift in the level of the bosonic worldsheet $\sut$ current algebra.}  Our interest here is in the character of the short multiplet $h = \ell$. In this case, the additional factor in the function $\mu_h$ is present to account for null states in the module, and does not
have a simple expression in terms of the bosonic $\sut$ current algebra.
We work with this character and primarily focus on the trace with the fermion number insertion.
Accordingly, we define the character for the short multiplet,\footnote{Generic characters can be recovered from the expressions in~\eqref{eq:genericchars}.}
\begin{equation}\label{eq:N44vacNS}
\begin{split}
\etchar(\ell) \= \etchar(\ell;r,q)
&\=
	\Tr_{_\text{NS}} \left( (-1)^F\,q^{\Ls}\, r^{\Js} \right)_{h=\ell}  \\
&\=
	-i\, q^{\ell+\frac{1}{4}}\,
	\frac{\vartheta_4^2(\frac{\rho}{2},\tau)}{\eta(\tau)^3\,\vartheta_1(\rho, \tau)} \Bigl( \mu_\ell(r,q) - \mu_\ell(r^{-1},q)\Bigr) \,, \\
\mu_\ell(r,q)
&\=
	\sum_{m\in \mathbb{Z} } \, \frac{r^{(\ks+1)\, m +  \,\ell + \frac{1}{2}}\, q^{(\ks+1)\, m^2 + (2\ell+1)\,m}}{(1 - r^{\frac{1}{2}}\, q^{m+\frac{1}{2}})^2 } \,.
\end{split}
\end{equation}
The terms with $m\neq 0$ in the above expression for~$\mu(r,z)$ scale as $q^{m^2\, \ks} \sim q^{\frac{c}{6}\,m^2}$.
From the trace interpretation, we see that these
states are heavy (with $\Ls = \order{\ks}$ as $\ks \to \infty$) compared to terms with $\Ls = \order{1}$ from the~$m=0$ term.
We will later characterize the~$m \neq 0$ terms as contributions from other saddles with non-trivial holonomy
for the R-symmetry background gauge field.

Now we proceed to discuss the~$m=0$ term, which captures the light fluctuations of supergravity and strings around thermal \AdS{3}.
We write this term as the factored expression
\begin{equation} \label{eq:chiNSfactors}
\etchar(\ell;r,q)\Big|_{m=0}
\= \,
C_\text{osc}(r,q) \, \left(\schol(\ell;r,q) - q\, \delta_{\ell,0}\right)  \,. \\
\end{equation}
The first factor $C_\text{osc}$, given by
\begin{equation}\label{eq:Coscdef}
C_\text{osc}(r,q) \; \equiv \;
-i\, \underbrace{{\frac{q^{\frac{1}{4}}}{\eta(\tau)}}}_{\Ls_{-n}} \,
	\underbrace{\frac{r^\frac{1}{2} - r^{-\frac{1}{2}}}{\vartheta_1(\rho, \tau)}}_{\Js_{-n} } \,
	\underbrace{\frac{\vartheta_4^2(\frac{\rho}{2},\tau)}{\eta(\tau)^2 \, (1-r^{\frac{1}{2}}\,q^{\frac{1}{2}})^2\, (1-r^{-\frac{1}{2}}\,q^{\frac{1}{2}})^2}}_{\Gs_{-r}} \,,
\end{equation}
captures the one-loop determinant of the oscillator states.
The factors have been assembled in~\eqref{eq:Coscdef} based on the action of the $\mathcal{N}=4$ raising generators.
This contribution would have been the entire content of a module
if it were freely generated by the action of $\Ls_{-n}$, $\Js_{-n}$, $n \ge 1$ and $\Gs_{-r}$, $r \ge 3/2$ respectively.

When $\ell \geq 0$ the second factor in~\eqref{eq:chiNSfactors} is the {\it reduced $\mathfrak{psu}(1,1|2)$ character}
\begin{equation}\label{eq:redchar}
\schol(\ell) \= \schol(\ell;r,q)
\=
	q^\ell  \, \chi_{\ell}(r) -2\,q^{\ell+\frac{1}{2} } \, \chi_{\ell-\frac{1}{2}}(r) +q^{\ell+1}\, \chi_{\ell-1}(r) \,,
\end{equation}
which is written in terms of the usual~$\sut$ characters,
\begin{equation}
    \chi_\ell(r) \=
    \frac{r^{\ell+ \frac{1}{2}} - r^{-\ell- \frac{1}{2}}}{r^{\frac{1}{2}}- r^{- \frac{1}{2}}} \,, \qquad \ell \geq 0 \qquad \qquad ( \text{and} \; \; \chi_\ell(r)= 0 \,, \; \;  \ell <0) \,.
\end{equation}
This simply encodes the observation that the supermultiplet is filled out by acting on a bottom
component by the supercharges $Q\indices{_\alpha^a}$ (which are in a R-symmetry doublet).  Starting with a state $\ket{\ell,\ell}$, this reduced character encodes the structure of the  multiplet with the $\mathrm{SU}(2)$ spin decreasing by~$\frac{1}{2}$ at each step: $\{\ket{\ell,\ell},\,  Q\indices{_\alpha^a} \ket{\ell,\ell}, Q\indices{_\alpha^a}\,Q\indices{_\beta^b} \ket{\ell,\ell}\}$.

The situation described above is for a generic multiplet. There are some special features at small values of $\ell$, owing to the existence of further truncations (or additional null states). For example,
\begin{equation}\label{eq:redcharlow}
\schol(0) \= 1  \,, \qquad
\schol\bigl(\tfrac{1}{2}\bigr)
\=  q^\frac{1}{2}\,\chi_{\frac{1}{2}}(r) -2\,q  \,.
\end{equation}
The first equation above states that the vacuum module contains only the identity state $\ket{0,0}$, and furthermore $\Ls_0\ket{0,0} =0$.
This is consistent with the extra term~$\delta_{\ell,0}$ in~\eqref{eq:chiNSfactors}.
The second equation captures the fact that multiplet built atop $\ket{\frac{1}{2}, \frac{1}{2}}$ truncates after the action of a single supercharge.

Returning to our discussion of the vertex operators on the worldsheet, note that there is a natural map between the reduced characters $\schol(j)$
and the multiplets built atop the operators recorded in~\eqref{eq:wsvops}. Specifically,
\begin{equation}
\begin{split}
\schol(j+1)
&\; \mapsto \;
    \{\mathcal{B}_j , Q\indices{_\alpha^a} \mathcal{B}_j , Q\indices{_\alpha^a} \, \indices{_\beta^b}\,\mathcal{B}_j \}\,,  \\
\schol(j)
&\; \mapsto \;
    \{\mathcal{A}_j, Q\indices{_\alpha^a} \mathcal{A}_j , Q\indices{_\alpha^a}\, Q\indices{_\beta^b}\,\mathcal{A}_j\} \,, \\
\schol\Bigl(j+\frac{3}{2}\Bigr)
&\; \mapsto \;
    \{\mathcal{R}\indices{_{\dot{\alpha}}^{\dot{a}}} , \;
    Q\indices{_\alpha^a}\mathcal{R}\indices{_{\dot{\alpha}}^{\dot{a}}} ,\; Q\indices{_\alpha^a}\, Q\indices{_\beta^b}\,\mathcal{R}\indices{_{\dot{\alpha}}^{\dot{a}}}
    \} \,.
\end{split}
\end{equation}
This identification will be extremely useful when we discuss the partition function, as we can quickly identify the characters and thence the states they correspond to.

Finally, we put together the left and right movers to obtain the complete set of chiral primaries. Working with the reduced characters to encode this data, we can identify three essential combinations, which can be labeled by the value of the spin $|\hs-\hsb|$. Define\footnote{
	Note that $\overline{\sccft}^{(2)}(j)$ and $\overline{\sccft}^{(\frac{3}{2})}(j)$ are complex conjugates of $\sccft^{(2)}(j)$ and $\sccft^{(\frac{3}{2})}(j)$, respectively,  with  opposite values of helicity.}
\begin{equation}\label{eq:cporeps}
\begin{split}
\sccft^{(1)}(j)
&\=
	\schol(j)\; \scholb(j) \,,  \\
\sccft^{(\frac{3}{2})}(j) \,
&\=
	\schol(j+\frac{1}{2})\; \scholb(j) \,,\\
\sccft^{(2)}(j)
&\=
	\schol(j+1)\; \scholb(j) \,. \\
\end{split}
\end{equation}
These {\it reduced $\mathfrak{psu}(1,1|2) \oplus \overline{\mathfrak{psu}(1,1|2)}$ characters},
which are labeled by the maximum value of the spin in the multiplet (given by the superscript)
and are functions of~$(r,q)$,
should be interpreted in the following manner.
The bottom component of the supermultiplets are NS-NS sector states obtained from combining $\mathcal{A}_j$ and $\mathcal{B}_{j+1}$ or R-NS states that take either
of these and pair them with $\mathcal{R}\indices{_{\dot{\alpha}}^{\dot{a}}}$.
At low values of the spin, we have again some simplifications
\begin{equation}\label{eq:lowjxi}
\begin{split}
\sccft^{(1)}(0)
&\=
	\schol\left( 0 \right) \, \scholb (0)
	\= 1 \,, \\
\sccft^{(\frac{3}{2})}(0)
&\=
	\schol\left( \tfrac{1}{2} \right) \, \scholb (0)
	\=  q^\frac{1}{2}\, \chi_{\frac{1}{2}}(r) - 2\, q    \,, \\
\sccft^{(2)}(0)
&\=
	\schol\left( 1 \right) \, \scholb (0)
	\=   q\, \chi_1(r) - 2\,q^{\frac{3}{2}}\, \chi_{\frac{1}{2}}(r) +q^2 \,.
\end{split}
\end{equation}
%

\section{Revisiting the one-loop superstring partition function}\label{sec:superZ}

We now turn to the computation of the partition function of the superstring on \asx{T^4}. The six dimensional geometry obtained after reducing on $T^4$ has the following line element,
\begin{equation}\label{eq:ads3s3metric}
ds^2 \= k \Bigl(\cosh^2\rho\, d\tE^2 + d\rho^2 + \sinh^2\rho\, d\varphi^2 + d\theta^2 + \cos^2\theta\, d\phi_1^2 + \sin^2\theta\, d\phi_2^2   \Bigr) \, .
\end{equation}
The NS-NS B-field supporting the geometry is given by
\begin{equation}
B \= k\Bigl( \sinh^2\rho\, d\tE \wedge d\varphi + \sin^2\theta\, d\phi_1 \wedge d\phi_2 \Bigr) \, .
\end{equation}
We impose the following identifications on the worldsheet fields:
\begin{equation}\label{eq:ads3s3identify}
\begin{split}
(\tE,\varphi,\phi_1, \phi_2)
&
	\;\sim \; (\tE,\varphi +2\pi,\phi_1, \phi_2)
\;\sim \; (\tE,\varphi,\phi_1 + 2\pi, \phi_2) \; \sim \; (\tE,\varphi,\phi_1, \phi_2 +2\pi) \,,\\
(\tE,\varphi,\phi_1, \phi_2)
&\; \sim \;
(\tE +\beta,\varphi + i\,\beta\,\mu,\phi_1 +i\,\beta\, \nu_1, \phi_2 + i\,\beta\, \nu_2)\,.
\end{split}
\end{equation}
The first line simply ensures that the coordinates parameterize a smooth \AdS{3} $\times \,\mathbf{S}^3$ geometry.
The second line contains the non-trivial identification,
i.e., the specification of thermal monodromies for the worldsheet fields.
The periodicity in the Euclidean time and the chemical potential for angular momentum around \AdS{3} are
parameterized by~$\beta$ and~$i \beta \mu$, respectively, as in the bosonic theory.
We also turn on R-charge chemical potentials corresponding to the two Cartan generators of~$\mathbf{S}^3$ characterized by $(\nu_1,\nu_2)$.
Equivalently, the $\mathbf{S}^3$ is fibered over the \AdS{3} as governed by $(\nu_1,\nu_2)$. Note that the
identification does not have a fixed point due the thermal shift, and the resulting manifold is smooth.

\subsection{Construction of the partition function }\label{sec:buildingblocks}

Let us first record the building blocks for constructing the worldsheet partition function on \asx{T^4}. The bosonic contributions are from $\slt_{k+2}$ WZW model, $\sut_{k-2}$ and four free compact bosons.

The contribution of the $\slt_{k+2}$ is the result from~\cite{Maldacena:2000kv} given earlier in~\eqref{eq:Zbosads3} with the replacement $k\to k+2$:
\begin{equation}\label{eq:Zsl2bos}
\begin{split}
\mathcal{Z}_{\slt}(\ts,\tsb;\tau,\overline{\tau})
\=
	\frac{\beta\, \sqrt{k}}{2\pi\, \sqrt{\ts_2}}\,
	\sum_{n,m \in \mathbb{Z}}\,  \frac{\exp \left( -\frac{k\, \beta^2}{4\pi\,\ts_2} \, \abs{m- n\,\ts}^2 + \frac{2\pi}{\ts_2} \, (\Im(\us_{n,m}))^2\right)}{\abs{\vartheta_1(\us_{n,m},\ts)}^2}\,,
\end{split}
\end{equation}
where the holonomies $\us_{n,m}$ are as defined earlier in~\eqref{eq:unmdef}. Note that the classical winding mode contribution remains proportional to $k$ and is insensitive to the shift of the bosonic level.\footnote{
	One can see this directly by analyzing the sigma model, as we illustrate in~\cref{sec:superpart}.}
The contribution of the torus bosons is given by the standard lattice sum over momentum and winding states, as well as the oscillators corresponding to four free left- and right-moving bosonic fields, i.e.,
\begin{equation}\label{eq:Ztorus}
\begin{split}
\mathcal{Z}_\text{torus}(\ts,\tsb)
\=
	\frac{1}{\abs{\eta(\ts)}^8}
	\, \sum_{\Gamma^{4,4}}\, \zs^{p_L^2}\, \overline{\zs}^{p_R^2}\,.
\end{split}
\end{equation}
We also have the $bc$ ghost contribution which is
\begin{equation}\label{eq:Zbcghosts}
\mathcal{Z}_{bc}(\ts,\tsb) \= \abs{\eta(\ts)}^4\,.
\end{equation}
The $\slt$ and torus partition are modular invariant on their own.
The $bc$-ghost contribution combined with an additional factor of~$\ts_2$ is modular invariant,
and the resulting extra factor of~$1/\ts_2$ is precisely what is needed to make the measure modular invariant.

The contribution of the $\sut_{k-2}$  is given by the following expression
\begin{equation}\label{eq:Zsusys3}
\begin{split}
    &\mathcal{Z}^{(n,m)}_{\sut} (\ts,\tsb;\rho, \overline{\rho})  \= \\
&\qquad \qquad e^{\frac{\pi\,k\,\Im(\rho)^2}{\ts_2}\, \abs{n\,\ts-m}^2  -\frac{\pi}{\ts_2}\,(k-2)\, (\Im(\rho_{n,m}))^2  }
	\, \sum_{\ell=0}^{\frac{k}{2}-1} \, \abs{\frac{\Theta_{2\ell+1}^{(k)}(\rho_{n,m},\ts) - \Theta_{-2\ell-1}^{(k)}(\rho_{n,m},\ts)}{\vartheta_1 \left( \rho_{n,m}, \ts \right)} }^2 \,,
\end{split}
\end{equation}
with
\begin{equation}
\Theta_l^{(k)}(\rho,\ts) \= \sum_{n\in \mathbb{Z} + \frac{l}{2k}}\, \zs^{k\,n^2}\, r^{n \,k} \,, \qquad \zs = e^{2\pi i \ts} \,, \qquad r = e^{2\pi i \rho} \,.
\end{equation}
The parameter $\rho = \beta\, \nu_2 - \beta\, \nu_1$ parameterizes the $\mathrm{SU}(2)$ chemical potential. It appears in the combination
\begin{equation}
\rho_{n,m} \= \rho \, (n\,\ts -m)\,,
\end{equation}
exactly as the \AdS{3} chemical potentials appear in the combination~$\us_{n,m}$.

A few comments are in order, as there are elements of this result that appear not to have been appreciated hitherto. Firstly, given the
monodromies~\cref{eq:ads3s3identify}, there is a non-trivial classical contribution. Modes that wind around the thermal circle are twisted by the monodromies along $\phi_1$ and $\phi_2$. Working with a different basis of coordinates to parallel the $\slt$ discussion (cf.~~\cref{sec:bosSWZW}) we deduce this to be
\begin{equation}
\frac{\pi\, k\,\Im(\rho)^2}{\ts_2}\, \abs{n\,\ts-m}^2\,,
\end{equation}
which is the first term in the pre-factor in~\eqref{eq:Zsusys3}.
The fluctuations around this classical solution can be argued to give the A-series modular invariant~\cite{Cappelli:1986hf}.
The elliptic parameter of the theta functions again is sensitive to the monodromies, and appears as the  combination $\rho_{n,m}$ introduced above.
We also note that the second piece of the exponential prefactor is necessary for modular invariance. The sum over the $\sut$ representations, which are capped off at, $\frac{k}{2}-1$ is only modular covariant.
The prefactor proportional to $k-2$ is made up of two contributions: the numerator gives a factor $k$, while
the shift of $2$ arises from the $\vartheta_1$ in the denominator (part of this will cancel when we include the fermions).
All in all, the bosonic $\sut$ partition function is invariant under modular transformations (with respect to the worldsheet torus), as well as
elliptic transformations of the $\mathrm{SU}(2)$ chemical potential. We provide a justification for these statements
in~\cref{sec:superpart}.

Next, we turn to the fermion contribution. The fermions making up the $\slt_k$  and $\sut_k$ super-WZW models are charged under the respective bosonic symmetries.
Since we have thermal boundary conditions, they transform in the $(n,m)$ sector by a phase rotation.
Effectively, they perceive a background gauge field $\us_{n,m}$ or $\rho_{n,m}$.
Specifically, the $\psi^\pm$ transform with charges $\pm 1$ under $\slt$, while $\chi^\pm$ have charges $\pm 1$
under $\sut$, respectively. Thus, the fermion characters will be twisted by the holonomies $\us_{n,m}$ and $\rho_{n,m}$.

For a fixed spin structure, the trace over the fermion Hilbert space leads to the holomorphic fermion character, $a=1,\dots, 4$, depending on~$\ts$, and~$(\tau,\rho)$,
\begin{equation}\label{eq:ZfermionsT4}
\begin{split}
\mathcal{Z}^{(n,m)}_a
&\=
	\underbrace{\frac{e^{-\frac{\pi}{\ts_2}\, (\Im(\us_{n,m}))^2}\,\vartheta_a(\us_{n,m},\ts)\, \sqrt{\vartheta_a(0,\ts)}}{\eta(\ts)^{\frac{3}{2}}}}_{\slt}
	\times
	\underbrace{e^{-\frac{\pi}{\ts_2}\, (\Im(\mathfrak{\rho}_{n,m}))^2}\frac{\vartheta_a(\rho_{n,m},\ts)\, \sqrt{\vartheta_a(0,\ts)}}{\eta(\ts)^{\frac{3}{2}}}}_{\sut} \\
&\qquad \qquad
	\times
	\underbrace{\frac{\vartheta_a(0,\ts)^2}{\eta(\ts)^2}}_{T^4}
	\times
	\underbrace{\frac{\eta(\ts)}{\vartheta_a(0,\ts)}}_{\beta\gamma} \,.
\end{split}
\end{equation}
The exponential prefactors in the $\slt$ and $\sut$ characters originate from the holomorphic anomaly for chiral fermion determinants, cf.~\cite{DHoker:1986hhu,Alvarez-Gaume:1986rcs}.
When we combine left and right movers, as with the bosonic determinants, we demand that the result preserves elliptic invariant and modular symmetries.
A cleaner way to organize the calculation is to combine the determinants of the bosons and fermions,
since ratios such as $\vartheta_a(\us_{n,m},\ts)/\vartheta_1(\us_{n,m},\ts)$ are well-behaved under these symmetries.
As it stands, one can see that the left-right combination of fermion characters will have a factor that cancels against the boson determinants
(specifically, the~$k$-independent part of the $\sut$ partition function~\eqref{eq:Zsusys3}, as noted above).

We need to put together the fermion characters, and sum over spin structures to implement the GSO projection~\cite{Seiberg:1986by}.
Before doing so, we have to specify the boundary conditions for the fermions in spacetime.
The simplest choice is the periodic boundary conditions for the spacetime fermions around the thermal circle.
The  chiral GSO projection for the IIA/B superstring then involves the following combination:
\begin{equation}\label{eq:pfermion}
\begin{split}
\bigl( \mathcal{Z}^\pm_{_\text{PF}} \bigr)^{(n,m)}
&\=
	\Tr_{_\text{NS}}\left(\frac{1-e^{i\pi F}}{2}\, e^{2\pi i \, \ts\, L_0}  \right) -
		\Tr_{_\text{R}}\left(\frac{1 \pm e^{i\pi F}}{2}\, e^{2\pi i \, \ts\, L_0}  \right) \\
&\= \frac12 \Bigl(
	\mathcal{Z}^{(n,m)}_3 - \mathcal{Z}^{(n,m)}_4 - \mathcal{Z}^{(n,m)}_2  \mp \mathcal{Z}^{(n,m)}_1 \Bigr) \,.
\end{split}
\end{equation}
In the second line, we have written out the characters using the notation of~\eqref{eq:ZfermionsT4}.

On the other hand, we could impose antiperiodic boundary conditions for the fermions around the thermal circle.
In this case, the spin-structure sum is performed with a specified set of phases~\cite{Atick:1988si}, which gives the thermal fermion partition sum
\begin{equation}
\begin{split} \label{eq:AWtwist}
\bigl( \mathcal{Z}^\pm_{_\text{AF}}\bigr)^{(n,m)}
&\=
	\Tr_{_\text{NS}} \left(\left[ \frac{1-e^{i\pi F}}{2} \frac{1+e^{i\pi\,n}}{2} + \, \frac{1+e^{i\pi F}}{2} \, e^{i\pi\,m}\, \frac{1-e^{i\pi\,n}}{2} \right] e^{2\pi i \, \ts\, L_0} \right) \\
&\qquad
	- \Tr_{_\text{R}} \left(\left[ \frac{1\pm e^{i\pi F}}{2} \frac{1-e^{i\pi\,n}}{2} + \, \frac{1+e^{i\pi F}}{2} \, e^{i\pi\,m}\, \frac{1-e^{i\pi\,n}}{2} \right] e^{2\pi i \, \ts\, L_0} \right)\\
&\=
	\frac{ 1 + (-1)^n + (-1)^m - (-1)^{n+m}}{2}\, \mathcal{Z}^{(n,m)}_3 - \frac{1 + (-1)^n - (-1)^m + (-1)^{n+m}}{2}\, \mathcal{Z}^{(n,m)}_4  \\
& \qquad
	- \frac{1 - (-1)^n + (-1)^m + (-1)^{n+m}}{2}\, \mathcal{Z}^{(n,m)}_2 \mp
	\frac{ -1 + (-1)^n + (-1)^m + (-1)^{n+m}}{2}\, \mathcal{Z}^{(n,m)}_1 \,.
\end{split}
\end{equation}
Note that, since~$\mathcal{Z}_1=\Tr_{_\text{R}} (e^{i\pi F}\, e^{2\pi i\,\ts L_0})=0$, the Type IIA and IIB partition functions are equal.
Henceforth, we denote
$\mathcal{Z}^+_{_\text{PF}}=\mathcal{Z}^-_{_\text{PF}}=\mathcal{Z}_{_\text{PF}}$ and
$\mathcal{Z}^+_{_\text{AF}}=\mathcal{Z}^-_{_\text{AF}}=\mathcal{Z}_{_\text{AF}}$.
We will therefore talk about the IIB partition function for the sake of definiteness.

Upon using the fermion characters given in~\eqref{eq:ZfermionsT4} and using Riemann identities
for theta functions, we obtain the following explicit expressions for the characters of the periodic  fermions and the thermal fermions,
\begin{equation} \label{eq:ZPFAFpm}
\begin{split}
\mqty(\mathcal{Z}^{(n,m)}_{_\text{PF}} (\ts;\tau,\rho)
\\  \mathcal{Z}^{(n,m)}_{_\text{AF}} (\ts;\tau,\rho) )
& \=
	 \frac{1}{\eta^4(\ts)} \, e^{-\frac{\pi}{\ts_2}\, \left(\Im(\us_{n,m})^2 + \Im(\mathfrak{\rho}_{n,m})^2 \right)}
  \mqty(\mathsf{Z}^{(n,m)}_\text{PF} (\ts;\tau,\rho) \\ \mathsf{Z}^{(n,m)}_\text{AF} (\ts;\tau,\rho) ) \, .
\end{split}
\end{equation}
Here we have defined
\begin{equation} \label{eq:ZPFAFdef}
\begin{split}
\mathsf{Z}^{(n,m)}_\text{PF} (\ts;\tau,\rho)
&\=
	\vartheta_1^2(\vs_{n,m}^+,\ts) \, \vartheta_1^2(\vs_{n,m}^-,\ts) \,, \\
\mathsf{Z}^{(n,m)}_\text{AF} (\ts;\tau,\rho)
&\=
	\vartheta_1^2(\vs^+_{n,m},\ts)\, \vartheta_1^2(\vs^-_{n,m},\ts)
	+ (-1)^n \, \vartheta_2^2(\vs^+_{n,m},\ts)\, \vartheta_2^2(\vs^-_{n,m},\ts) \\
&\qquad \qquad
	+ (-1)^m \, \vartheta_4^2(\vs^+_{n,m},\ts)\,
	\vartheta_4^2(\vs^-_{n,m},\ts)
	-(-1)^{n+m}  \vartheta_3^2(\vs^+_{n,m},\ts)\, \vartheta_3^2(\vs^-_{n,m},\ts) \,,
\end{split}
\end{equation}
where
\begin{equation}
    \vs_{n,m}^\pm \= \frac12 \bigl(\us_{n,m} \pm \rho_{n,m} \bigr) \= \frac12(n\ts - m) (\tau \pm \rho) \,.
\end{equation}
The piece that is factored out in~\eqref{eq:ZPFAFpm} turns out to cancel against a corresponding factor from the bosonic characters.

The partition function for the superstring is obtained by
putting together the $\slt$, the $\sut$, the torus, the ghost, and the fermion pieces.
Using the explicit characters given in~\eqref{eq:ZPFAFdef}, we arrive at our final expression
\begin{equation}\label{eq:Zads3s3t4}
\begin{split}
\mathcal{Z}_\text{IIB}
&=
	\frac{\beta\, \sqrt{k}}{2\pi} \, \int_{\mathcal{F}}\, \frac{d^2\ts}{\ts_2^\frac{3}{2}}
	\left( \sum_{\Gamma^{4,4}} \zs^{p_L^2} \, \overline{\zs}^{p_R^2} \right)
	\sum_{n,m} \, 	\frac{\exp \left( -\frac{k}{\ts_2}\left(\frac{\beta^2}{4\pi} + \pi\, \Im(\rho)^2 \right) \, \abs{m- n\,\ts}^2 + \frac{2\pi\,k\,\Im (\rho_{n,m})^2}{\ts_2} \right)}{\abs{\eta(\ts)}^{12} \, \abs{\vartheta_1(\us_{n,m},\ts)}^2} \\
&\qquad \times
	\sum_{\ell=0}^{\frac{k}{2}-1} \abs{\frac{\Theta_{2\ell+1}^{(k)}(\rho_{n,m},\ts) - \Theta^{(k)}_{-2\ell-1}(\rho_{n,m},\ts)}{\vartheta_1 \left( \rho_{n,m}, \ts \right)}}^2 \times
\begin{dcases}
	\abs{\mathsf{Z}^{(n,m)}_\text{PF}}^2, &\text{ periodic fermions},\\
	\abs{\mathsf{Z}^{(n,m)}_\text{AF}}^2 &\text{ antiperiodic fermions}.
\end{dcases}
\end{split}
\end{equation}
The integrand is modular invariant in both cases
once we sum over the integers $m$ and $n$ that characterize the winding
of the worldsheet torus around the Euclidean time circle.
Thus, we have a well-defined integral over the fundamental domain of the string worldsheet.

\subsection{Swapping fermion boundary conditions with chemical potential shifts}\label{sec:rhoFsswap}

The spacetime CFT partition function with $\mathrm{SU}(2)_R$ chemical potential~$\rho$ is
\begin{equation}
\mathscr{Z}_{_\text{CFT}}(\tau,\overline{\tau}, \rho, \overline{\rho})
\= \text{Tr} \, \exp \Bigl( 2 \pi i \, \tau \, L_0^\text{CFT} -2 \pi i \, \overline{\tau} \, \overline{L}_0^\text{CFT} + 2 \pi i \,
\rho \, J_\text{CFT} - 2 \pi i \, \overline{\rho} \, \overline{J}_\text{CFT} \Bigr) .
\end{equation}
Here~$J_\text{CFT}$, $\overline{J}_\text{CFT}$ are quantized in integer multiples of~$\frac12$, so that the
spacetime fermion number is given by~$(-1)^{F_s} = e^{2 \pi i (J_\text{CFT}+\overline{J}_\text{CFT})}$.
Thus, we see the shift in the spacetime $\sut$ chemical potential $\rho \to \rho+1$ implements a change from
periodic to antiperiodic conditions around the thermal circle for the fermions.

We can now check the effect of this shift on the worldsheet partition function, using either the
representations~\eqref{eq:ZfermionsT4}--\eqref{eq:AWtwist} or, equivalently,~\eqref{eq:ZPFAFdef}.
We illustrate it using the latter.
The effect of the shift~$\rho \to \rho+1$ is
\begin{equation}
\begin{split}
    \mathsf{Z}_\text{PF} \; \to & \;
    \vartheta_1^2 \bigl(\vs_{n,m}^+ + \tfrac12 (n\ts -m), \ts\bigr) \,
    \vartheta_1^2 \bigl(\vs_{n,m}^- - \tfrac12 (n\ts -m) , \ts\bigr)  \\
& \=
	\tfrac14 \bigl(1+(-1)^m \bigr) \bigl(1+(-1)^n \bigr) \, \vartheta_1^2(\vs^+_{n,m},\ts)\, \vartheta_1^2(\vs^-_{n,m},\ts) \\
& \quad + \tfrac14 \, \bigl(1-(-1)^m \bigr) \, \bigl(1+(-1)^n \bigr) \, \vartheta_2^2(\vs^+_{n,m},\ts)\, \vartheta_2^2(\vs^-_{n,m},\ts) \\
& \quad + \tfrac14 \, \bigl(1+(-1)^m \bigr) \, \bigl(1-(-1)^n \bigr) \, \vartheta_4^2(\vs^+_{n,m},\ts)\, \vartheta_4^2(\vs^-_{n,m},\ts) \\
& \quad + \tfrac14 \, \bigl(1-(-1)^m \bigr) \, \bigl(1-(-1)^n \bigr) \, \vartheta_3^2(\vs^+_{n,m},\ts)\, \vartheta_3^2(\vs^-_{n,m},\ts) \,.
\end{split}
\end{equation}
Upon collecting the various factors of $1, (-1)^m, (-1)^n, (-1)^{m+n}$ and using the Riemann quartic theta relations,  we obtain precisely~$\mathsf{Z}_\text{AF}$ in~\eqref{eq:ZPFAFdef} on the right-hand side.
We summarize this as
\begin{equation}
    \mathsf{Z}_\text{PF}(\ts; \tau, \rho+1) \=
    \mathsf{Z}_\text{AF}(\ts; \tau, \rho) \,, \qquad
      \mathsf{Z}_\text{PF}(\ts; \tau, \rho+2) \=
        \mathsf{Z}_\text{PF}(\ts; \tau, \rho) \,.
\end{equation}

that is to say, the implementation of~$(-1)^{F_s}$ in the spacetime CFT through a shift of the boundary
$\mathrm{SU}(2)_R$ chemical potential leads us precisely to the expected Atick-Witten spacetime twists in
fermionic string theory~\cite{Atick:1988si}, which was imposed from considerations of worldsheet modular invariance by summing over spin structures.

\subsection{The single string spectrum}\label{sec:superspec}

We now have at hand the worldsheet partition functions for either choice of boundary conditions for the fermions around the spacetime thermal circle. We can exploit the modular invariance of the integrand and unfold the integration domain to the half-strip. The manipulation parallels the bosonic discussion. When the dust settles, we arrive at the single string free energy\footnote{
    For simplicity, we will leave implicit the antiholomorphic dependence in the single string
    free energy.}
\begin{equation}\label{eq:rnsper1}
\begin{split}
f_{_\text{PF}} (\tau,\overline{\tau},\rho, \overline{\rho})
&\=
	\frac{\sqrt{k}\,\beta}{2\pi} \,
	\int_0^\infty\, \frac{d\ts_2}{\ts_2^\frac{3}{2}} \int_{-\frac{1}{2}}^{\frac{1}{2}}\, d\ts_1\,
	 e^{- \frac{k\,\beta^2}{4\pi\,\ts_2}} \; \mathcal{I}_{_\text{PF}}\bigl(\ts,\tsb;\tau,\overline{\tau},\rho, \overline{\rho} \bigr) \,, \\
&\=
	\frac{4}{i\,k} \, \int_{-\infty}^{\infty}\, \zeta\,d\zeta \int_{0}^\infty\, d\ts_2 \, \int_{-\frac{1}{2}}^{\frac{1}{2}}\, d\ts_1\, e^{2i\,\beta\,\zeta - \frac{4\pi}{k} \, \ts_2\,\zeta^2 }\, \mathcal{I}_{_\text{PF}}\bigl(\ts,\tsb;\tau,\overline{\tau},\rho, \overline{\rho} \bigr)\,,
\end{split}
\end{equation}
with
\begin{equation}\label{eq:IIBintegrand}
\begin{split}
& \mathcal{I}_{_\text{PF}} \bigl(\ts,\tsb;\tau,\overline{\tau},\rho, \overline{\rho} \bigr)
\=\\
&\qquad 	\Biggl(\, \sum_{\Gamma^{4,4}} \, \zs^{p_L^2} \,
	\overline{\zs}^{p_R^2}\Biggr) \abs{
	\frac{ \vartheta_1^2\left(\frac{\tau-\rho}{2},\ts\right) \,
		\vartheta_1^2\left(\frac{\tau+\rho}{2},\ts\right)}{\vartheta_1(\tau,\ts )\, \vartheta_1 ( \rho, \ts )\, \eta(\ts)^6}}^2
	\,  \sum\limits_{\ell=0}^{\frac{k}{2}-1} \abs{
		\sum_{n\in \mathbb{Z}}\, \zs^{\frac{(nk+\ell+ \frac{1}{2})^2}{k}} \,(r^{nk + \ell + \frac{1}{2}} -r^{-(nk + \ell + \frac{1}{2})} )}^2 \, .
\end{split}
\end{equation}
Note that the trace is taken over the NS sector for the spacetime fermions.
We have written out the $\sut_{k-2}$ theta functions as an infinite sum.
The final result is consistent with the earlier analysis of~\cite{Raju:2007uj}, which directly quotes the single string free energy.
In obtaining this, it was important to have accounted for the factors associated with the holomorphic
anomaly in the fermion characters, the modular covariance of the $\sut$ characters, and the classical
winding contribution from the holonomy around the Hopf angles in $\mathbf{S}^3$. It is the combination of all
of these effects that cancels out the additional phase factors that depend on $\Im(\rho)$ in the final expression.

\medskip

To extract the spectrum from the single string free energy, we need to integrate over the worldsheet, which can be done as in the bosonic string discussion in~\cref{sec:bosonicZ}.
In order to isolate the various contributions, it is useful to decompose the integrand into three pieces\footnote{The factor of $(q\,\qb)^{-\frac{1}{2}}$ has been factored out to simplify $I_0$; it will
cancel against a similar contribution from the worldsheet integral.}
\begin{equation}
\mathcal{I}_{_\text{PF}}
\=
	(q\,\qb)^{-\frac{1}{2}}\, I_0 \times I_{_\text{primaries}} \times I_{_\text{oscillators}} \,.
\end{equation}
The first piece is the contribution from the zero modes of the $\vartheta_1$ functions, and is given by
\begin{equation}\label{eq:I0}
I_0
\=
	\abs{q^\frac{1}{2}\, \frac{  \left((q\,r)^{\frac{1}{4}} -  (q\,r)^{-\frac{1}{4}} \right)^2 \left( \left( \frac{q}{r} \right)^{\frac{1}{4}} -  \left( \frac{q}{r} \right)^{-\frac{1}{4}} \right)^2}{q^\frac{1}{2} - q^{-\frac{1}{2}} }}^2 \,.
\end{equation}
The second contribution is from the primaries of the $\sut$ and $T^4$ sigma models, and is given by
\begin{equation}\label{eq:Iprimary}
I_{_\text{primaries}}
\=
	\left( \sum_{\Gamma^{4,4}} \, \zs^{p_L^2} \,
	\overline{\zs}^{p_R^2}\right)
	\left(\sum_{\ell=0}^{\frac{k}{2}-1} \abs{\sum_{n\in\mathbb{Z}} \zs^\frac{(nk+\ell+ \frac{1}{2})^2}{k} \chi_{nk + \ell}(r)}^2 \right).
\end{equation}
We have used the zero mode contribution from $\vartheta_1(\rho,\ts)$ to complete the $\sut$ characters in the above expression.
The third and final contribution is that of the string oscillators
\begin{equation}\label{eq:Iosc}
I_{_\text{oscillators}}
\=
	\abs{\prod_{n=1}^\infty \,
	\frac{(1-(qr)^\frac{1}{2}\, \zs^n)^2 \,(1-(qr)^{-\frac{1}{2}}\,\zs^n)^2\,
	(1-(\frac{q}{r})^\frac{1}{2}\, \zs^n)^2\, (1- (\frac{q}{r})^{-\frac{1}{2}}\, \zs^n)^2}{(1-\zs)^4\, (1-q\,\zs^n)\,(1-q^{-1}\,\zs^n)\, (1-r\,\zs^n) \, (1-r^{-1}\,\zs^n)}}^2 .
\end{equation}

\medskip

The zero mode contribution captured by~$I_0$ is important for us,
and can be rewritten as
\begin{equation}\label{eq:I0simp}
\begin{split}
I_0
&\=
	\abs{\frac{1 - 2\, \chi_{\frac{1}{2}}(r) \, \sqrt{q}+ \chi_1(r)\, q +  3\,q - 2\, \chi_{\frac{1}{2}}(r) \, q^{\frac{3}{2}}+ q^2}{1-q}}^2
\end{split}
\end{equation}
in terms of $\sut$  characters.
This will be useful momentarily in deciphering the spacetime currents, and in organizing the spectrum.
The oscillator contribution can also be simplified.
The best way to proceed is to extract the piece $\abs{ \prod_{n=1}^\infty\frac{1-\zs^n}{(1- q\,\zs^n)\,(1- q^{-1} \,\zs^n)}}^2$, which is the non-trivial contribution from the $\slt$ and $bc$ ghost oscillators, as in the bosonic string.
This is the only factor that needs care in its Taylor series expansion, and is sensitive to the spectral flow
sum.
The remaining factors in $I_{_\text{oscillators}}$ can be expressed in terms of $\sut$ characters, as we note
in~\cref{sec:superpart}. For now, we will simply note that the oscillator contribution can be
brought to the form (see~\eqref{eq:IoscQ})
\begin{equation}\label{eq:IoscP}
I_{_\text{oscillators}} \= \abs{\sum_{w=0}^\infty \,\sum_N\, \mathbf{Q}_{N}^w  (q,r)\, q^w  \, \zs^{-\frac{1}{2}\, w (w+1) + N} }^2\,.
\end{equation}

With this decomposition, we can write the integrand~\eqref{eq:IIBintegrand} as
\begin{equation}
\begin{split}
\mathcal{I}_{_\text{PF}}
&\=
	(q\,\qb)^{-\frac{1}{2}}\, I_0 \;
	\widehat{\sum} \,
	(q\, \qb)^w \, \mathbf{Q}_{N}^w  (q,r) \overline{\mathbf{Q}}_{_{\overline{N}}}^w  (\qb,\overline{r})\;
	\chi_{nk + \ell} (r) \, \chi_{\overline{n} k + \ell }(\overline{r}) \\
& \qquad \qquad \qquad
	\times 	 (\zs\, \zsb)^{-\frac{1}{2}\, w (w+1) } \,
	\zs^{p_L^2 +\frac{(nk+\ell+ \frac{1}{2})^2}{k} + N} \, \zsb^{p_R^2 +\frac{(\overline{n} k+\ell+ \frac{1}{2})^2}{k} + \overline{N}} \,,
\end{split}
\end{equation}
where for the sake of brevity we introduced the compact summation notation
\begin{equation}
\widehat{\sum} \; \equiv \;
	\sum\limits_{w=0}^\infty\,\sum\limits_{N,\overline{N}}\,
	\sum\limits_{n,\overline{n}\in \mathbb{Z}} \,  \sum_{\Gamma^{4,4}}  \, \sum\limits_{\ell=0}^{\frac{k}{2}-1}   \;.
\end{equation}

\medskip

We can now carry out the integrals over the worldsheet moduli.
Firstly, the $\ts_1$ integral leads to the level matching constraint
\begin{equation}\label{eq:IIBlevelmatch}
\delta_\text{LM} :\quad
p_L^2 + n \, (n\, k + 2\, \ell+ 1) + N \= p_R^2 +  \overline{n} \, (\overline{n}\, k+2\ell+ 1) + \overline{N}\,.
\end{equation}
The integral over $\ts_2$ is now straightforward and results in
\begin{equation}\label{eq:susyfsp}
\begin{split}
f_{_\text{PF}}\bigl(\tau,\overline{\tau},\rho, \overline{\rho} \bigr)
&\=
	  I_0\, (q\,\qb)^{-\frac{1}{2}}\, \widehat{\sum} \,  (q\, \qb)^w \, \mathbf{Q}_{N}^w  (q,r) \overline{\mathbf{Q}}_{_{\overline{N}}}^w  (\qb,\overline{r})\;
	\chi_{nk + \ell} (r) \, \chi_{\overline{n} k + \ell }(\overline{r})
	\, \delta_\text{LM}  \\
&\qquad \qquad
	\times \int_{-\infty}^{\infty}\, \frac{d\zeta}{i\pi} \, \frac{\zeta\,e^{2i\,\beta\,\zeta}}{\zeta^2 + \Delta_w^2} \Bigl( e^{-\frac{2\beta}{w+1}\,\frac{\zeta^2  + \Delta_w^2}{k}} - e^{-\frac{2\beta}{w}\, \frac{\zeta^2 + \Delta_w^2}{k}} \Bigr) \,,
\end{split}
\end{equation}
where $\delta_\text{LM}$ is an insertion of the level matching constrain as a delta function~\eqref{eq:IIBlevelmatch}, and
\begin{equation}\label{eq:DeltaSS}
\Delta_{w}^2 \= \Bigl(\ell+ \frac{1}{2}\Bigr)^2 + k\, \Bigl( n\, ( n \, k + 2\,\ell +1) + p_L^2 + N - \frac{1}{2}w(w+1) \Bigr) \, .
\end{equation}

Upon shifting the contour and picking up the residues from the zeros of $\zeta^2 + \Delta_{w}^2$,
we end up with a sum of a discretuum and a continuum of states
\begin{equation}\label{eq:fspSS}
\begin{split}
f_{_\text{PF}}\bigl(\tau,\overline{\tau},\rho, \overline{\rho} \bigr)
& \=
	f_{_\text{PF,disc}}\bigl(\tau,\overline{\tau},\rho, \overline{\rho} \bigr)  +
	f_{_\text{PF,cont}}\bigl(\tau,\overline{\tau},\rho, \overline{\rho} \bigr)\,.
\end{split}
\end{equation}
The contribution from the set of states with discrete spacetime conformal weights is
\begin{equation}\label{eq:fdSS}
f_{_\text{PF,disc}}\bigl(\tau,\overline{\tau},\rho, \overline{\rho} \bigr)
\=
I_0 \, \widehat{\sum} \, \delta_\text{LM}\, \mathbf{Q}_{N}^w  (q,r)\, \overline{\mathbf{Q}}_{_{\overline{N}}}^w  (\qb,\overline{r})\;
	\chi_{nk + \ell} (r) \, \chi_{\overline{n} k + \ell }(\overline{r}) \,  (q\, \qb)^{w -\frac{1}{2} + \Delta_{w}} \,.
\end{equation}
The oscillator level is constrained in the $w^\textrm{th}$ spectral flowed sector to satisfy
\begin{equation}\label{eq:Deltarange}
 \frac{k}{2}\, w  \; \leq \; \Delta_{w}
 \;\leq \; \frac{k}{2}\, (w +1) \,.
\end{equation}
The continuum spectrum is in turn given by a density of states $\bm{\varrho}(s; q,\qb,r,\overline{r})$, and takes the form
\begin{equation}
 f_{_\text{PF,cont}}\bigl(\tau,\overline{\tau},\rho, \overline{\rho} \bigr)
 \=
    I_0\,\sum_{w=1}^\infty\, \int_{-\infty}^\infty\,\frac{ds}{i\pi}\, \frac{\bm{\varrho}(s; q,\qb,r,\overline{r})}{\abs{1-q}^2}\, (q\,\qb)^{\frac{k\,w}{4} + \frac{1}{w\,k}\, \left(s^2 + \Delta_w^2\right)}\,.
\end{equation}
The derivation of the density of states proceeds similarly to the bosonic string. Its precise form will not play a role in our analysis, and so we refrain from writing it out explicitly.

\subsection{Spacetime currents}\label{sec:supercurrents}

Now that we have the single string free energy, let us examine the contributions closely.
Our first task is to isolate the spacetime currents. To this end, note that
we can focus on exciting no $\sut$ and $T^4$ primaries ($p_L = p_R = \ell =0$), nor the oscillators and their spectral flow images.
This implies we restrict to $P_{_0}^0(q,r) \, \overline{P}_{_0}^0(\qb,\overline{r}) \, \chi_0(r) \,\chi_0(\overline{r}) =1$.
Then, we are left with only the contribution from $I_0$, which we have written out in~\eqref{eq:I0simp}.

To interpret this, it is useful to rewrite $I_0$ in terms of the reduced $\mathfrak{psu}(1,1|2) \oplus \overline{\mathfrak{psu}(1,1|2)}$ characters introduced in~\eqref{eq:cporeps}. A short calculation reveals
\begin{equation}\label{eq:I0currents}
\begin{split}
I_0
&\=
	1 + \frac{\sccft^{(2)}(0)  -2\, \sccft^{(\frac{3}{2})}(0)}{1-q}  + \frac{\overline{\sccft}^{(2)}(0)  -2\, \overline{\sccft}^{(\frac{3}{2})}(0)}{1-\overline{q}} \\
&\qquad
    + \frac{\sccft^{(1)}(1) -2\, \sccft^{(\frac{3}{2})}\left( \frac{1}{2} \right) - 2\, \overline{\sccft}^{(\frac{3}{2})}\left( \frac{1}{2} \right) + 4\, \sccft^{(1)}\left( \frac{1}{2} \right)}{\abs{1-q}^2} \,.
\end{split}
\end{equation}
We now see the spacetime currents explicitly in the first line, while the second line comprises states which are current-current bilinears.
We will see later that they assemble into the chiral primary states of supergravity (after including states with all~$\ell$).

Let us focus therefore on the first line. The contribution proportional to unity is that of the spacetime CFT vacuum.
The holomorphic part can be
written in terms of the characters~\eqref{eq:lowjxi} as follows,
\begin{equation}
I_{0,\text{hol}} (\tau,\rho) \=
\frac{\sccft^{(2)}(0)  -2\, \sccft^{(\frac{3}{2})}(0)}{1-q}
=
    \frac{q\, \chi_1(r) - 2\,q^{\frac{3}{2}}\, \chi_{\frac{1}{2}}(r) +q^2 }{1-q}
  -2\, \frac{q^\frac{1}{2}\, \chi_{\frac{1}{2}}(r) - 2\, q }{1-q} \,.
\end{equation}
The interpretation of the various pieces is as follows:
\begin{itemize}
\item $\sccft^{(2)}(0)$ is the contribution from the $\sut_R$ symmetry generators, the spin-$\frac{3}{2}$ supercurrent generators, and the boundary gravitons.
\item $\sccft^{(\frac{3}{2})}(0)$ on the other hand, captures the contribution of four spin-$\frac{1}{2}$ and spin-$1$ currents.
\end{itemize}
As noted in Equation~\eqref{eq:lowjxi} these multiplets are ultra-short owing to the low value of $\sut$-spin for the superconformal primary.

\bigskip

The corresponding multi-string partition function is given by
\begin{equation}\label{eq:stringvacchar}
\begin{split}
& \exp \biggl(\, \sum_{j=1}^\infty\, \frac{1}{j} \, I_{0,\text{hol}}(j\,\tau, j\,\rho) \biggr) \\
& \qquad \qquad \=
    \left[-i\, q^\frac{1}{4}\, \frac{\vartheta_4(\frac{\rho}{2},\tau)^2}{\eta(\tau)^3\, \vartheta_1(\rho,\tau)} \,
 \frac{(r^{\frac{1}{2}} -r^{-\frac{1}{2}})\, (1-q)}{(1 - r^{\frac{1}{2}}\, q^{\frac{1}{2}})^2\, (1 - r^{-\frac{1}{2}}\, q^{\frac{1}{2}})^2} \right]
\times
	\left[q^{\frac{1}{4}}\, \frac{\vartheta_4(\frac{\rho}{2},\tau)^2}{\eta(\tau)^6} \right] .
\end{split}
\end{equation}
The first factor is equal to~$\etchar(0)\big|_{m=0}$
defined in~\eqref{eq:chiNSfactors}.
This is the one-loop determinant of the generators of the superconformal algebra for the vacuum module (recall that there is an extra factor of~$(1-q)$ for~$\ell=0$ compared to the generic module).
The second factor is the contribution of 4 spacetime $\mathrm{U}(1)$ currents and their superpartners,
arising from the reduction on $T^4$ of the metric and B-field in the NS sector.\footnote{In the dual boundary
CFT$_2$ this corresponds to the overall~$T^4$, as we discuss in the Introduction.}

Let us compare this to the expectation from the spacetime CFT. The vacuum superconformal
character~$\etchar(0)$
involves the infinite sum~\eqref{eq:N44vacNS}, which we repeat here for convenience,
\begin{equation} \label{eq:mu0avg}
\mu_0(r,q) -\mu_0(r^{-1},q)
\=
\sum_{m\in \mathbb{Z} } \, \left[\frac{r^{(\ks+1)\, m + \frac{1}{2}}\, q^{(\ks+1)\, m^2 + m}}{(1 - r^{\frac{1}{2}}\, q^{m+\frac{1}{2}})^2}\,
- (r\to r^{-1}) \right] .
\end{equation}
As noted above, the worldsheet result~\eqref{eq:stringvacchar} captures precisely the $m=0$ term of this infinite sum. The~$m^\text{th}$ term in this sum corresponds to an image of~$m=0$ under~$m$ units of spectral flow of
the~$\mathcal{N}=2$ spacetime superconformal algebra.\footnote{For the second sum involving~$\mu_0(r^{-1})$, we relabel~$m \to -m$ in the sum.}

We recall that the sum over the spectral flow images maps any function~$f(r)$ to an average over the elliptic parameter (see e.g.,~\cite[Sec.~8.2]{Dabholkar:2012nd})
\begin{equation}\label{eq:ellipticaverage}
f(r) \; \mapsto \;    \sum_{m \in \mathbb{Z}} q^{\mathsf{k}\,m^2} \, r^{\mathsf{k}\,m} \, f(q^{2m} r) \,,
\end{equation}
where $\mathsf{k}$ is the level of the superconformal algebra.
This average can be explained in supergravity by using an argument given in~\cite{Kraus:2006nb}.
First, we observe that the spacetime theory has
an~$\mathrm{SU}(2)_R$ gauge field governed by a Chern-Simons interaction~\cite{Henneaux:1999ib}.
The vacuum solutions for this theory include all flat connections.
The boundary conditions for a Chern-Simons gauge field~$A$ on~\AdS{3} is that,
at the asymptotic boundary,
$A_{\overline w}$ is held fixed, while~$A_{w}$ is allowed to fluctuate.
In our problem, only the Cartan component of the $\mathrm{SU}(2)_R$ gauge field is non-vanishing at the asymptotic boundary, and
it is fixed to be the chemical potential of the boundary CFT$_2$
as~$A^3_{\overline w}=\rho$.
Now, smoothness of the gauge field in spacetime means that it should have trivial holonomy around the contractible circle.
This fixes the value~$A^3_{w} = (\rho + m), \, m \in \mathbb{Z}$, so that~$e^{i \oint A^3_{w}} = r$.

The partition function is expressed as a sum over these flat gauge
field configurations, labelled by~$m$. The bulk CS action vanishes because the gauge fields
are flat, but there is a boundary action given by~$ \mathsf{k} \int A^i_w A^i_{\overline{w}}$,
where,~$\mathsf{k}=c/6$ is the level of the CS theory fixed by supersymmetry in terms of the central charge.
The value of the action in the holonomy sector~$m$ is such as to precisely implement
the action of spectral flow by~$m$ units in the $\mathcal{N}=2$ superconformal algebra~\cite{Kraus:2006nb}.
Under such a spectral flow, the charge of a state shifts as~$Q \to Q + \mathsf{k} \,m$, and the energy changes to~$L_0 \to L_0 + \mathsf{k}\, m^2 + 2\, Q\, m$.
This is precisely what is necessary to implement the averaging procedure described in~\eqref{eq:ellipticaverage}.

Let us go back to our worldsheet result and examine the first factor in~\eqref{eq:stringvacchar}. Using it as the seed function $f(r)$, one recovers the vacuum $\mathcal{N}=4$ superconformal character $\etchar(\ell=0)$ upon implementing~\eqref{eq:ellipticaverage}.\footnote{Note added in v2: In v1 of this paper we claimed erroneously that the average~\eqref{eq:ellipticaverage} was to be done with a shifted level $\mathsf{k}+1$ instead of $\mathsf{k}$, which is at odds with the supergravity discussion. We are grateful to Joaquin Turiaci for helping us clear this up.}
Note that the infinite sum~\eqref{eq:mu0avg} transforms under this elliptic average with level $\mathsf{k}+1$, but the one-loop determinant factor $\frac{\vartheta_4(\rho/2,\tau)^2}{\vartheta_1(\rho,\tau)}$ also transforms at level $-1$, leading to the final result. In performing this elliptic average, we do not include the contribution from the fermionic partners of the 4 spacetime $\mathrm{U}(1)$ currents (the $\vartheta_4(\rho/2, \tau)$ in the second factor of~\eqref{eq:stringvacchar}, which transforms at level $+1$).

\subsection{The supergravity spectrum}\label{sec:sugraspec}

Having understood the origin of the currents from the worldsheet partition function, we now turn to the
spectrum of chiral primary operators. These are the states which are visible in supergravity.

To anchor the discussion, let us record the spectrum of the supergravity states. This was obtained by
Kaluza-Klein reduction of 10d Type II supergravity on $\mathbf{S}^3 \times T^4$ in~\cite{Deger:1998nm,Larsen:1998xm}. The data can be
expressed directly in terms of $\mathfrak{psu}(1,1|2) \oplus \overline{\mathfrak{psu}}(1,1|2)$
characters. One finds that the representations of interest are the following:
\begin{equation}\label{eq:sugraSM}
\begin{split}
s=2: &\qquad
	\sum_{j\geq 0}\,\sccft^{(2)}(j) + \overline{\sccft}^{(2)}(j) \,, \\
s=\frac{3}{2}: &\qquad
	2\,\sccft^{(\frac{3}{2})}(0)+ 2 \,\overline{\sccft}^{(\frac{3}{2})}(0)+ \sum_{j\geq \frac{1}{2}} \mathbf{4}_5\, \sccft^{(\frac{3}{2})}(j) + \mathbf{4}_5\,\overline{\sccft}^{(\frac{3}{2})}(j) \,,\\
s=1: &\qquad
 \sum_{j\geq 1/2}\, \mathbf{5}_5\,  \sccft^{(1)}(j) + \sccft^{(1)}\bigl(j+\tfrac{1}{2}\bigr) \,.
\end{split}
\end{equation}
We have organized the spectrum based on the highest helicity state in the multiplet.
The numerical coefficients represent the degeneracy labels. We have decorated some degeneracy factors with a subscript, referring to the representation
under $SO(5)$ U-duality symmetry in six dimensions (see below).

A few comments are in order for comparison to the aforementioned literature.
The original analysis of~\cite{Deger:1998nm} focussed on reducing six dimensional supergravity on $\mathbf{S}^3$.
They assumed that the gravitational sector was coupled to $n$ tensor multiplets, and derived the resulting spectrum in \AdS{3}.
The torus reduction of Type II theories results in 5 six dimensional tensor multiplets (this is the origin
of the $SO(5)$ symmetry).  However, there are additional gravitinos and vector fields, owing to all the 10d supercharges being preserved under $T^4$
compactification. This was separately analyzed in~\cite{Larsen:1998xm}.

\medskip

Let us now see how to recover~\eqref{eq:sugraSM} from the worldsheet.
Since these states are obtained in supergravity, we can ignore string oscillators i.e.,~put $N= \overline{N} = 0$, and keep the spectral flow number to be $w=0$.
Furthermore, Kaluza-Klein reduction on $T^4$ implies that $p_R = p_L =0$.
The level matching condition~\eqref{eq:IIBlevelmatch} now implies $n = \overline{n}$, and the constraint~\eqref{eq:Deltarange} further forces $n =0$, and reduces to the bound on the spin of the $\sut$ primaries $0\leq \ell \leq \frac{k-1}{2}$ (this is weaker than the cut-off in $\sut$ representations in the partition sum). Taking these into account, we find
\begin{equation}
f_{_\text{PF}}(\tau, \rho, \overline{\tau}, \overline{\rho}) \, \Big|_{\text{sugra}} \= I_0\, \sum_{\ell =0}^{\frac{k}{2}-1} \, q^{\ell} \, \abs{\chi_\ell(r)}^2 \,.
\end{equation}
Note that the $\ell=0$ contribution is included here for convenience, even though a part of it was already analyzed in~\cref{sec:supercurrents}.

To proceed, we use the following simple identity of the chiral $\mathfrak{psu}(1,1|2)$ characters
\begin{equation}\label{eq:schprod}
q^j \,\chi_j(r)
	\left( 1 -q+ \sccft^{(2)}(0)  -2\, \sccft^{(\frac{3}{2})}(0) \right)
\=
	\schol(j+1,j+1) +  \schol(j,j) - 2\,  \schol\bigl(j+\tfrac{1}{2},j+\tfrac{1}{2} \bigr) \,.
\end{equation}
It is then a simple matter to put together the above expression~\eqref{eq:schprod} with the analogous expression in anti-holomorphic sector, which results in
\begin{equation}\label{eq:j0charid}
\begin{split}
f_{_\text{PF}}(\tau, \rho, \overline{\tau}, \overline{\rho}) \Big|_{\text{sugra}}
&\=
	1 + \frac{\sccft^{(2)}(0)  -2\, \sccft^{(\frac{3}{2})}(0)}{1-q}  + \frac{\overline{\sccft}^{(2)}(0)  -2\, \overline{\sccft}^{(\frac{3}{2})}(0)}{1-\overline{q}} \\
&\qquad\qquad
	+ \frac{\sccft^{(1)}(1) -2\, \sccft^{(\frac{3}{2})}\left( \frac{1}{2} \right) - 2\, \overline{\sccft}^{(\frac{3}{2})}\left( \frac{1}{2} \right) + 4\, \sccft^{(1)}\left( \frac{1}{2} \right)}{\abs{1-q}^2} \\
&\qquad
	+ \sum\limits_{\frac{1}{2} \; \leq \, j \, \leq \; \frac{k}{2}-1}\,
	\frac{\sccft^{(2)}(j) + \overline{\sccft}^{(2)}(j)
	+ \sccft^{(1)}(j) + 4\,\sccft^{(1)}\left(j+ \frac{1}{2}\right)+ \sccft^{(1)}(j+1)}{\abs{1-q}^2} \\
&\qquad
	-2\, \sum\limits_{\frac{1}{2} \; \leq \, j \, \leq \; \frac{k}{2}-1}\,  \frac{\sccft^{(\frac{3}{2})}(j)+  \overline{\sccft}^{(\frac{3}{2})}(j) + \sccft^{(\frac{3}{2})}\left(j+\frac{1}{2}\right) +\, \overline{\sccft}^{(\frac{3}{2})}\left(j+\frac{1}{2}\right)}{\abs{1-q}^2} \,.
\end{split}
\end{equation}
The first line is the contribution from the identity and the boundary currents, which we have already discussed.
The rest of the spectrum, as one can easily check, corresponds precisely to the supergravity chiral primary states recorded in~\eqref{eq:sugraSM}.

It is interesting to examine certain aspects of the spectrum.
First, note that the helicity~$\frac{3}{2}$ and helicity~$1$ multiplets are staggered.
While we expect four helicity~$\frac{3}{2}$ multiplets at each value of $\sut$ spin, say $\ell=j$,
two of them come from the spin  $\ell=j -\frac{1}{2}$ part of the spectrum,
which combine with 2 more states from the spin $\ell=j$ to complete the spectrum. A similar phenomenon happens for the helicity-$1$ states.
The reason for this is that certain chiral primary states originate from the $\ell =0$ sector, as we saw in~\eqref{eq:I0currents}.
These operators lie in truncated multiplets owing to low spin values, cf.~~\eqref{eq:lowjxi}.
They originate from left-right combinations of the current multiplets; indeed it is easy to verify
\begin{equation}\label{eq:redcharprod}
\begin{split}
\sccft^{(2)}(0) \times \overline{\sccft}^{(2)}(0)
&\=
	\sccft^{(1)}(1) \,,\\
\sccft^{(2)}(0) \times \overline{\sccft}^{(\frac{3}{2})}(0)
&\=
	\sccft^{(\frac{3}{2})}\left( \frac{1}{2} \right) ,\\
\sccft^{(\frac{3}{2})}(0) \times \overline{\sccft}^{(\frac{3}{2})}(0)
&\=
	\sccft^{(1)}\left( \frac{1}{2} \right) .
\end{split}
\end{equation}
Therefore these states should be viewed as current bilinears, and are the analog of the weight $(2,2)$ operator we encountered in the bosonic string in~\cref{sec:Zwsspace}.
This set contains not only the spacetime dilaton operator from  $\sccft^{(2)}(0) \times \overline{\sccft}^{(2)}(0)$,
but also bilinears of the superconformal currents and the $T^4$ currents.

\subsection{Strings beyond supergravity}\label{sec:beyondsugra}

We have thus far focussed on a particular subset of states, and have switched off the contributions from the worldsheet oscillators, torus primaries, etc.
In the limit where the $\lads \gg \ell_s$, viz., $k \gg 1$, it is clear from~\eqref{eq:fdSS} that these states are massive. In the $w^\textrm{th}$ spectral
flowed sector, the conformal weights are bounded from below by
\begin{equation}
\hs \; \geq \; w - \frac{1}{2} + \Delta_w  + \text{order}_q (\mathbf{Q}_N^w(q,r) )\; \sim \; \order{k}\,.
\end{equation}
where $\text{order}_q$ here refers to the least power of $q$ in the power series $(\mathbf{Q}_N^w(q,r)$.  The lower bound follows from~\eqref{eq:Deltarange} similar to the bosonic discussion (see~\cref{sec:bosZeval}).
This is the case for all the primaries on the torus, as well as for the states of the $\sut$ sigma model with $n\neq 0$ in~\eqref{eq:fdSS}.

It is interesting to examine the spectrum in the regime where the curvature length scales are comparable to the string scale.
In this case, there is no real hierarchy between the chiral primary states, discussed above, and the states with conformal weights of $\order{\sqrt{k}}$.
As an extreme example, consider $k=2$, which is the smallest value attainable within the RNS formalism. In this limit, $\mathcal{Z}_{\sut}=1$;
the shift of the level after decoupling the bosons implies that only the identity representation contributes.
The contribution of the discrete states simplifies to
\begin{equation}
\begin{split}
  &  f_{_\text{PF,disc}}(\tau, \rho, \overline{\tau}, \overline{\rho})
\= \\
& \qquad I_0 \, \sum_{\Gamma^{4,4}} \,\sum\limits_{N,\overline{N},w=0}^\infty \, \delta_{N+(p_L^2- p_R^2)  , \overline{N} }\, \mathbf{Q}_{N}^w  (q,r) \overline{\mathbf{Q}}_{_{\overline{N}}}^w  (\qb,\overline{r})\;
	 (q\, \qb)^{w - \frac{1}{2} + \sqrt{\frac{1}{4} + 2\, p_L^2 + 2\,N - w \,(w+1)}} \,,
\end{split}
\end{equation}
with
\begin{equation}
w \; \leq \; \sqrt{\frac{1}{4} + 2\, p_L^2 + 2\,N - w \,(w+1)} \; \leq \; w+1 \,.
\end{equation}
The result simplifies for $p_L =0$. We only get contributions from string oscillators only at certain levels specified by the amount of spectral flow, resulting in
\begin{equation}
f_{_\text{PF,disc}}(\tau, \rho, \overline{\tau}, \overline{\rho}) \= I_0 \biggl( \,\sum_{w=0}^\infty\,
	\abs{ \mathbf{Q}_{{w(w+1)}}^w  (q,r)}^2   \, (q\,\qb)^{2w} + (q\,\qb)^{-\frac{1}{2} + \sqrt{\frac{1}{4} + p_L^2} }+ \cdots
	\biggr)
\end{equation}
where the ellipsis includes modes with all parameters $N, w, p_L \neq 0$.
The main point to note is that the spectrum has a non-trivial twist gap; the spacetime currents discussed
in~\cref{sec:supercurrents} are the only chiral states of the theory.
Because of the current-current bilinears the twist gap is of order~1
for all $k \geq 2$.

\section{Comments on near-extremal BTZ black holes}\label{sec:nhbtz}

We now have at hand the one-loop determinant around the thermal \AdS{3} geometry for both the bosonic string and the superstring. The former is given by~\eqref{eq:bosexpZws}, and we write down an analogous expression for the latter below.
These results can now be used to infer aspects of the BTZ black hole thermodynamics, which was one of the motivations behind our analysis.

We recall that the Euclidean manifold $\mathbb{H}^+_3/\mathbb{Z}$ can be interpreted either as the thermal \AdS{3} geometry, as we have done in~\eqref{eq:EucAdS3} and \eqref{eq:thermalid},
or as the Euclidean BTZ spacetime. The latter is obtained by declaring the contractible cycle in the geometry to the Euclidean time circle. In the boundary~$T^2$ geometry
this is achieved by an S-modular transform, $\tau \to -\frac{1}{\tau}$. One therefore obtains the well-known result\footnote{In this section we will use $\tau$ to indicate the modular parameter of the boundary torus in the BTZ frame.
}
\begin{equation}\label{eq:BTZfromAdS}
\mathscr{Z}_{_\text{CFT}}(\tau,\overline{\tau}) \,\Big{|}_\text{BTZ} \=
\mathscr{Z}_{_\text{CFT}}\left(-\frac{1}{\tau},-\frac{1}{\overline{\tau}}\right) \,\Big{|}_\text{thermal \AdS{3}} \,.
\end{equation}
Therefore, using~\eqref{eq:Zwssp},~\eqref{eq:bosexpZws}, we can immediately obtain the partition function for strings around the BTZ spacetime.
In the case of the superstring, there are additional considerations related to the spin structure, that we elaborate below.

\paragraph{Bosonic strings on \AdS{3}:}
Let us first discuss the bosonic string on BTZ $\times\, X$. Using~\eqref{eq:BTZfromAdS},  \eqref{eq:Zwssp}, \eqref{eq:bosexpZws} we have
\begin{equation}\label{eq:Zbtzbos}
\mathscr{Z}_{_\text{CFT}}(\tau,\overline{\tau}) \,\Big{|}_\text{BTZ}
\=
e^{S_\text{tree}-\frac{2\pi \,i}{24} \left(\frac{1}{\tau} - \frac{1}{\overline{\tau}} \right) \,+ \, \cdots} \times  \abs{\frac{\left(1-e^{-\frac{2\pi\,i}{\tau}}\right)}{\eta(-\frac{1}{\tau})}}^2
\times \prod_{n_1,n_2=0}^\infty \frac{1}{1-e^{- 2\pi\,i \left( \frac{2+n_1}{\tau} - \frac{2+n_2}{\overline{\tau}}\right)}} \times \cdots \,.
\end{equation}
The worldsheet analysis does not capture the tree level piece, but we can estimate it at large~$k$ from
semiclassical gravity, where one obtains $S_\text{tree} \= \frac{c}{24} \left(\frac{1}{\tau} - \frac{1}{\overline{\tau}}\right)$,
cf.~\cite{Maloney:2007ud}. This is obtained by evaluating the on-shell action of Einstein-Hilbert gravity with the Gibbons-Hawking boundary term.

We can map the BTZ black hole to a configuration contributing to the grand canonical partition function of the spacetime CFT, with fixed temperature $T$ and angular chemical potential $\Omega$, with the identifications
\begin{equation}
T \= \frac{1}{\pi \,i} \,\frac{1}{\overline{\tau}-\tau} \,, \qquad
\Omega \= \pi\, (\tau + \overline{\tau}) \,.
\end{equation}
The near-extremal limit is obtained as follows:\footnote{A more precise characterization of the limit involves the spacetime CFT central charge $c$~\cite{Ghosh:2019rcj,Pal:2023cgk}. Specifically, we require, as~$c \to \infty$, $\tau \sim \order{c}$
and~$\overline{\tau} \sim \order{1/c}$
in the complex directions as shown. These statements refer to a complex saddle-point in the Euclidean theory wherein~$\tau$, $\overline{\tau}$ are independent complex numbers that are not necessarily complex conjugate.}
\begin{equation}\label{eq:BTZnext}
\tau \to i\,\infty \,, \qquad  \overline{\tau} \to -i\,0, \qquad \text{holding }\;  \abs{\tau} <1 \,.
\end{equation}
The latter condition ensures that the BTZ black hole is the dominant saddle for the gravitational problem. In this limit we can evaluate~\eqref{eq:Zbtzbos}.
First, note that the contribution from the current bilinears (the third term in the product) is exponentially close to unity, and the same is true for all the terms we have elided over owing to the
non-vanishing twist gap for any $k>3$. Upon dropping these, and assuming that we can use the  semiclassical
expression~$S_\text{tree} \sim \frac{c}{24}(\frac{1}{\tau} - \frac{1}{\overline\tau})$,
the modular properties of the $\eta$ function leads to
\begin{equation}
\begin{split}
\mathscr{Z}_{_\text{CFT}}(\tau,\overline{\tau}) \,\Big{|}_\text{BTZ}
&\; \;
\underset{\scriptstyle{\overline{\tau} \to -i\, 0}}{\sim} \; \;
	e^{\frac{2\pi \,i}{24} \,(c-1)\, \left(\frac{1}{\tau} - \frac{1}{\overline{\tau}} \right) } \times
	 \frac{\left(1-e^{-\frac{2\pi\,i}{\tau}}\right)}{(-i\,\tau)^{\frac{1}{2}} \;\eta(\tau)} \\
&\; \; \underset{\scriptstyle{\tau \to i\, \infty}}{\sim} \; \;
(-i\,\tau)^{-\frac{3}{2}} \; \exp\Bigl(-\frac{2\pi \,i \,(c-1) }{24\,  \overline{\tau}} - \frac{2 \pi \, i\, \tau}{24} \Bigr) \, .
\end{split}
\end{equation}
The Legendre transform of~$\mathscr{Z}$ to a mixed ensemble with fixed large angular momentum~$J$ and temperature $T \to 0$ is given by~\cite{Ghosh:2019rcj},
\begin{equation}\label{eq:bosBTZT32}
\widetilde{\mathscr{Z}}_{_\text{CFT}}(T,J) \,\Big{|}_\text{BTZ}
\;\;\sim\; \;
    T^\frac{3}{2}\, J^{-\frac{3}{4}}\, e^{2\pi\, \sqrt{\frac{c}{6}\, J}}\, e^{-\frac{1}{T}\, \left(J-\frac{1}{12}\right)}\,.
\end{equation}
In the following we study this limit, combined with the limit~$c \to \infty$. More precisely, the temperature range of interest is~$T_\text{gap} \, e^{-S_\text{tree}} \ll T \ll T_\text{gap}$ where~$T_\text{gap} \sim 1/c$.

The contribution from $S_\text{tree}$ reproduces the Bekenstein-Hawking entropy of the extremal black hole,
while the one-loop determinant captures the expected $T^\frac{3}{2}$ contribution to the partition function.
In the semiclassical limit, one attributes this result to the Schwarzian modes which are supported in the near-
horizon region~\cite{Ghosh:2019rcj,Iliesiu:2020qvm}.\footnote{One can equivalently understand this contribution as
arising from low-lying eigenvalues of the quadratic fluctuation operator in the full geometry~\cite{Kolanowski:2024ta}.}
We see here that the result continues to hold for finite string length $\lads \gtrsim \ell_s$.
This is explained by the presence of a non-trivial twist gap~\cite{Ghosh:2019rcj,Pal:2023cgk}, which is present for~$k>3$.

\paragraph{Superstrings on \AdS{3}:}
Turning to the superstring, we can carry out a similar exercise.
The choice of boundary conditions for the fermions in spacetime entails a corresponding choice of the boundary spin structure.
The partition functions we have computed with periodic and antiperiodic boundary conditions in the thermal \AdS{3} geometry, correspond to the Neveu-Schwarz sector trace in the spacetime CFT  with and without the insertion of $(-1)^{F_s}$, respectively.
When we use the S-modular transform to evaluate the result for the BTZ geometry, these  turn into the
Neveu-Schwarz and Ramond sector
traces without any insertions. Specifically,
\begin{subequations}\label{eq:btzNSRsusy}
\begin{align}
\mathscr{Z}^\text{NS}_{_\text{CFT}}(\tau, \rho, \overline{\tau},\overline{\rho}) \Big{|}_\text{BTZ}
&\=
\Tr_{_\text{NS}}(q^{L_0}\, r^{J_0^3}\, \overline{q}^{\overline{L}_0}\, \overline{r}^{\overline{J}_0^3})\Big{|}_\text{BTZ}
\=
    \Tr_{_\text{NS}}(q_{_\text{AdS}}^{L_0}\, r_{_\text{AdS}}^{J_0^3}\, \overline{q}_{_\text{AdS}}^{\overline{L}_0}\, \overline{r}_{_\text{AdS}}^{\overline{J}_0^3})  \Big{|}_\text{AdS}  \label{eq:btzNS}\\
\mathscr{Z}^\text{R}_{_\text{CFT}}(\tau, \rho, \overline{\tau},\overline{\rho}) \Big{|}_\text{BTZ}
&\=
    \Tr_{_\text{R}}(q^{L_0}\, r^{J_0^3}\, \overline{q}^{\overline{L}_0}\, \overline{r}^{\overline{J}_0^3})\Big{|}_\text{BTZ}
\=
    \Tr_{_\text{NS}}( (-1)^{F_s}\, q_{_\text{AdS}}^{L_0}\, r_{_\text{AdS}}^{J_0^3}\, \overline{q}_{_\text{AdS}}^{\overline{L}_0}\, \overline{r}_{_\text{AdS}}^{\overline{J}_0^3})\Big{|}_\text{AdS}  \label{eq:btzR}
\end{align}
\end{subequations}
with
\begin{equation}\label{eq:taurhomapbtz}
q_{_\text{AdS}} \= e^{-\frac{2\pi i}{\tau }}  \,,
\qquad
r_{_\text{AdS}} \= e^{-2\pi i\,\frac{\rho}{\tau}}\,.
\end{equation}

Let us first analyze the NS sector in the BTZ background. Using Equation~\eqref{eq:btzNS} we can equivalently analyze the AdS calculation.
From the worldsheet analysis around the AdS space~\eqref{eq:Zads3s3t4}, we obtain
(we read the variables below as~$q_{_\text{AdS}}$ and~$r_{_\text{AdS}}$)
\begin{equation}\label{eq:AFTAdS}
\begin{split}
\Tr_{_\text{NS}}(q_{_\text{AdS}}^{L_0}\, r_{_\text{AdS}}^{J_0^3}\,\overline{q}_{_\text{AdS}}^{\overline{L}_0}\, \overline{r}_{_\text{AdS}}^{\overline{J}_0^3} )\Big{|}_\text{thermal \AdS{3}}
&\=
    e^{S_\text{tree}} \times \abs{\etchar^\text{AF}(0)}^2\, \times
    \abs{\frac{\vartheta_3\left(\frac{\rho}{2},\tau \right)^2}{\eta^6(\tau) }} \times \cdots \,,
\end{split}
\end{equation}
with
\begin{equation}\label{eq:afcharl0}
\begin{split}
\etchar^\text{AF}(0)
&\=
    (1-q)\, \frac{\vartheta_3\left(\frac{\rho}{2},\tau \right)^2}{\eta(\tau)^3\, \vartheta_1(\rho,\tau)}
    \sum_{m\in \mathbb{Z}}\, q^{(k+1)\,m^2 }\, r^{(k+1)\,m}\,
    \frac{q^m\,r^{\frac{1}{2}} - q^{-m}\, r^{-\frac{1}{2}}}{(1+ q^{m+\frac{1}{2}}\ r^{\frac{1}{2}})^2 \, (1+ q^{-m+\frac{1}{2}}\ r^{-\frac{1}{2}})^2 } \,.
\end{split}
\end{equation}
We have directly given the final answer above.
To obtain it, we carried out the integral over the worldsheet modulus using the explicit form of $\mathsf{Z}^{(n,m)}_\text{AF}$.
This leads to a single string free energy  $f_{_\text{AF}}$, which has a discrete and a continuous spectrum.
The low-lying discrete states are chiral currents (the analog of~\eqref{eq:I0currents}), whose contribution after exponentiating  to obtain the multi-string partition function is the $m=0$ term of~\eqref{eq:afcharl0}.
Finally, the spectral flow described in~\cref{sec:supercurrents} leads to the terms with $m\neq0$, which are geometries with non-trivial $\mathrm{SU}(2)$ gauge field turned on at the boundary.
Our focus here is on the chiral states of the spacetime CFT. In~\eqref{eq:AFTAdS} we have indicated the contribution of the chiral states.
The non-chiral states have finite twist for $k\geq 2$, and are not explicitly indicated in the formula.

We want to take the near-extremal limit
as in~\eqref{eq:BTZnext}.
Using the modular properties of the theta functions and the asymmetric scaling between left and right movers, we find,
\begin{equation}\label{eq:NSBTZfinal}
\mathscr{Z}^\text{NS}_{_\text{CFT}}(\tau, \rho, \overline{\tau},\overline{\rho}) \Big{|}_\text{BTZ}
\; \sim \;
    e^{S_\text{tree} +  \beta + \frac{T}{T_\text{gap}}\, \rho^2}\, \frac{T^5}{T_\text{gap}^5} \, \sum_{m\in \mathbb{Z}}
    \frac{\left( m+ \frac{\rho}{2} \right)\, e^{\frac{T}{T_\text{gap}} \bigl(1-4 \,\left(m +\frac{\rho}{2}\right)^2\bigr)}}
    {\sin(\pi\rho)} \,.
\end{equation}
In the above expression $T_\text{gap} = \frac{4}{\pi^2}\, E_\text{gap}  \sim 1/c$ refers to the temperature scale at which the quantum effects around the classical saddle become important. We have folded in the tree-level contribution which comes from the right-movers, and the zero-point energies do not vanish owing to our antiperiodic fermion boundary conditions.

We can interpret the result as follows: in the near-extremal limit, the BTZ black hole develops a long \AdS{2} throat, which supports a set of zero modes.
One set of these arises from the Schwarzian mode described in the bosonic analysis, and leads to a factor of $T^{\frac{3}{2}}$. For the compactification on $\mathbf{S}^3 \times T^4$, we have additional bosonic zero modes, each of which contributes an additional $T^\frac{1}{2}$.
The $\mathbf{S}^3$ isometries lead to three zero modes associated with the $\mathrm{SU}(2)$ R-symmetry, while there are four zero modes associates with the $\mathrm{U}(1)^4$ gauge fields coming from $T^4$. Altogether, the one-loop determinant has an overall $T^5$  factor.

The $\mathrm{SU}(2)$ chemical potential has been kept at a  generic value in~\eqref{eq:NSBTZfinal}. One can rewrite the answer as the sum over fixed charge sectors by Poisson summation or, equivalently, by expressing the sum explicitly in terms of theta functions and using a modular transformation.\footnote{
The sum is a derivative of a theta function with respect to the elliptic variable which is itself a weight-$\frac32$ theta function,
\[
 \sum_{m\in \mathbb{Z}}
      \left( m+ \frac{\rho}{2} \right)
    e^{-\frac{4\,T}{T_\text{gap}} \left(m +\frac{\rho}{2}\right)^2}
    = \pdv{\rho}( e^{-\frac{T}{T_\text{gap}} \,\rho^2}\, \vartheta_3\left(e^{-\frac{4\,T}{T_\text{gap}}\, \rho}, e^{-\frac{8\,T}{T_\text{gap}}}\right) )    = \pdv{\rho} \vartheta_3\left(e^{i\pi\, \rho}, e^{-\frac{\pi^2\,T_\text{gap}}{2\,T}}\right) .
\]
\label{fn:AFsum}
}
Doing so, we obtain,
\begin{equation}\label{eq:AFmodular}
\begin{split}
\mathscr{Z}^\text{NS}_{_\text{CFT}}(\tau, \rho, \overline{\tau},\overline{\rho}) \Big{|}_\text{BTZ}
\; \sim \;
    \left(\frac{T}{T_\text{gap}}\right)^{\frac{7}{2}}\, e^{S_\text{tree} + \beta + \frac{T}{T_\text{gap}}\, \rho^2}  \sum_{n=1}^\infty\,  n\, \chi_{\frac{n-1}{2}}(r)\, e^{- \frac{T_\text{gap}}{T}\, n^2  } \,.
\end{split}
\end{equation}
Notice that the $T^\frac{3}{2}$ factor from the $\mathrm{SU}(2)$ zero modes disappears in the fixed charge sector.
A similar story applies for the $\mathrm{U}(1)^4$ currents (cf.~\cite{Mertens:2019tcm}), although in this case, we did not turn on the associated chemical potentials in our worldsheet analysis.
In any event, in the canonical ensemble where all the charges are fixed, we reproduce the result~\eqref{eq:bosBTZT32} encountered earlier in the bosonic string, which we can write for low temperatures $T_\text{gap} \, e^{-S_\text{tree}} \ll T \ll T_\text{gap}$  as
\begin{equation}
\widetilde{\mathscr{Z}}^\text{NS}_{_\text{CFT}}(T,Q_{_{\mathrm{SU}(2)}}, Q_{_{\mathrm{U(1)^4}}} =0) \Big{|}_\text{BTZ}
\; \sim \;
   Q_{_{\mathrm{SU}(2)}}\,  \left(\frac{T}{T_\text{gap}} \right)^{\frac{3}{2}}\,
    e^{S_\text{tree}+ T\,S_1 -\beta\, \left(T_\text{gap} \, Q_{_{\mathrm{SU}(2)}}^2-1 + E_1\right)} \,.
\end{equation}
%
%
Here $S_1$ is the classical correction to the thermodynamic entropy, and $E_1$ the shift in the ground state energy which includes the $J-\frac{1}{12}$ factor observed in the bosonic string discussion~\eqref{eq:bosBTZT32}. We have left explicit an additional contribution from the charged states, the factor $T_\text{gap} \, Q_{_{\mathrm{SU}(2)}}^2$, which shifts the energy in the fixed charge sector (a similar statement applies for the case of the $\mathrm{U(1)}^4$ charges).
Note that this suppression means that even in the grand canonical ensemble, the dominant contribution comes from the state with the lowest value of charge (in this case~$n=1$, since~$n=0$ has vanishing contribution).
Laplace transforming the above will result in a microcanonical density of states with a continuum spectrum in a fixed charge sector, as expected for the partition function in the absence of supersymmetry~\cite{Iliesiu:2020qvm}.

\bigskip

Finally, let us turn to the R sector in the BTZ background. In this case, we invoke~\eqref{eq:btzR} to equivalently analyze the AdS calculation, where we have  periodic boundary conditions for the spacetime fermions around the time circle.
Then, using the single string free energy~\eqref{eq:fspSS} and the analysis of~\cref{sec:sugraspec}, we find
\begin{equation}\label{eq:Zadssuper}
\mathscr{Z}^{\text{NS},(-1)^F}_{_\text{CFT}}(\tau,\overline{\tau},\rho,\overline{\rho}) \,\Big{|}_\text{thermal \AdS{3}}
\=
e^{S_\text{tree}} \times  \abs{\etchar(0)}^2
\times \abs{ q^{\frac{1}{4}}\, \frac{\vartheta_4(\frac{\rho}{2},\tau)^2}{\eta(\tau)^6} }^2 \times \cdots \,.
\end{equation}
where, $\etchar(0)$ is the vacuum character (in NS sector) of the $\mathcal{N} =4$ superconformal algebra~\eqref{eq:N44vacNS}.
The second factor involving the theta function incorporates the contribution from the zero-twist currents from the $T^4$.
We are again assuming that the elliptic average described in~\eqref{eq:ellipticaverage} has been carried out in obtaining this vacuum character. The terms we are dropping are the contributions from the other states of the theory which, as we have argued in~\cref{sec:beyondsugra},
are separated from the vacuum by a finite twist gap for $k\geq 2$.

We can now specialize the general answer~\eqref{eq:Zadssuper} to the low-temperature limit. Let us do so  first for a generic value of the chemical potential. We end up with,
\begin{equation}\label{eq:RBTZfinal}
\mathscr{Z}^{R}_{_\text{CFT}}(\tau, \rho, \overline{\tau},\overline{\rho}) \Big{|}_\text{BTZ}
\; \sim \;
    e^{S_\text{tree} + \frac{T}{T_\text{gap}} \,(\rho^2-1)}\, \frac{T}{T_\text{gap}} \, \frac{\cos^4(\frac{\pi}{2}\,\rho)}{\sin(\pi \rho)}
    \sum_{m\in \mathbb{Z}}
    \frac{\left( m+ \frac{\rho}{2} \right)\, e^{\frac{T}{T_\text{gap}} \left(1- 4\,\left(m +\frac{\rho}{2}\right)^2\right)}}
    { \left(1- 4\,\left(m +\frac{\rho}{2}\right)^2\right)^2} \,.
\end{equation}
We once again can understand the factors of $T$ in this calculation by counting the zero modes. As before, we have 10 bosonic zero modes (3 from \AdS{3}, 3 from $\mathbf{S}^3$, and 4 from $T^4$), but now we have 8 fermionic zero modes in addition (from the gravitini). This gives an overall factor of $T\sim T^5/T^4$.
Note that compared to the analysis of~\cite{Heydeman:2020hhw} there is  an extra factor of  $T^2\, \cos^2(\frac{\pi}{2}\,\rho)$  originating from the $U(1)^4$ currents.

We can specialize the above to obtain the index of BPS black holes. The strategy as in~\cite{Heydeman:2020hhw} is to work in the near-extremal near-BPS regime.\footnote{One can approach the BPS point in a manner which manifestly preserves supersymmetry. This leads to complex gravitational saddles for the supersymmetric index~\cite{Cabo-Bizet:2018ehj,Cassani:2019mms,Bobev:2019zmz, Bobev:2020pjk,Larsen:2021wnu,Iliesiu:2021are,BenettiGenolini:2023rkq,Boruch:2023gfn}.}
The BPS BTZ black hole is the ground state in the R-R sector of the spacetime CFT.  To obtain the index in the full spacetime, we will need to set $\rho \to 1$.

Before doing to, let us understand the behavior for general values of $\rho$ and low temperatures, by estimating the sum appearing in~\eqref{eq:RBTZfinal}. Consider,
\begin{equation}\label{eq:Fyrhodef}
F(T,\rho)
\= \cot(\frac{\pi\,\rho}{2}) \, \sum_{m\in \mathbb{Z}}  \,
    \frac{\left( m+ \frac{\rho}{2} \right) \,e^{ \frac{T}{T_\text{gap}}\, \left(1-4\,\left[m +\frac{\rho}{2}\right]^2\right)} }{\left(1-4\,\left[m+\frac{\rho}{2}\right]^2 \right)^2} \\
\end{equation}
One can check that the function is, in fact, independent of $\rho$, and satisfies
\begin{equation}
\pdv{\rho}F(T,\rho) \=  0 \,, \qquad
F(0,\rho)  \= 0 \,,     \qquad
\pdv{T}F(T,\rho)\bigg|_{T =0} \= \frac{\pi}{4\, T_\text{gap}} \,.
\end{equation}
The first statement follows because the summands in $\partial_\rho F$ are odd under simultaneous reflection, $m \to -m$ and $\rho \to -\rho$. The leading behavior at low $T$ can be deduced by expanding the exponential and completing the sums at some specific value of $\rho$. This will suffice for our purposes, but note that $\pdv[2]{T} F$ reduces to the sum appearing in~\eqref{eq:NSBTZfinal} which implies that the function is an integral of a theta-function.

Therefore, at low temperatures, $T_\text{gap} e^{-S_\text{tree}} \ll T \ll T_\text{gap}$, at a generic value of the chemical potential we obtain
\begin{equation}
\mathscr{Z}^{R}_{_\text{CFT}}(\tau, \rho, \overline{\tau},\overline{\rho}) \Big{|}_\text{BTZ}
\; \sim \;
    e^{S_\text{tree} } \,
  \frac{T^2}{T_q^2} \, \cos[2](\frac{\pi\rho}{2})
     + \cdots \,.
\end{equation}
The leading degeneracy captured by $S_\text{tree}$ arises from the right movers, as in the bosonic case.
Switching to mixed ensemble,  this would give $ e^{2\pi\, \sqrt{\frac{c}{6}\, J} }$ in terms of the angular momentum. In the present situation, we have also fermionic and bosonic zero modes coming from the $T^4$,  which lead to the left-movers receiving an additional contribution of $\vartheta_4(\rho/2,\tau)^2/\eta(\tau)^6$.
In the limit of interest~$\rho \to 1$ for evaluating the index, these currents lead to a factor of  $T^2\, (\rho-1)^2$. This results in a vanishing index. Since these fermion zero modes are delocalized in the spacetime, they need to be  soaked up if we want to focus on the index of the black hole~\cite{Dabholkar:2010rm}.  Upon doing so, and upon removing the bosonic zero modes by passing to the fixed charge ensemble, we obtain a finite, $T$-independent result as expected.
Had there been no fermionic or current zero modes, then the result would have been finite and independent of $\rho$, demonstrating that in the R-sector the index equals the degeneracy. This should be the case for \asx{K3}.

\bigskip

We therefore see that the string partition function, even at finite $k$, captures the thermodynamic properties of near-extremal BPS black holes correctly.
There is, however, one caveat worth pointing out.
Our analysis assumes that the tree level result, $S_\text{tree}$, agrees with the semiclassical gravity computation. In the superstring, one may argue that this ought to be guaranteed given the worldsheet computation of the spacetime CFT conformal anomaly~\cite{Eberhardt:2023lwd}.
Without this assumption, we cannot argue that the string computation sees the gap $E_\text{gap} = \frac{3}{c}$,  which separates the BPS states from the non-BPS states in the spectrum.
In particular, demonstrating that the index matches the black hole degeneracy is tied intimately to recovering the Bekenstein-Hawking result from a worldsheet computation.

\section{Discussion}\label{sec:discuss}

The thrust of our analysis was to re-investigate the computation of the 1-loop partition function for strings in \AdS{3}. One motivation was to argue that the lessons learnt from the semiclassical gravitational path integral computations continue to hold even when there is no hierarchical separation between the \AdS{} and string length scales.
However, as noted in the Introduction, there are other reasons to undertake this exercise, such as understanding the spacetime symmetries from a worldsheet partition function.
We have shown
how the worldsheet torus partition function captures the information about the asymptotic symmetries. Using this, we have demonstrated that the results for near-extremal black hole thermodynamics hold for finite string length.

\bigskip

There are several interesting directions for future exploration, especially in the context of the superstring, which we outline below.

\paragraph{Type II superstrings on \asx{K3}:}
The backgrounds \asx{T^4} and \asx{K3} have served as paradigmatic examples of the \AdS{3}/CFT$_2$ correspondence.
Both these cases admit pure NS-NS solutions, which should be amenable to RNS analysis of the superstring.
However, as mentioned in the Introduction, there is a tension between the GSO projection
of~\cite{Giveon:1998ns} and realizing~$K3$ as the orbifold~$T^4/\mathbb{Z}_2$ where the~$\mathbb{Z}_2$ acts as a reflection on the four worldsheet bosons as well as the four superpartner RNS fermions.
While this action preserves worldsheet modular invariance by construction, it breaks spacetime supersymmetry.
It would be interesting to modify the orbifold action so as to be consistent with spacetime supersymmetry and modular invariance.
Another approach to resolve this tension would be to exploit the hybrid formalism for strings developed in~\cite{Berkovits:1999im}.
Once this is resolved, it will be interesting to explicitly verify the Type IIA/heterotic duality at the level of the one-loop partition function.
This would give an example of a spacetime CFT with~$(0,4)$ supersymmetry, which has
both supersymmetric as well as non-supersymmetric extremal BTZ black holes.

\paragraph{Superstrings on \asx{\,\mathbf{S}^3\, \times \,\mathbf{S}^1}:}
A second example, which is worth analyzing, given the issues encountered in the $K3$ compactification of Type II theories, is to consider the geometry \asx{\mathbf{S}^3 \times \mathbf{S}^1}.
This is known to give a dual 2d CFT with {\it large} $\mathcal{N} =4$ superconformal algebra~\cite{deBoer:1999gea}. Some aspects of this background, which were confusing from the holographic perspective~\cite{Gukov:2004ym}, have now been clarified~\cite{Eberhardt:2017fsi,Eberhardt:2017pty,Witten:2024yod}.
In an upcoming work~\cite{Murthy:2024s3s1}, we will argue that the one-loop string partition function exactly reproduces the supergravity spectrum derived in~\cite{Eberhardt:2017fsi}, and describe some implications for near-extremal black holes. As a further generalization, one could consider orbifolds of these theories preserving lower amount of supersymmetry (e.g., $\mathcal{N} =3$)~\cite{Eberhardt:2018sce}.\footnote{Related investigations with an aim of understanding the full spectrum in theories with more than 4 supercharges are being pursued in~\cite{Heydeman:2024tba}. We thank Joaquin Turiaci for discussions on related issues. }

\paragraph{Strings on BTZ:} In our analysis of the BTZ black holes, we have exploited the fact that the Euclidean geometry is simply related to the thermal \AdS{3} background.
This absolves us from directly having to compute the spectrum of strings around the BTZ geometry.
While aspects of the spectrum have been analyzed directly in the Lorentzian BTZ geometry~\cite{Rangamani:2007fz,Parsons:2009si}, the near-extremal regime has not been investigated thoroughly.
We expect to see imprints of the near-gapless modes localized in the near-horizon region in the spectrum and observables. It would be interesting to explore these explicitly from the worldsheet away from the semiclassical gravity regime.

\paragraph{Beyond the RNS formalism:} The superstring theories are constrained to have $k\geq 2$ within the RNS formalism. However, in the hybrid formalism there is no obstruction to probing the regime $1 \leq k <2$, which allows one to approach the tensionless limit (at the lower end~\cite{Eberhardt:2020bgq}).\footnote{Another regime of interest is the sub-stringy regime, $k<1$, where black holes are argued not to dominate the ensemble~\cite{Balthazar:2021xeh}.}
We are unaware of an explicit construction of the worldsheet torus amplitude in the hybrid formalism for the $\slt$ WZW models with $k>1$,
though~\cite{Chandrasekhar:2008qx} analyzes the worldsheet theory in \AdS{2}. It would be interesting to develop this further and examine the nature of near-extremal black holes in this near-tensionless regime of string theory.

The essence of the arguments we use in this paper relies on (a) the thermodynamics of the near-horizon BTZ geometry and
(b) the modular transformation of AdS3. Both these arguments should apply to string theory with mixed-flux (NS-NS+RR) backgrounds as well, cf.~\cite{Berkovits:1999im}.
While the global $\mathfrak{psu}(1,1|2)$ charges are understood in other formalisms (cf.~\cite{Seibold:2024qkh}), the origin of the full currents is as yet unclear. This poses an interesting problem for those approaches with a clear target.

\section*{Acknowledgements}

It is a pleasure to thank Ofer Aharony, Matthias Gaberdiel, Rajesh Gopakumar, Lorenz Eberhardt, Per Kraus, Luca Iliesiu, Juan Maldacena, Emil Martinec, Shiraz Minwalla, Ashoke Sen, Joaquin Turiaci,  and Edward Witten for illuminating discussions. We would also like to thank Ofer Aharony and Joaquin Turiaci for comments on a draft of the manuscript.

C.F.~and M.R.~were supported by U.S.~Department of Energy grant DE-SC0009999 and funds from the University of California.
S.M.~acknowledges the support of the J.~Robert Oppenheimer
Visiting Professorship at the Institute for Advanced Study, Princeton and the STFC grants ST/T000759/1,  ST/X000753/1. We would like to thank KITP for hospitality during the program, “What is string theory? Weaving perspectives together”, which was supported by the grant NSF PHY-2309135 to the Kavli Institute for Theoretical Physics (KITP). S.M.~and M.R.~would also like to thank the Aspen Center for Physics, which is supported by National Science Foundation grant PHY-2210452 where this work was completed.
Finally, M.R.~is also grateful to the long-term workshop YITP-T-23-01 held at YITP, Kyoto University, TSIMF, and  Institut Pascal at Universit\'e Paris-Saclay  (supported by the program ``Investissements d'avenir" ANR-11-IDEX-0003-01), during the course of this work.

\appendix

\section{Bosonic string partition functions}\label[appendix]{sec:appreviews}

We review some ingredients entering the bosonic string calculations. In~\cref{sec:sl2part} we quickly sketch the derivation of $\mathcal{Z}_{\slt_k}$ and detail its modular properties. In~\cref{sec:bosZeval} we provide the details for the evaluation of the worldsheet modular integral, which leads to the final result for the spectral decomposition studied in~\cref{sec:spectrum}.

\subsection{The \texorpdfstring{$\slt_k$}{sl2} partition sum}\label[appendix]{sec:sl2part}

To compute the partition function of the $\slt_k$ WZW model, it is convenient to
use a different coordinate system, where the metric takes the form
\begin{equation}\label{eq:H3coord}
ds^2 \= k \left[ d\xi^2 + (dv + v\, d\xi) \, (d \overline{v} + \overline{v} \, d\xi)  \right] ,
\end{equation}
This can be obtained by parameterizing the $\mathrm{SL}(2)$ group element as
\begin{equation}\label{eq:H3g}
g \=
\begin{pmatrix}
	e^\xi \, (1+\abs{v}^2) & v \\
	\overline{v} & e^{-\xi}
\end{pmatrix} .
\end{equation}
The identifications of these coordinates are
\begin{equation}\label{eq:H3identify}
\xi \sim \xi + \beta\,, \qquad
v \sim v \, e^{i\mu\beta} \,,
\qquad
\overline{v} \sim \overline{v} \, e^{-i \mu \beta} \,.
\end{equation}
The worldsheet action taking into account the NS-NS B-field in these variables is
\begin{equation}\label{eq:H3ws}
S \= \frac{k}{\pi}\, \int\, d^2 z\,
	\left( \partial \xi \overline{\partial} \xi + \abs{\partial \overline{v} + \overline{v}\, \partial \xi}^2 \right) .
\end{equation}
We now have solitonic sectors labelled by two integers $(n,m)$ owing to the identification, characterizing the maps from the worldsheet torus to the spacetime circle
\begin{equation}\label{eq:H3monodromy}
\begin{aligned}
\xi(z+2\pi) & \= \xi(z) + n\,\beta \,, &\qquad \xi (z  + 2\pi\ts) & \= \xi(z) + m\,\beta \,, \\
v(z+ 2\pi) & \= v(z)\, e^{in\mu\beta} \,, &\qquad  v(z+2\pi\ts) & \= v(z)\, e^{im\mu\beta} \,.
\end{aligned}
\end{equation}

To evaluate the partition function, one as usual, separates the fields into a classical part capturing the maps above, and a fluctuating quantum piece, viz.,
\begin{equation}\label{eq:H3cq}
\begin{split}
\xi(z, \overline{z})
&\=
	\xi_q(z,\overline{z}) +
	 i \frac{\beta}{4\pi \ts_2}
	\left[ z\, (n\, \tsb -m ) -  \overline{z} \ (n\,\ts - m)\right] ,\\
v(z, \overline{z})
&\=
	v_q(z,\overline{z}) \, \exp\left( -\frac{\mu\beta}{4\pi \ts_2}
	\left[ z\, (n \,\tsb -m ) -  \overline{z} \ (n\,\ts - m)\right] \right) .
\end{split}
\end{equation}
The sigma model action now reads
\begin{equation}\label{eq:H3sigma}
\begin{split}
S \= \sum_{n,m\in \mathbb{Z}}
	\bigg(\frac{k\,\beta^2}{4\pi\ts_2} \abs{n\ts -m}^2 + \frac{k}{\pi} \int\, d^2 z\,
	\left[ \abs{\partial \xi_q}^2 + \abs{ \left( \partial + \frac{1}{2\ts_2} \, \usb_{n,m}
	+ \partial \xi_q \right) \overline{v}_q}^2 \right]  \bigg) \,.
\end{split}
\end{equation}

The final task is to carry out the functional integral over the fluctuating fields $\{\xi_q, v_q, \overline{v}_q\}$. This is achieved using techniques similar to those used to compute the chiral anomaly~\cite{Gawedzki:1991yu}. As demonstrated in~\cite{Maldacena:2000kv} this results in the final answer~\eqref{eq:Zbosads3}, reproduced here for convenience
\begin{equation}\label{eq:Zbosads3A}
\begin{split}
\mathcal{Z}_{\slt_k}\bigl(\ts,\tsb;\tau,\overline{\tau}\bigr)
\=
	\frac{\beta\, \sqrt{k-2}}{2\pi\, \sqrt{\ts_2}}\,
	 \sum_{n,m \in \mathbb{Z}}\,  \frac{\exp \left( -\frac{k\, \beta^2}{4\pi\,\ts_2} \, \abs{m- n\,\ts}^2 + \frac{2\pi}{\ts_2} \, (\Im(\us_{n,m}))^2\right)}{\abs{\vartheta_1(\us_{n,m},\ts)}^2}\,.
\end{split}
\end{equation}

\paragraph{Comments on the derivation:} To arrive at~\eqref{eq:Zbosads3A} from~\eqref{eq:H3sigma} we proceed as follows. First, one decouples the fields $\{v_q, \overline{v}_q\}$ from $\xi_q$. Since $\xi_q$ couples in the form of a background gauge potential, this is achieved by a chiral rotation. This results in a Gaussian integral over $\xi_q$, which can be carried out. For the  fields $\{v_q, \overline{v}_q\}$ one is left with a constant background gauge field set by the holonomy $\us_{n,m}$. We therefore need to evaluate the determinant
\begin{equation}
    \det[\left(\partial + \frac{1}{2\,\ts_2} \,\us_{n,m}\right)\left(\overline{\partial} + \frac{1}{2\,\ts_2} \,\usb_{n,m}\right)] .
\end{equation}

Consider the holomorphic piece. The eigenvalues of the differential operator are
\begin{equation}\label{eq:bostwists}
    \lambda_{k_1,k_2} \= k_1\, \ts + k_2 + \frac{1}{2\,\ts_2} \,\us_{n,m} \,.
\end{equation}
Equivalently, because of the background field, the moding of the eigenvalues gets shifted to $j_1 + \nu_1$ and $j_2 + \nu_2$, respectively. Here the twisted boundary condition is characterized by complex twists $(\nu_1, \nu_2) =  \frac{\tau }{2\,\ts_2}\, (n, -m)$. One can evaluate the determinant using standard zeta function regularization~\cite{DHoker:1986hhu} or using operator methods~\cite{Alvarez-Gaume:1986rcs}. Either choice leads to an ambiguity owing to the holomorphic/chiral anomaly.

The trick is to evaluate the determinant for real twist, fixing the overall phase of the holomorphic determinant, and then analytic continue the result to complex shifts. For instance, with real twists $\nu_1, \nu_2$,  one finds~\cite{DHoker:1986hhu,Alvarez-Gaume:1986rcs}
\begin{equation}
 \prod_{k_1,k_2}   \, \lambda_{k_1,k_2} \=
 e^{i\pi\,\nu_1^2 \,\ts}\,  \frac{\vartheta_1(\nu_2 - \nu_1\, \ts, \ts)}{\eta(\ts)}\,.
\end{equation}
Combining the left and right movers, we have a contribution of $\abs{e^{-i\pi\,\nu_1^2\,\ts}}^2$ in the boson determinant. For real $\nu$, this factor can be rewritten as  $\exp(\frac{2\pi}{\ts_2} \, (\Im(\nu_1\,\ts- \nu_2))^2 )$. Using this form, and the specific values of the twists~\eqref{eq:bostwists}, and including the contribution from $\xi_q$, leads to the final answer in the form quoted in~\eqref{eq:Zbosads3A}.

\paragraph{Modular invariance:} To check the modular invariance of $\mathcal{Z}_{\slt_k}$ we first note that under the $T: \ts \to \ts+1$ and $S: \ts \mapsto -1/\ts$ generators of the modular transformations, the holonomy field $\us_{n,m} = \tau\, (n\, \ts - m)$ transforms as
\begin{equation}
T: \us_{n,m} \; \mapsto \; \us_{n,m-n} \,, \qquad
S: \us_{n,m} \; \mapsto \; \frac{1}{\ts}\, \us_{-m,n} \,.
\end{equation}

To check invariance of $\mathcal{Z}_{\slt_k}$ under $T$, note that the prefactor is invariant. Furthermore,  both the classical  contribution  $\abs{m-n\ts}^2$, and the $\us_{n,m}$ imply the doublet $(n,m)$ transforms as $(n,m) \mapsto (n,m-n)$.  Hence,  the summation over both sets of windings ensures invariance of the partition sum. For the $S$ generator, we instead have,
\begin{equation}
\begin{split}
& \frac{2\pi}{\beta\, \sqrt{k-2}}\, \mathcal{Z}_{\slt_k}
\left(-\frac{1}{\ts},-\frac{1}{\tsb}; \tau, \overline{\tau} \right)
 \\
& \qquad \qquad  \= \, \frac{\abs{\ts}}{\sqrt{\ts_2}}\,
	 \sum_{n,m}\,  \frac{\exp \left( -\frac{k\, \beta^2\,\abs{\ts}^2}{4\pi\,\ts_2} \, \abs{m- n/\ts}^2 + \frac{2\pi\,\abs{\ts}^2}{\ts_2} \, (\Im\left(\frac{\us_{-m,n}}{\ts})\right)^2\right)}{\abs{\vartheta_1(\frac{\us_{-m,n}}{\ts},-\frac{1}{\ts})}^2}\,,\\
&\qquad \qquad \=
	\frac{\abs{\ts}}{\sqrt{\ts_2}}\,
	 \sum_{n,m}\,  \frac{\exp \left( -\frac{k\, \beta^2}{4\pi\,\ts_2} \, \abs{m \ts-n}^2 + \frac{2\pi\,\abs{\ts}^2}{\ts_2} \, (\Im\left(\frac{\us_{-m,n}}{\ts})\right)^2\right)}{\sqrt{\ts\, \tsb}\, \exp\left( -2\pi\,\Im\left( \frac{\us_{-m,n}^2}{\ts} \right) \right)\abs{\vartheta_1(\us_{-m,n},\ts)}^2}\,,\\
&\qquad \qquad \=
	\frac{1}{\sqrt{\ts_2}}\,
	 \sum_{n,m}\,\frac{\exp \left( -\frac{k\, \beta^2}{4\pi\,\ts_2} \, \abs{m \ts-n}^2 + \frac{2\pi\,\abs{\ts}^2}{\ts_2} \, (\Im(\us_{-m,n}))^2 \right)}{\abs{\vartheta_1(\us_{-m,n},\ts)}^2}\,,
\end{split}
\end{equation}
which is the original expression for~$\mathcal{Z}_{\slt_k}
\left(\ts,\tsb; \tau, \overline{\tau} \right) $ with the summation variables shifted as $(n,m) \mapsto (-m,n)$. Once again, the sum over $m$ and $n$ serves to render the answer modular invariant.

One way to state the result is to record that the combination
\begin{equation}
\frac{\exp \left(\frac{2\pi}{\ts_2} \, (\Im(\us_{n,m}))^2\right)}{\sqrt{\ts_2}\, \abs{\vartheta_1(\us_{n,m},\ts)}^2}\,,
\end{equation}
which arises from the fluctuating quantum fields transforms like the classical contribution, mapping the doublet $(n,m)$ appropriately. This will be useful for writing down fermion characters in the superstring.

\subsection{Evaluation of the worldsheet integral}\label[appendix]{sec:bosZeval}

Our starting point for the evaluation of the worldsheet integral is~\eqref{eq:fsingleB},
which we reproduce here for convenience:
\begin{equation}
f(\tau, \overline{\tau})
\=
	\frac{4}{i\beta\,(k-2) } \,
	\int\limits_{-\infty}^\infty\, d\zeta\,\zeta\,
	\int_{-\frac{1}{2}}^{\frac{1}{2}} \, d\ts_1\, \int_0^\infty\, d\ts_2 \,
	e^{4\pi\, \left(a_k - \frac{\zeta^2}{k-2}\right) \ts_2 + 2i\,\beta\,\zeta} \, \mathcal{P}_\text{bos}(\zs,\zsb;q,\qb) \,, \\
\end{equation}
\begin{equation}
\mathcal{P}_\text{bos}(\zs,\zsb;q,\qb)
\=
	\frac{Z_X(\zs,\zsb)}{\abs{q^{\frac{1}{2}} - q^{-\frac{1}{2}}}^{2}}
	\;\abs{ \prod_{n=1}^\infty\frac{1-\zs^n}{(1- q\,\zs^n)\,(1- q^{-1} \,\zs^n)}}^2 \,.
 \qquad \qquad \qquad \qquad \qquad
\end{equation}
We would like to expand out the infinite product so that we can isolate terms of the form $\zs^N\, \zsb^{\overline{N}}$, which can then be integrated over.
Focusing on the holomorphic piece, two of the three infinite products, viz., the numerator factor of $1-\zs^n$ and the factor $1-q\, \zs^n$ are straightforward, while the last
the term with $q^{-1}$ (and its conjugate) need a bit more care. Following~\cite{Maldacena:2000kv}, we divide the domain of $\ts_2$ integration into cells
\begin{equation}
[0,\infty) \=  \bigcup_{w=1}^\infty\, \frac{\beta}{2\pi}\,\left[\frac{1}{w+1}, \frac{1}{w} \right] \; \cup \; \frac{\beta}{2\pi}\,[1,\infty)  .
\end{equation}
For a given $w$, we can parameterize the cell as
\begin{equation}
\ts_2 \= \frac{\beta}{2\pi} \left( \frac{1-\alpha}{w+1} + \frac{\alpha}{w}\right) \,, \qquad \alpha \in[0,1]
\end{equation}
and so
\begin{equation}
\abs{q}^{-1}\,\abs{\zs}^n \= \exp\left( \beta \left[1-\frac{n\,(w+\alpha)}{w\,(w+1)}  \right] \right) \ \rightarrow \
\begin{dcases}
	>\; 1, & n \leq w \,,\\
	< \; 1, & n\geq w+1  \,.
\end{dcases}
\end{equation}
Therefore, we expand these terms as
\begin{equation}\label{eq:qinvzexpand}
\frac{1}{1-q^{-1}\, \zs^n} \=
\begin{dcases}
	\ \sum_{j=0}^\infty \, \left( q^{-1}\, \zs^{n}\right)^j, &  n\geq w+1\,,\\
	-\sum_{j=0}^\infty \left( q^{-1}\, \zs^{n}\right)^{-(j+1)} , & n\leq w \,.
\end{dcases}
\end{equation}

Now we need to assemble these expansions of the individual factors
and write the infinite product
\begin{equation}
    g(\zs;q)
\=
	\prod_{n=1}^\infty\frac{1-\zs^n}{(1- q\,\zs^n)\,(1- q^{-1} \,\zs^n)} \,,
\end{equation}
as a series in $\zs$ and $\zsb$ with coefficients which are functions of $q$ and $\qb$, respectively.
Indeed, the discussion of~\cite{Maldacena:2000kv} implicitly assumes such a series in order to proceed. However, naively multiplying an infinite number of the individual expansions~\eqref{eq:qinvzexpand} is not a well-defined procedure.
Instead, we proceed by exploiting an expansion of the inverse theta function due to D.~Zagier,
which allows us to write a double series of the form
\begin{equation} \label{eq:gexp}
    g(\zs;q)
\= \sum_{\substack{n \in \mathbb{Z} \\ m \ge m_{n,w}}} \, A^w_{m,n} \, q^n \, \zs^m \,,
\end{equation}
for a given range
\begin{equation}\label{eq:DZconstraint}
\abs{\zs}^{w+1} < \abs{q} < \abs{\zs}^w \,.
\end{equation}
Here, the sum over~$m$ is bounded below by an integer, possibly negative, depending on~$n, w$.

\bigskip

The formula for the inverse theta function reads as follows~\cite{Zagier:InvTheta}. In this paragraph (only) we use the notation as in~\cite{Eichler:1985ja}, namely~$q=e^{2 \pi i \tau}$
with~$\tau \in \mathbb{H}$ for the modular parameter, and~$\zeta = e^{2 \pi i z}$ with $z \in \mathbb{C} \,  \backslash \, \mathbb{Z} \tau + \mathbb{Z}$ for the elliptic parameter. The convention for the odd Jacobi theta function $\theta \, (= i \vartheta_1)$ is
\begin{equation}
    \theta(z,\tau) \= q^{1/8}\, \zeta^{1/2} \,
    \prod_{n=1}^\infty \, (1-q^n) \, (1-q^n \zeta)\,
    (1-q^{n-1} \zeta^{-1}) \,.
\end{equation}
There are two formulas of our interest.
The first one is the expression of the inverse theta function as a sum over poles,
\begin{equation} \label{eq:DZFormula1}
 \frac{\eta(\tau)^3}{\theta(z,\tau)}   \= \zeta^{\frac{1}{2}}\, \sum_{n\in \mathbb{Z}}\, \frac{(-1)^{n+1}\, q^{n\,(n+1)/2}}{1-\zeta\, q^n} \,.
\end{equation}
The second formula is obtained by expanding each term in the sum~\eqref{eq:DZFormula1} and reassembling the terms. One obtains the following series expansion
for~$w \in \mathbb{Z}$, $z \in \mathbb{C}$, with~$w < \frac{\text{Im}(z)}{\text{Im}(\tau)} < w+1$,
\begin{equation}\label{eq:DZexpand}
\frac{\eta(\tau)^3}{\theta(z,\tau)} \=
\sum_{n\in \mathbb{Z}}\, (-1)^{n}\, F_{n+w+1} \, q^{-\frac{n\,(n+1)}{2}}\, \zeta^{-n-\frac{1}{2}}\,,
\end{equation}
with the coefficients given by the infinite series
\begin{equation}\label{eq:Fsdef}
F_s \= F_s(\tau) \= \sum_{r\geq s}\, (-1)^r\, q^{\frac{r\,(r+1)}{2}}\, \qquad (s \in \mathbb{Z}) \,,
\end{equation}
which is~$\text{O} \bigl( q^{(s^2 + |s|)/2} \bigr)$ as~$q \to 0$.

\bigskip


Upon switching back to the notation of the present paper, we obtain the required double series expansion for~$g(\zs;q)$.
Now we can expand the infinite product in~$\mathcal{P}_\text{bos}$ as a power series in~$q, \qb, \zs, \zsb$ with the ranges of the expansions as above.
It is convenient to collect terms to define the series~$\mathbf{P}_N^w(q)$ as follows,
\begin{equation}\label{eq:gbosexp}
g(\zs;q)
\=
 q^{1+w} \, \sum\limits_{N\in \mathbb{Z}}  \,
	\mathbf{P}_N^w(q)\; \zs^{-\frac{w\,w(+1)}{2}}\, \zs^{N} \,,
\end{equation}
and similarly its complex conjugate, in the range~\eqref{eq:DZconstraint} defined by~$w$.
Folding in the rest of the factors, we finally obtain the desired expansion for $\mathcal{P}_\text{bos}$, viz.,
\begin{equation}\label{eq:Pbosexp}
\begin{split}
   & \mathcal{P}_\text{bos}(\zs,\zsb;q,\qb)
\= \\
& \qquad 	\sum\limits_{w=0}^\infty\, \sum\limits_{N,\overline{N} \in \mathbb{Z}} \sum\limits_{\hx,\hxb} \, \frac{(q\,\qb)^{\frac{1}{2}+w}}{\abs{1-q}^2}  \, D(\hx,\hxb)\,
	\mathbf{P}_N^w(q)\, \mathbf{P}_{\overline{N}}^w(\qb)\; \abs{\zs}^{-w\,w(+1)}\, \zs^{N+\hx} \, \zsb^{\overline{N}+\hxb} \,.
\end{split}
\end{equation}
We now make some comments about this expansion.
\begin{enumerate}[wide,left=0pt]
    \item
In the $w^\mathrm{th}$ cell we have pulled out the factor,  $\abs{q}^{2w}\, \abs{\zs}^{- w(w+1)}$ from the offset in the expansion described in~\eqref{eq:qinvzexpand}, as in~\cite{Maldacena:2000kv}.
This is useful to compare the result from the partition function with the spectrum obtained in canonical quantization of the worldsheet sigma model in~\cite{Maldacena:2000hw}.
\item
As mentioned above, the powers of $\zs$ and $\zsb$ in the above expansion are not necessarily positive, for $w\neq 0$, but are bounded below.
As we see below, the~$\ts_2$ integral in the $w^\mathrm{th}$ cell has a finite range in the middle of the fundamental domain and there is no divergence associated with these negative powers.
\item Despite there being negative powers of $q$ and $\qb$ in the above expansion, it will become clear at the end after performing the worldsheet modular integral that the spacetime CFT conformal dimensions are non-negative.
\item However, we will not demonstrate here that the spectrum respects all the unitarity constraints. Showing that requires that the coefficients in the $q$, $\qb$ are also non-negative integers, which while we believe is true, does not appear obvious to prove from the expressions above.
\end{enumerate}

\bigskip

Armed with~\eqref{eq:Pbosexp} we can perform the integrals over the modular parameters. The $\ts_1$ integral gives the level matching condition,
\begin{equation}\label{eq:levelmatch}
\int_{-\frac{1}{2}}^{\frac{1}{2}} \,d\ts_1\, e^{2\pi i\, \ts_1 (N+ \hx -\overline{N}-\hxb)} \= \delta_{N + \hx,\overline{N}+\hxb}\,.
\end{equation}
Imposing the constraint the $\ts_2$ integral can be done in each cell. Working with the $\alpha$ variable, in the $w^\mathrm{th}$ cell we need to carry out the integral
\begin{equation} \label{eq:NNwrel}
\int_0^1\, d\alpha \,
	\frac{\beta}{2\pi} \left( \frac{1}{w}-\frac{1}{w+1} \right)  \, \exp\left[-2\beta\left(\frac{\zeta^2}{k-2} + \Nwx\right) \left( \frac{1-\alpha}{w+1}+\frac{\alpha}{w} \right)\right],
\end{equation}
which is, of course, straightforward. $\Nwx$ captures the dependence on the level $N$, internal CFT weight $\hx$, and spectral flow parameter $w$, and is defined to be
\begin{equation}\label{eq:NXApp}
\Nwx \= N + \hx -a_k - \frac{1}{2}\, w(w+1) \,.
\end{equation}
Here $a_k$ is the zero-point energy defined in~\eqref{eq:zeroptB}.

The final result as derived in~\cite{Maldacena:2000kv} is then
\begin{equation}\label{eq:fPformula}
\begin{split}
f(\tau,\overline{\tau})
&\=
	\frac{1}{\beta}  \,
	\sum\limits_{N,\overline{N}} \sum\limits_{\hx,\hxb} \,
	\sum\limits_{w=0}^\infty \, \frac{(q\,\qb)^{\frac{1}{2}+w}}{ \abs{1-q}^2 } \,
	\delta_{N + \hx,\overline{N}+\hxb} \, D(\hx,\hxb)\,
	\mathbf{P}^w_N(q)\, \mathbf{P}^w_{\overline{N}}(\qb) \,  \times  \mathfrak{I}_w(q,\qb) \,.
\end{split}
\end{equation}
The function $\mathfrak{I}_w(q,\qb)$ is given as an integral over $\zeta$,
\begin{equation}\label{eq:Iwdef}
\mathfrak{I}_w(q,\qb)
\=
	\int\limits_{-\infty}^\infty\, \frac{d\zeta}{i\pi}  \;
	\frac{ \zeta \,e^{2i\beta \,\zeta}}{ \zeta^2 + (k-2)\, \Nwx}
	\left[ e^{-\frac{2\beta}{w+1} \left(\frac{\zeta^2}{k-2} + \Nwx\right)}  - e^{-\frac{2\beta}{w} \left(\frac{\zeta^2}{k-2} + \Nwx\right)}  \right] .
\end{equation}

The last step is to  explicitly compute $\mathfrak{I}_{w}(q,\qb)$ and extract the contribution in a manner that makes the spacetime CFT interpretation manifest. To do this, we note that the integrand has poles in the $\zeta$ plane, which are located at
\begin{equation}
\zeta_*^\pm  \=  \pm\, i\, \sqrt{(k-2) \, \Nwx} \,.
\end{equation}
It is therefore efficacious to shift the contour up from  the $\Re(\zeta)$ axis. We do so by defining $\zeta = s + i\, (k-2) \frac{w+1}{2} $ or $\zeta =   s + i\,(k-2)\,\frac{w}{2} $, respectively, for the two terms in~\eqref{eq:Iwdef}. This is also suggested by the $\zeta $ dependence in the exponent.

Carrying out this shift, we find that only the poles $\zeta_*^+$ with $\Im(\zeta) \in \frac{k-2}{2}[w,w+1]$ contribute to the difference of the two terms, and that too provided $N+\hx$  satisfies the constraint
\begin{equation}\label{eq:Nbound}
\frac{k-2}{4}\, w^2 \; \leq \; \Nwx \leq
\frac{k-2}{4}\, (w+1)^2 \,.
\end{equation}
Therefore, $\mathfrak{I}_w(q,\qb) $ has a piece which is a sum of terms of the above form, which we can interpret as a localized contribution. In addition, we  still have the integration over $s\in \mathbb{R}$ to perform.
All in all, we can write
\begin{equation}
\begin{split}
\mathfrak{I}_w(q,\qb)
\= \mathfrak{I}_{w,\text{disc}}(q,\qb)  + \mathfrak{I}_{w,\text{cont}}(q,\qb) \,,
\end{split}
\end{equation}
where
\begin{equation}
\mathfrak{I}_{w,\text{disc}}(q,\qb)
\=
	\oint_{\zeta_*^+} \frac{d\zeta}{i\pi}\, \frac{ \zeta \,e^{2i\beta \,\zeta}}{\zeta^2 + (k-2)\,\Nwx}
	\left[
    e^{-\frac{2\beta}{w+1} \left(\frac{\zeta^2}{k-2} + \Nwx\right)}
    - e^{-\frac{2\beta}{w} \left(\frac{\zeta^2}{k-2} + \Nwx\right)} \right] .
\end{equation}

From the spacetime CFT perspective, the residues from the poles comprise the  discrete part of the spectrum. The residue at these poles is readily computed to be
\begin{equation}
(q\,\qb)^{\sqrt{(k-2) \, \Nwx}} \,.
\end{equation}
They capture the discrete series representation
$f_\text{disc}$ reported in~\eqref{eq:bosshort}. For simplicity, we are also including the contribution from the tachyon $f_\text{tachyon}$ here.

While the powers of $q$ in $\mathbf{P}^w_N(q)$ are not necessarily positive, in the final discrete series spectrum we can show that only non-negative spacetime conformal dimensions appear.
To do this straightforwardly, it is useful to go back to~\eqref{eq:DZexpand}, focus on the coefficient of $q^{n+\frac{1}{2}}$, with $n\in \mathbb{Z}$ and ask what the additional contribution to the conformal weight from the worldsheet modular integral is.
This way of reorganizing the calculation does not distinguish primaries from descendants in the spacetime CFT, but that is immaterial if we wish to show that all conformal weights are non-negative.

Following similar steps as above, one finds the discrete part of~\eqref{eq:fPformula} can be written as a sum of terms of the following form,
\begin{equation}\label{eq:qexpalternate}
\begin{split}
 &\sum_{m,\overline{m}=0}^\infty\, \sum_{\hx,\hxb}\, \sum_{n,\overline{n} \in \mathbb{Z}} \, \sum_{\substack{r\; \geq\; n + w +1 \\ \overline{r} \;\geq \; \overline{n}+w+1}}\,   (-1)^{r+ n + \overline{r} + \overline{n}} \, D(\hx,\hxb)\, p_m\, p_{\overline{m}}\,\\
 &\qquad \qquad  \times q^{n+\frac{1}{2} }\, \qb^{\overline{n}+\frac{1}{2} } (q\,\qb)^{\sqrt{(k-2) \left(m + \frac{r\,(r+1)}{2} - \frac{n\,(n+1)}{2} + \hx -a_k\right)}} \,,
\end{split}
\end{equation}
where $r\geq n+w+1$ and $n\in \mathbb{Z}$ are the summation variables introduced in~\eqref{eq:DZexpand}, and~$p_m$, $m \in \mathbb{Z}_{\geq0}$ is the coefficient of the inverse Dedekind eta function (and similarly  for the right-movers). Essentially, we have replaced (likewise for $\overline{N}$)
\begin{equation}
    N - \frac{1}{2} \, w\, (w+1) \rightarrow m + \frac{r\,(r+1)}{2} - \frac{n\,(n+1)}{2} \,.
\end{equation}

The full spectrum consists of terms of the form $\mathbf{P}^w_N(q)\, \mathbf{P}^w_{\overline{N}}(\qb) \, (q\,\qb)^{\frac{1}{2} +w + \sqrt{(k-2) \, \Nwx}}$.
The key point is that in a particular spectral flowed sector parameterized by~$w$, only a finite range of $N$'s
(related to~$\Nwx$ by~\eqref{eq:NNwrel}) contribute, owing to the constraint~\eqref{eq:Nbound}. Level matching as before correlates the holomorphic and anti-holomorphic parts, specifically
\begin{equation}
  m + \frac{r\,(r+1)}{2} - \frac{n\,(n+1)}{2} + \hx
  \= \overline{m} + \frac{\overline{r}\,(\overline{r}+1)}{2} - \frac{\overline{n}\,(\overline{n}+1)}{2} + \hxb\,.
\end{equation}

Now, while a priori $n, \overline{n}\in\mathbb{Z}$ can be negative, the constraint~\eqref{eq:Nbound} in the $w^{\mathrm{th}}$ cell
implies that too negative values of $n$ are forbidden from appearing. In particular, taking $m=\hx =0$, which gives the
most stringent constraint, and using $r\geq n+w + 1$, we find~$n$ to be bounded from below by
\begin{equation} \label{eq:nbound}
    n \; \geq \; \frac{1}{w+1} \left(\frac{k-2}{4}\, w^2 + 1 -  \frac{1}{4\,(k-2)}\right) -1 - \frac{w}{2}\,.
\end{equation}
The spacetime CFT conformal weights $\hs$, i.e.~the exponents of~$q$, $\qb$ in~\eqref{eq:qexpalternate}, are themselves bounded from below, since the expression under the square root
in~\eqref{eq:qexpalternate} is also constrained by~\eqref{eq:Nbound}. This implies
\begin{equation}
\hs \= n + \frac{1}{2} + \sqrt{(k-2) \, \Nwx}
\; \geq \;  n + \frac{1}{2} + \frac{k-2}{2}\, w \; \geq
0\,.
\end{equation}
To see the last inequality we note that the combination appearing in $\hs$ is constrained by~\eqref{eq:nbound} as
\begin{equation}
 n + \frac{1}{2} + \frac{k-2}{2}\, w \geq    \frac{1}{4} \left(\frac{k^2-5}{(k-2)\, (w+1)}+(3\, k-8) \,w- k\right),
\end{equation}
which is non-negative for $k\geq 3$, $w \in \mathbb{Z}_{\geq 0}$. In fact, for large~$k$, the above inequality shows that~$\hs \sim \order{k}$.
Finally, to show unitarity we need to also demonstrate that the coefficients in~\eqref{eq:qexpalternate} are non-negative, which we, however, do not undertake here.

\bigskip

The rest captured in $\mathfrak{I}_{w,\text{cont}}$ is a continuum integral over $s\in \mathbb{R}$
\begin{equation}
\begin{split}
\mathfrak{I}_{w,\text{cont}}(q,\qb)
 &\=
 	 (q\,\qb)^{\frac{\Nwx}{w+1} + \frac{k-2}{4}\,(w+1)}
		\int\limits_{-\infty}^\infty\, \frac{ds}{i\pi}
		\frac{\left(s+ \frac{i\,(k-2)\,(w+1)}{2} \right) (q\,\qb)^{ \frac{s^2}{(k-2)\,(w+1)}} }{ \left(s+\frac{i\,(k-2)\,(w+1)}{2} \right)^2 +\Nwx } \\
&\qquad \qquad
	-(q\,\qb)^{\frac{\Nwx }{w}+ \frac{k-2}{4}\,w}
		\int\limits_{-\infty}^\infty\, \frac{ds}{i\pi}
		\frac{\left(s+ \frac{i\,(k-2)\,w}{2}\right) (q\,\qb)^{\frac{s^2}{(k-2)\,w}} }{ \left(s+\frac{i\,(k-2)\,w}{2}  \right)^2 +   \Nwx }
		\,.
\end{split}
\end{equation}
This gives rise to a continuum of states. This contribution is best dealt with  by considering the sum over the spectral flow parameter $w$, and combining terms in the sum suitably. Consider, therefore,
\begin{equation}
\begin{split}
& \sum_{w=0}^\infty\,  (q\,\qb)^{\frac{1}{2}+w} \,\mathbf{P}^w_N(q)\, \mathbf{P}^w_{\overline{N}}(\qb) \; \mathfrak{I}_{w,\text{cont}}(q,\qb) \\
& \=
	\sum_{w=1}^\infty \,
	\int\limits_{-\infty}^\infty\, \frac{ds}{i\pi} \, (q\,\qb)^{E(s) } \, \left(s+\frac{i\,(k-2)\,w}{2}\right) \left[
		\frac{  \mathbf{P}^{w-1}_N(q)\, \mathbf{P}^{w-1}_{\overline{N}}(\qb) }{ \left(s+\frac{i\,(k-2)\,w}{2}\right)^2 + \mathsf{N}^{^\text{int}}_{w-1} }
    - \, 	\frac{\mathbf{P}^w_N(q)\, \mathbf{P}^w_{\overline{N}}(\qb)}{ \left(s+\frac{i\,(k-2)\,w}{2}\right)^2 +\Nwx } \right] ,
\end{split}
\end{equation}
with
\begin{equation}
\begin{split}
E(s)
&\=
    \frac{1}{2} + w +(k-2)\, \frac{w}{4}+ \frac{1}{w}\left(\frac{s^2}{k-2} + \Nwx \right), \\
&\=
    \frac{k}{4}\, w + \frac{1}{w}  \left(\frac{s^2+ \frac{1}{4}}{k-2} + N+ \hx -1\right) .
\end{split}
\end{equation}
To obtain this result, we have combined terms from $\mathfrak{I}_{w-1,\text{cont}}(q,\qb) $ and $\mathfrak{I}_{w,\text{cont}}(q,\qb) $ to isolate the contribution proportional to $(q\,\qb)^{E(s) }$. Let us  define the continuum density of states as
\begin{equation}
\begin{split}
\bm{\varrho}(s;q,\qb)
& \=
    \sum\limits_{N,\overline{N}} \sum\limits_{\hx,\hxb}  \,
	\delta_{N + \hx,\overline{N}+\hxb} \, D(\hx,\hxb)\, \left(s+\frac{i\,(k-2)\,w}{2}\right) \\
&\qquad\qquad \times  \left[
		\frac{  \mathbf{P}^{w-1}_N(q)\, \mathbf{P}^{w-1}_{\overline{N}}(\qb) }{ \left(s+\frac{i\,(k-2)\,w}{2}\right)^2 + \mathsf{N}^{^\text{int}}_{w-1} }
    - \, 	\frac{\mathbf{P}^w_N(q)\, \mathbf{P}^w_{\overline{N}}(\qb)}{ \left(s+\frac{i\,(k-2)\,w}{2}\right)^2 +\Nwx } \right] .
\end{split}
\end{equation}
As described in~\cite{Maldacena:2000kv}, one can  extract an expression for the density $\bm{\varrho}(s;q,\qb)$ after suitably regulating divergences implicit in the above expression.

\section{Superstring partition function}\label[appendix]{sec:superpart}

In this appendix, we describe some salient features in the computation of the superstring partition function.

\subsection{Contributions from the bosonic modes}\label[appendix]{sec:bosSWZW}

The classical action of an $\mathcal{N}=1$ super-WZW model can be described in terms of a group valued superfield $G = g + i\, \overline{\theta}\, \psi + \frac{i}{2}\, \overline{\theta} \,\theta \, F$. Here $\psi$ is the fermionic field, and $F$ the auxiliary field in the multiplet. The super-WZW model action at level $k$ written out in component form after integrating out the auxiliary field takes the form~\cite{DiVecchia:1984nyg}
\begin{equation}\label{eq:SWZW}
    S_{\text{super-WZW}}[G] = S_{\text{WZW}}[g] + i\, \frac{k}{16\pi} \, \int d^2x\, \Tr(\overline{\psi}^\dagger \left(\slashed{\partial} + \sigma_3\, \slashed{\partial} g g^\dagger \right) \psi ) .
\end{equation}
By a chiral rotation
\begin{equation}\label{eq:fermidecouple}
    \chi = g^\dagger \, \frac{1+\sigma_3}{2} \, \psi
    + \frac{1-\sigma_3}{2}\, \psi \, g^\dagger \,,
\end{equation}
one can decouple the fermions.

There are two features here that are worth highlighting. First, when we impose boundary conditions, the non-trivial classical solutions, viz., the winding modes around the thermal circle in $\mathbb{H}_3^+/\mathbb{Z}$, only receive contribution from $S_{\text{WZW}}[g]$ in~\eqref{eq:SWZW}. We will evaluate this below, but note that these classical contributions scale with the level of the algebra, and do not perceive the shift in the level of the currents when we decouple the fermions.
Second, the background holonomies twist the boundary conditions for the fermions, just as they do in the bosonic $\slt$ WZW discussed in~\cref{sec:sl2part}.

Armed with this knowledge, let us now address the classical contribution to the string partition function from the \AdS{3} $\times \, \mathbf{S}^3$ part. We work with the coordinates introduced in~\cref{sec:sl2part} for the \AdS{3} directions. For the $\mathbf{S}^3$
we can similarly introduce coordinates $\{\hat{\xi}, \hat{v}, \hat{\overline{v}}\}$. Their relation to the coordinates $\{\theta,\phi_1, \phi_2\}$ used in~\eqref{eq:ads3s3metric} can be inferred from the parameterization of the $\mathrm{SU}(2)$ group element
\begin{equation}
g_{\mathrm{SU}(2)} =
\begin{pmatrix}
 e^{\hat{\xi}}\, \left(1-\abs{\hat{v}}^2\right) & i\, \hat{v} \\
 i\, \hat{\overline{v}} & e^{-\hat{\xi}}
\end{pmatrix}
=
\begin{pmatrix}
 e^{-i\phi_1}\, \cos\theta & i\, e^{i \phi_2}\,\sin \theta \\
  i\, e^{-i \phi_2}\,\sin \theta &  e^{i\phi_1}\, \cos\theta
\end{pmatrix} \,.
\end{equation}
Turning on the R-symmetry chemical potential as in~\eqref{eq:ads3s3identify} implies,
\begin{equation}\label{eq:sutwist}
  g_{\mathrm{SU}(2)} \sim
  \mqty(e^{\frac{i}{2}\,\rho} & 0 \\ 0 & e^{-\frac{i}{2}\,\rho}) \, g_{\mathrm{SU}(2)}\,
  \mqty(e^{-\frac{i}{2}\,\overline{\rho}} & 0 \\ 0 & e^{\frac{i}{2}\,\overline{\rho}}) \,,
\end{equation}
where
\begin{equation}\label{eq:nurhodef}
    \rho = \beta\,(\nu_2 - \nu_1) \,, \qquad
    \overline{\rho} = \beta\,(\nu_2 + \nu_1) \,.
\end{equation}
The classical Polyakov action then takes the form
\begin{equation}
S = \underbrace{\frac{k}{\pi}\, \int\, d^2 z\,
	\left( \partial \xi \overline{\partial} \xi + \abs{\partial \overline{v} + \overline{v}\, \partial \xi}^2 \right) }_{\slt_k}
 \;+\; \underbrace{\frac{k}{\pi} \,
 \int d^2z \left[ \left( \partial \hat{\xi} \overline{\partial} \hat{\xi} - \abs{\partial \hat{\overline{v}} + \hat{\overline{v}}\, \partial \hat{\xi}}^2 \right)  \right]}_{\sut_k} .
\end{equation}

The identifications in~\eqref{eq:ads3s3identify} can be implemented by splitting the fields into a classical part, and a fluctuating part, exactly as before.  The classical action from the $\slt$ WZW model remains unchanged from the bosonic theory; we again end up with~\eqref{eq:H3sigma}.  Note that the classical action from the thermal winding modes comes with a coefficient of $k$, and not $k+2$. The latter is the shifted level of the bosonic current algebra after we have decoupled the fermions. However, since the fermions do not contribute at the classical level, we should directly use $S_\text{WZW}[g]$ from~\eqref{eq:SWZW}, which the reader can verify has a coupling set by $k$.

Turning to the $\sut_k$ WZW model, we now want to demonstrate that it too has a non-trivial classical contribution. To see this, note that the twisted boundary conditions~\eqref{eq:sutwist} require
\begin{equation}\label{eq:S3cq}
\begin{split}
\hat{\xi}(z, \overline{z})
&\=
	\hat{\xi}_q(z,\overline{z}) -
	 i \,\frac{\Im(\rho)}{4\pi \ts_2}
	\left[ z\, (n\, \tsb -m ) -  \overline{z} \ (n\,\ts - m)\right] ,\\
\hat{v}(z, \overline{z})
&\=
	\hat{v}_q(z,\overline{z}) \, \exp\left( i\frac{\Re(\rho)}{4\pi \ts_2}
	\left[ z\, (n \,\tsb -m ) -  \overline{z} \ (n\,\ts - m)\right] \right) .
\end{split}
\end{equation}
The winding numbers $(n,m)$  are unchanged, since the twisted boundary conditions on $\mathbf{S}^3$ are correlated with those around the thermal circle~\eqref{eq:ads3s3identify}. Again, notice that this action is proportional to $k$, and is not sensitive to the shifted level  (which is $k-2$ for $\sut$) of the bosonic currents after decoupling the fermions.

Once we have implemented the boundary conditions, we can immediately read off the classical contribution from the $\sut$ WZW by comparing against the analogous expression for the $\slt$ case~\eqref{eq:H3sigma}. The classical piece we seek is
\begin{equation}\label{eq:su2class}
\frac{\pi \, k\,\Im(\rho)^2}{\ts_2}\, \abs{n\,\ts-m}^2\,.
\end{equation}
Note that this is rescaled relative to the $\slt$ result by a factor of $4\pi^2$, which originates from the normalization of $\Im(\tau) =\frac{\beta}{2\pi}$. In addition, the fluctuating fields $\hat{v}_q$ have a background holonomy given by
\begin{equation}
    \frac{1}{2\,\ts_2}\, \overline{\rho}_{n,m} \,, \qquad
    \rho_{n,m} = \rho\, (n\,\ts -m )\,.
\end{equation}
These twists lead to the elliptic parameter of in the  $\mathbf{S}^3$ partition function, one hopes. It is important for modular invariance of the worldsheet partition function that the  chemical potential does not involve $\rho$ alone, but is modulated by $n \,\ts -m$, and appears as $\rho_{n,m}$.

Having understood the classical contributions, let us now turn to the functional determinants. The analysis for $\slt_k$ is as reviewed in~\cref{sec:sl2part}. For the bosonic $\sut_k$ WZW model, we simply need some facts about the characters, and modular invariants.

We recall that the  $\sut$ affine algebra characters are (using $\rho$ for the present to denote the $\mathrm{SU}(2)$ chemical potential)
\begin{equation}\label{eq:su2charcacter}
\begin{split}
\chi_{\ell}^{(k)}(\rho,\ts)
&\=
	\frac{\Theta_{2\,\ell+1}^{(k+2)}(\rho,\ts) - \Theta_{-2\,\ell-1}^{(k+2)}(\rho,\ts)}{\Theta_{1}^{(2)}(\rho,\ts)-\Theta_{-1}^{(2)}(\rho,\ts)} \,, \\
&\=
	\frac{i}{\vartheta_1(\rho,\ts)}\,  \sum_{n\in \mathbb{Z}}\, \zs^\frac{\left((k+2) n +\ell+ \frac{1}{2}\right)^2}{k+2}\left[ r^{n\, (k+2) +\ell+ \frac{1}{2}}
		- r^{-n\, (k+2) -\ell- \frac{1}{2}} \right] .
\end{split}
\end{equation}
These characters have the following modular properties:
\begin{equation}\label{eq:su2TSmod}
\begin{split}
\chi_{\ell}^{(k)}(\rho,\ts + 1)
&\=
	e^{2\pi \,i \left( \frac{\ell\,(\ell+1)}{k+2}- \frac{k}{8\,(k+2)} \right)} \, \chi_\ell^{(k)}(\rho,\ts ) \\
\chi_\ell^{(k)}\left(\frac{\rho}{\ts},-\frac{1}{\ts} \right)
&\=
	\sqrt{\frac{2}{k+2}}\, e^{ \frac{i\,\pi\,k}{2}\, \frac{\rho^2}{\ts}} \,
	\sum_{j=0}^{\frac{k}{2}} \, \sin(\frac{\pi\, (2\,\ell+1)\,(2\,j+1)}{k+2})\, \chi_{j}^{(k)}(\rho,\ts ) .
\end{split}
\end{equation}
The modular S-matrix
\begin{equation}
S_{\ell j} \= \sqrt{\frac{2}{k+2}}\, \sin(\frac{\pi\, (2\,\ell+1)\,(2\,j+1)}{k+2})
\end{equation}
is unitary $S\, S^\dagger = 1$, so it follows that the S-transform of combined left and right movers can be made modular invariant. The simplest one is the diagonal modular invariant (the A-series):
\begin{equation}\label{eq:sutdiag}
Z(\rho,\ts) \= e^{ -\pi\,k \, \frac{\Im(\rho)^2}{\ts_2} } \,
\sum_{\ell=0}^{\frac{k}{2}}\; \chi_\ell^{(k)}(\rho,\ts )\, \overline{\chi}_\ell^{(k)}(\overline{\rho},\overline{\ts} )  \,.
\end{equation}

The exponential prefactor in~$Z(\rho,\ts)$ is present to ensure its invariance  under the modular group as well as under elliptic transformations~$\rho \to \rho + m \ts + n$, $m,m \in \mathbb{Z}$.
For elliptic transformations, the invariance under~$\rho \to \rho +1$ is trivial and the invariance under~$\rho \to \rho + \ts$ is easy to check. For modular transformations, the invariance under~$\ts\to \ts+1$ is trivial.
Under~$(\ts,\rho) \to (-1/\ts,\rho/\ts)$, it is clear from the transformations~\eqref{eq:su2TSmod} that the
sum over~$\chi_\ell^{(k)}$ remains invariant up to the prefactor.
To see the invariance of the prefactor, it is useful to write~$\rho=\rho_1+i \rho_2$, $\ts=\ts_1+i\ts_2$. The factor~$\rho_2 \to (\rho_2 \ts_1-\rho_1 \ts_2)/|\ts|^2$, so that the exponent of the prefactor in~\eqref{eq:sutdiag} transforms according to
\begin{equation}
   \frac{\rho_2^2}{\ts_2} \; \to \; \frac{1}{|\ts|^2}\frac{(\rho_2 \,\ts_1-\rho_1\,\ts_2)^2}{\ts_2}
   \= \frac{\rho_1^2 \,\ts_2 -2\, \rho_1 \, \rho_2 \, \ts_1 +\rho_2^2 \, \ts_1^2/\ts_2}{|\ts|^2} \,.
\end{equation}
Including the prefactor~$e^{i\,\frac{\pi}{2}\,k\, \frac{\rho^2}{\ts}}$ in the transformation~\eqref{eq:su2TSmod} of~$\chi_\ell^{(k)}$, we see that the overall prefactor in~$Z(\rho,\ts)$ transforms to
\begin{equation}
    - \pi\, k \, \frac{\rho_2^2}{|\ts|^2}  \, \biggl(\frac{\ts_1^2}{\ts_2} + \ts_2 \biggr)
    \= - \pi\, k \, \frac{\rho_2^2}{\ts_2} \,,
\end{equation}
so that, indeed, $Z(\rho,\ts)$ is invariant.

Given this information, we can now write down the bosonic $\sut$ WZW model contribution, by putting together the classical piece~\eqref{eq:su2class} and the modular invariant~\eqref{eq:sutdiag}, with an elliptic parameter $\rho_{n,m}$ for the latter to correctly account for the twisted boundary conditions.
This is the result quoted in~\eqref{eq:Zsusys3}.

\subsection{Evaluating the worldsheet integral}\label[appendix]{sec:ssIosc}

In evaluating the worldsheet integral, we need an expansion of $I_{_\text{oscillators}}$ given in~\eqref{eq:Iosc}.
To do so, we write the infinite product as follows
\begin{equation}
I_{_\text{oscillators}}
\= \prod_{n=1}^\infty\, \frac{(1-\zs^n) }{(1-q\, \zs^n)\, (1-q^{-1}\, \zs^n) } \, \frac{\mathfrak{B}_n(\zs;q,r)}{1-\zs^n}\,,
\end{equation}
where we introduced
\begin{equation}
\mathfrak{B}_n (\zs;q,r)
\= \frac{(1-(qr)^\frac{1}{2}\, \zs^n)^2 \,(1-(qr)^{-\frac{1}{2}}\,\zs^n)^2\,
	(1-(\frac{q}{r})^\frac{1}{2}\, \zs^n)^2\, (1- (\frac{q}{r})^{-\frac{1}{2}}\, \zs^n)^2}{(1-\zs)^4\,  (1-r\,\zs^n) \, (1-r^{-1}\,\zs^n)} \,.
\end{equation}
By virtue of the symmetry under $r\to r^{-1}$ in the expression for $\mathfrak{B}_n$, it can be expressed in terms of $\sut$ characters.
For instance, the numerator and denominator separately can be expressed as
\begin{equation}
\begin{split}
\text{Num}(\mathfrak{B}_n)
&=
	1- 2\,(q^{\frac{1}{2}} + q^{-\frac{1}{2}})\,\chi_{\frac{1}{2}}(r) \, (\zs^n + \zs^{7n}) + \left[4 (1+ \chi_1(r) ) + (3+\chi_1(r))(q+q^{-1}) \right] (\zs^{2n} + \zs^{6n}) \\
&\qquad
	-2 \left[ (q^{\frac{3}{2}} + q^{-\frac{3}{2}})\,\chi_{\frac{1}{2}}(r) + 2\,(q^{\frac{1}{2}} + q^{-\frac{1}{2}})\,(\chi_{\frac{3}{2}}(r) + 4\, \chi_{\frac{1}{2}}(r)) \right]\, (\zs^{3n}+\zs^{5n}) \\
&\qquad
	+  \left[ q^2 + q^{-2} + 4\,(q+q^{-1}) \, (1+\chi_1(r)) + (\chi_2(r) + 7\,\chi_1(r)+10)\right]\zs^{4n} + \zs^{8n} \\
\text{Den}(\mathfrak{B}_n)
&=
	1- (3+ \chi_1(r)) (\zs^n + \zs^{5n}) + (4\,\chi_1(r) +3 ) \,(\zs^{2n} + \zs^{4n}) - 2\, (1+3\,\chi_1(r))\,\zs^{3n}  + \zs^{6n} \,.
\end{split}
\end{equation}
%
We can therefore write
\begin{equation}
\prod_{n=1}^\infty \, \frac{\mathfrak{B}_n(\zs;q,r)}{1-\zs^n} \= \sum_{m=0}^\infty\, \mathfrak{d}_m(q,r)\, \zs^m\,,
\end{equation}
where the degeneracies~$\mathfrak{d}_m(q,r)$ are nicely organized into $\sut$ representations.

Now we combine this with the bosonic string results and write
\begin{equation}\label{eq:IoscQ}
\begin{split}
I_{_\text{oscillators}}
&\=
    \sum_{w=0}^\infty\,q^w\, \zs^{-\frac{1}{2}\, w\,(w+1)} \sum_N\, \mathbf{Q}_N^w(q,r)\, \zs^N\,,
\end{split}
\end{equation}
with
\begin{equation}\label{eq:Qoscdef}
    \mathbf{Q}_N^w(q,r) = \sum_{m=0}^\infty\, \mathfrak{d}_m(q,r)\, \mathbf{P}_{N-m}^w(q) \,,
\end{equation}
where, $\mathbf{P}_{N}^w(q)$ being given as before in~\eqref{eq:gbosexp}.

Therefore, $I_{_\text{oscillators}}$ has a series expansion in $\zs^n$ with the coefficients $\mathbf{Q}_{N}^w (q,r)$  being polynomials in $q$ with coefficients that are $\sut$ characters. For example,
\begin{equation}
\mathbf{Q}_{0}^0  (q,r) = 1 \,, \qquad
\mathbf{Q}_{1}^0  (q,r) = q + \frac{1}{q} + \chi_1(r) - 2 \left( q^{\frac{1}{2}} + q^{-\frac{1}{2}} \right)\, \chi_{\frac{1}{2}}(r) \,.
\end{equation}
With this decomposition of the oscillator contribution, we can address the evaluation of the string partition function. The calculation proceeds as in the bosonic case, with minor changes.


\providecommand{\href}[2]{#2}\begingroup\raggedright\endgroup

\end{document}